\documentclass[journal, onecolumn]{IEEEtran}
\usepackage{cite}
\usepackage[cmex10]{amsmath}
\usepackage{algorithmic}
\usepackage{cases}
\usepackage{array}
\usepackage{amssymb}
\usepackage{amsfonts, amscd, amsthm, xspace}
\usepackage{mathrsfs}
\usepackage{txfonts}
\allowdisplaybreaks
\usepackage{verbatim}
\usepackage{color,xcolor}
\usepackage[pdftex]{graphicx}
\usepackage{epstopdf}
\usepackage{tcolorbox}
\usepackage{subfigure}
\usepackage[font={small,it}]{caption}
\usepackage{microtype}
	\DisableLigatures[f]{encoding=T1}
\usepackage[T1]{fontenc}
\usepackage{url}
\usepackage{bbm}
\newtheorem{lemma}{Lemma}
\newtheorem{definition}[lemma]{Definition}
\newtheorem{theorem}[lemma]{Theorem}

\newtheorem{claim}[lemma]{Claim}

\newtheorem{remark}{Remark}

\newcommand{\upperRomannumeral}[1]{\uppercase\expandafter{\romannumeral#1}}

\usepackage{lipsum}

\newcommand\blfootnote[1]{%
  \begingroup
  \renewcommand\thefootnote{}\footnote{#1}%
  \addtocounter{footnote}{-1}%
  \endgroup
}

\newcommand{\C}{\mathbf{C}} 
\newcommand{\Ci}{\mathbf{C}^{(i)}} 
\newcommand{\cc}{C} 
\newcommand{\ci}{C^{(i)}} 
\newcommand{\T}{T} 
\newcommand{\That}{\hat{T}} 
\newcommand{\N}{N}
\newcommand{\B}{B}
\newcommand{\F}{\mathcal{F}}
\newcommand{\TT}{\mathcal{T}}
\newcommand{\M}{M}
\newcommand{\W}{W}
\newcommand{\wi}{w^{(i)}}
\newcommand{\Wi}{w^{(i)}}
\newcommand{\Mhat}{\hat{M}}
\newcommand{\What}{\hat{W}}

\newcommand{\LL}{l_1}

\newcommand{\n}{n}
\newcommand{\npri}{B} 
\newcommand{\pb}{p} 
\newcommand{\pw}{q} 
\newcommand{\smv}{\underline{\mathbf{w}}_s} 
\newcommand{\pmv}{\underline{\mathbf{w}}_p} 
\newcommand{\mv}{\underline{\mathbf{w}}}
\newcommand{\I}{\mathbb{I}} 
\newcommand{\x}{\underline{\mathbf{x}}}

\newcommand{\X}{\underline{\mathbf{X}}} 
 
\newcommand{\yw}{\underline{\mathbf{z}}}
\newcommand{\Yw}{\underline{\mathbf{Z}}} 
\newcommand{\zw}{\underline{\mathbf{N}}_z} 
\newcommand{\zb}{\underline{\mathbf{N}}_y} 
\newcommand{\Yb}{\underline{\mathbf{Y}}} 
\newcommand{\yb}{\underline{\mathbf{y}}}
\newcommand{\aywi}{\mathcal{A}^{\B}_{Z}}
\newcommand{\aywiL}{\left(\mathcal{A}^{\B}_{Z}\right)^{\otimes l_1}}
\newcommand{\tywi}{\mathcal{T}^{\B}_{Z}}

\newcommand{\aaybi}{\mathcal{A}^{\B,(0)}_{Y}}

\newcommand{\aybi}{\mathcal{A}^{\B}_{Y}}
\newcommand{\axywi}{\mathcal{A}^{\B}_{X| \underline{\mathbf{z}}^{(i)}}}
\newcommand{\axywiL}{\left(\mathcal{F}^{xz}_{B}\right)^{\otimes l_1}}
\newcommand{\axYwi}{\mathcal{A}^{\B}_{X| \underline{\mathbf{Z}}^{(i)}}}
\newcommand{\axybi}{\mathcal{A}^{\B}_{X | \underline{\mathbf{y}}^{(i)}}}

\newcommand{\axYbi}{\mathcal{A}^{\B}_{X | \underline{\mathbf{Y}}^{(i)}}}

\newcommand{\txywi}{\mathcal{T}^{\B}_{X | \underline{\mathbf{z}}^{(i)}}}

\newcommand{\txybi}{\mathcal{T}^{\B}_{X | \underline{\mathbf{y}}^{(i)}}}

\newcommand{\ixyw}{\I(\underline{\mathbf{x}}^{(i)}; \underline{\mathbf{z}}^{(i)})}

\newcommand{\pywi}{P_1^{(i)}(\underline{\mathbf{z}}^{(i)})}

\newcommand{\pywxi}{P(\underline{\mathbf{z}}^{(i)} | \underline{\mathbf{x}}^{(i)}_w)}

\newcommand{\ru}{r_u}
\newcommand{\rin}{\hat{r}}
\newcommand{\ez}[1]{{\color{red}#1}}

\begin{document}

\title{Computationally Efficient Covert Communication}

\author{
	   \IEEEauthorblockN{Qiaosheng (Eric) Zhang\IEEEauthorrefmark{1},
                     Mayank Bakshi\IEEEauthorrefmark{2},
	                     Sidharth Jaggi\IEEEauthorrefmark{1}}
                     
	   \IEEEauthorblockA{\IEEEauthorrefmark{1}%
	                     Department of Information Engineering,
	                     The Chinese University of Hong Kong,
	                     \{zq015, jaggi\}@ie.cuhk.edu.hk}
	   
	   \IEEEauthorblockA{\IEEEauthorrefmark{2}%
	   	Institute of Network Coding,
	                     The Chinese University of Hong Kong,
	                     mayank@inc.cuhk.edu.hk}
 }

\begin{comment}
\author{\ \ \ \ Qiaosheng Zhang\qquad\qquad \ Mayank Bakshi\qquad\qquad \ Sidharth Jaggi\qquad\qquad \vspace{0.5em} \\  zq015@ie.cuhk.edu.hk \ \ mayank@inc.cuhk.edu.hk \ \ jaggi@ie.cuhk.edu.hk \vspace{0.7em}\\  Institute of Network Coding, Chinese University of Hong Kong}

%\title{Computationally Efficient Deniable Communication}
%\author{Qiaosheng Zhang \\ \href{mailto:me@somewhere.com}{me@somewhere.com} 
   %\and Mayank Bakshi \\  \href{mailto:someone@somewhere.com}{someone@somewhere.com} 
   %\and Sidharth Jaggi \\ \href{mailto:someone@somewhere.com}{someone@somewhere.com}}

\end{comment}
\maketitle

%\IEEEpeerreviewmaketitle

\begin{abstract}

\textbf{In this paper, we design the first \emph{computationally efficient} codes for simultaneously \emph{reliable} and \emph{covert} communication over Binary Symmetric Channels (BSCs). Our setting is as follows --- a transmitter Alice wishes to potentially reliably transmit a message to a receiver Bob, while ensuring that the transmission taking place is covert with respect to an eavesdropper Willie (who hears Alice's transmission over a \emph{noisier} BSC). Prior works show that Alice can reliably and covertly transmit $\mathcal{O}(\sqrt{n})$ bits over $n$ channel uses without any shared secret between Alice and Bob. One drawback of prior works is that the computational complexity of the codes designed scales as $2^{\Theta(\sqrt{\n})}$. In this work we provide the first computationally tractable codes with provable guarantees on both reliability and covertness, while simultaneously achieving the best known throughput for the problem. }
\end{abstract}

\blfootnote{The work of Qiaosheng Zhang, Mayank Bakshi and Sidharth Jaggi described in this paper was partially supported by a grant from University Grants Committee of the Hong Kong Special Administrative Region, China (Project No. AoE/E-02/08).}
\blfootnote{A preliminary version of this work~\cite{zhang2016computationally} was presented at the 2016 IEEE International Symposium on Information Theory (ISIT), Barcelona, Spain.}

%%%%%%%%%%%%%%%%%%%%%%%%%%%%%%%%%%%%%%%%%%%%%%%%%%%%%%%%%%%%%%%%%%%%%%%%%%%%%%%%%%%%%%%%%%%%%%%%%%%%%%%%%%%%%%%%%%%%%%%%%%%%%%%%%%%%%%%%%%%%%%%%%%%%%%%%%%%%

\section{Introduction} \label{sec:intro}
Alice {\it may or may not} wish to communicate with a receiver Bob over a Binary Symmetric Channel with crossover probability $p$, denoted by BSC($\pb$). However, an adversary Willie is able to eavesdrop on their communication over a ``noisier'' Binary Symmetric Channel -- BSC($\pw$) (here $\pw$ is strictly larger than\footnote{Note that without this asymmetry, whenever Bob can decode reliably, so can Willie.} $\pb$), and only cares about whether Alice is transmitting or not. Therefore, Alice would like to use a novel communication scheme to prevent her transmission status from being detected by Willie (\emph{covert} with respect to Willie) and also ensure that her messages are received by Bob correctly.\footnote{For ease of exposition, in this work we focus on scenarios in which all channels are BSCs. However, following the lead of~\cite{7407378}, it is likely that these results can be directly generalized to other DMCs. }

We first give an overview of several problems related to our setup. Shannon first defined the concept of \emph{information-theoretic security}~\cite{Sha:49}, which requires the key rate to be as large as the message rate to achieve perfect secrecy. \emph{Kerckhoff's principle}~\cite{Ker:83}, however, states that a system should be secure even if everything about the system, except the key, is public knowledge. Wyner demonstrated that shared secrets can be replaced with asymmetry in channel noise~\cite{wyner_wire-tap_1975,ozarow1984wire} (as in this work). The reader is referred to~\cite{bloch2011physical, liang2009information} for recent surveys on physical-layer security. The classical \emph{steganography} problem, which considers how to hide a undetectable message in plain sight, has been well-studied --- see, for instance,~\cite{cox2007digital} for a survey. Cachin~\cite{cachin1998information} first focused on the problem of information-theoretic steganography, and Maurer\cite{maurer1996unified} drew connections between the problem of steganography and that of {\it hypothesis testing}. In~\cite{WanM:08}, Wang and Moulin gave an information-theoretic characterization of the capacity of the perfectly secure steganography problem (with unbounded-sized shared secrets between Alice and Bob).

We now turn to reliable and covert communication, which is the main focus of this work. Even though, the early literature on this topic used a plethora of terms such as ``covertness'', ``deniability'' and ``low probability of detection (LPD)'' to define essentially the same security requirement, of late, the term ``covertness'' has gained acceptance as the preferred nomenclature. Bash \emph{et al.} gave the first results on information-theoretically guaranteed covert communication over noisy AWGN channels~\cite{BasGT:12a,bash2014lpd,bash2015quantum,BasGT:12}. Noting that the result of Bash {\em et al.} relied critically on the presence of large shared secrets between Alice and Bob\footnote{In fact, the size of the keys required by their scheme is larger than the throughput from Alice to Bob.}. Che \emph{et al.} designed reliable and covert (and information-theoretically secure) communication schemes over BSCs without using {\it any} shared secrets, relying only on the asymmetry of level of channel noise on the two channels~\cite{CheBJ:13, CheBCJ:14a, CheBCJ:14b, CheSBCJA:14}. The work of~\cite{HouK:14} studied covert communication from a {\it channel resolvability} approach, while Wang \emph{et al.}~\cite{wang_limits_2015} and Bloch~\cite{7407378} first derived tight capacity characterizations for discrete memoryless channels (DMCs). We discuss the intuition behind these schemes in greater detail in Section~\ref{sec:intuition} below.

%The work on SNR walls for cognitive radio~\cite{tandra_snr_2008} also has bearing on this model, as do a variety of works that design covert communication schemes~\cite{LeeB:14,LeeBMS:14, deshotels_inaudible_2014,carrara_characterizing_2015,classen_practical_2015, korzhik_existence_2005}, albeit without information theoretic proofs of deniability. 
While the plethora of codes and bounds in the recent literature paint a clear picture of the limits of reliable communication possible between the transmitter Alice and the receiver Bob while remaining covert (or deniable/stealthy/LPD) with respect to the eavesdropper Willie, prior to this work there were still no computationally efficient communication schemes with information-theoretic proofs of covertness. Though a variety of computationally-efficient schemes\cite{LeeB:14,LeeBMS:14, deshotels_inaudible_2014,carrara_characterizing_2015,classen_practical_2015, korzhik_existence_2005} give good heuristics for such communication, they typically lack proofs that the proposed schemes do indeed provide information-theoretic covertness of such detectors that may be employed by the eavesdropper, regardless of the computational complexity. 

In this paper, we present the first coding scheme which has provable throughput and covertness guarantees while ensuring that the computational complexity for both encoding and decoding is at most polynomial in the number of transmitted message bits. Throughout this paper we use \emph{asymptotic notation}~\cite[Ch. 3.1]{cormen2009introduction} to describe the limiting behaviour of functions. 
The rest of this paper is organized as follows. We formally describe our model in Section~\ref{sec:model}. In Section~\ref{sec:result}, we give the main result of this paper, and provide a performance characterization of a specific class of computationally-efficient reliable and covert communication schemes. Section~\ref{sec:code} describes the corresponding codes in greater detail. We introduce the mathematical preliminaries and probability distributions of interest in Sections~\ref{sec:def}. Sections~\ref{sec:deniability} and~\ref{sec:reliability} provide the proofs of covertness and reliability respectively of our codes. Section~\ref{sec:conclusion} concludes this work and proposes several future directions that are worthy exploring.

\section{Intuition}
\label{sec:intuition}
We begin by first giving an intuitive description of our work and place it in the context of prior works.
\subsection{Challenges}
The intuition behind the covert schemes first presented in~\cite{BasGT:12a} and elaborated on in other works such as~\cite{bash2014lpd,BasGGT:13,BasGT:12,CheBCJ:14a,CheSBCJA:14,CheBJ:13,CheBCJ:14b,wang_limits_2015,7407378} is that most reasonable noise processes have, with non-zero probability, ``some deviation'' in the ``noise intensity''. For instance, a length-$n$ Bernoulli($q$) sequence (corresponding to the additive noise sequence in a BSC($q$) -- a {\it Binary Symmetric Channel with crossover probability $q$} -- the channel from the transmitter Alice to the eavesdropper Willie) has expected value $nq$, but has standard deviation $\sqrt{nq(1-q)}$. Hence, if Alice uses a carefully designed codebook containing codewords with low Hamming weight (about ${\cal O}(\sqrt{n})$) then the expected ``power density'' at the eavesdropper (about $nq + {\cal O}(\sqrt{n})$) may reasonably be attributed by Willie to natural variations in the noise-level he observes. Further, it is also known~\cite{CheBJ:13} that to ensure covertness in communication, one {\em must} use codes with very low average Hamming weight ({\em i.e.}, with weights no larger than\footnote{The results of~\cite{CheBCJ:14b} indicate an interesting phenomenon when there is uncertainty about the {\it level of noise of the channel}, and the coherence time is ``long'' -- then, in fact, the throughput can be shown to scale linearly with the number of channel uses, rather than as $\sqrt{n}$.} ${\cal O}(\sqrt{n})$). This restriction on codeword weights, along with the requirement that Bob be able to reliably decode, implies that the optimal reliable throughput from Alice to Bob that is simultaneously covert with respect to Willie scales only as a factor of $\sqrt{n}$, rather than linearly in the number of channel uses (as is the common paradigm in Shannon theory). Hence the capacity of such covert communication schemes converges to zero! The interesting ``first-order'' question, therefore, is how many bits can be communicated reliably (to Bob) and covertly (with respect to Willie) as a function of the square-root of the number of channel uses.

However, just choosing a codebook with low average Hamming weight does not suffice to guarantee covertness. For instance, suppose Alice chooses a codebook containing length-$n$ binary vectors such that about half of the first $\sqrt{n}$ locations are non-zero, but {\it all} the succeeding $n-\sqrt{n}$ bits in each codeword are zero. While such a codebook would satisfy the low average Hamming weight requirement, it is nonetheless still easy for Willie to detect whether or not Alice is transmitting in such a scenario. If Alice is silent, he would expect to see about $q\sqrt{n}$ non-zero values in the first $\sqrt{n}$ locations of his observation (with a standard deviation of about ${\cal O}( n^{1/4} )$), whereas if Alice were transmitting a non-zero codeword, he would expect to see about $\sqrt{n}/2$ non-zero values in the same locations (again with a standard deviation of about ${\cal O}( n^{1/4})$). By relatively standard analysis from the hypothesis-testing literature~\cite{cover2012elements}, it can be shown his estimate of Alice's transmission status would be correct with high probability (over the noise in the channel to him). Hence one needs ``good spreading'' of the bits in the codewords as well -- not all codewords can have their support concentrated in the same small set of locations.\footnote{Indeed, this is the intuition in some recent heuristic approaches~\cite{LeeB:14,LeeBMS:14, deshotels_inaudible_2014,carrara_characterizing_2015,classen_practical_2015, korzhik_existence_2005} to designing covert communication schemes -- codes designed via ``spread spectrum'' techniques are analyzed. However, an information-theoretically rigorous proof of the covertness of such schemes is lacking.}

While the above serves as good intuition for constructing covert communication schemes, providing mathematical guarantees for a given code can be extremely challenging -- one has to prove that two different probability distributions supported on an exponentially large set are ``very close''. Specifically, one distribution, denoted by $P_0$, corresponds to the scenario when Alice is silent, and corresponds to a Binomial($n,q$) distribution. The other, denoted by $P_1$, corresponds to the scenario when Alice is transmitting using some code ${\cal C}$. Both these distributions are supported on the set (of exponential size in the block-length $n$) of possible observations seen by the eavesdropper Willie. Since the structure of $P_1$ depends intimately on the structure of ${\cal C}$, characterizing the difference between $P_0$ and $P_1$ for any specific code, or specific ensembles of codes, can be quite complicated.

A second challenge is due to the fact that most computationally efficient code designs in the literature (see, for example,~\cite{gallager1962low, arikan2009channel}) naturally lead to codes such that the average Hamming weight of codewords in the code is tightly concentrated around half the block-length, $n/2$.  As noted above, simply designing codes of block-length about ${\cal O}(\sqrt{n})$ and embedding the codewords into a pre-specified and publicly known set of about ${\cal O}(\sqrt{n})$ locations in length-$n$ vectors padded with $0$s {\it also} does not work. To the best of our knowledge, prior to this work there were no binary \emph{constant composition codes}~\cite{CsiszarK:11} with such low Hamming weight, with good spreading properties, that enable communication at rates close to the optimal rates characterized in~\cite{CheBJ:13,wang_limits_2015,7407378}, and that are simultaneously computationally-efficient to encode and decode.
\subsection{Our approach}
Our approach is to use {\it concatenated-style} codes, that are inspired by Forney's classical work~\cite{forney1966concatenated} that gave the first computationally-efficient codes for arbitrary channels that also approached capacity. Forney noticed that since the computational cost of Shannon's random codes is exponential in the blocklength $n$, dividing the message into ${\Theta}(\log{n})$-sized chunks and applying Shannon's codes on each chunk would ensure that the overall complexity is only polynomial in the total blocklength, while still operating at rates close to the channel capacity. However, na\"{i}vely applying this ``divide-and-conquer'' idea would lead to an overall high decoding error probability owing to the small blocklength (and hence, relatively large decoding error probability) for each chunk and the large number of chunks. In order to overcome this, Forney's solution was to combine the ``inner code'' provided by Shannon with an ``outer code''. The purpose of the outer code -- typically a Reed-Solomon (RS) code -- is to computationally efficiently correct any chunks that are in error by paying a negligible rate penalty. 

We follow Forney's lead, but adapt our construction to the constraints imposed by covertness. Foremost, while Forney's construction operates with ${\Theta}(n)$ message bits, in our setting, at most ${\cal O}(\sqrt{n})$ bits of reliable transmission are possible. Thus, to ensure that each chunk contains ${ \Theta}(\log{n})$ message bits, the blocklength for each chunk is $\Theta(\sqrt{n}\log{n})$. First, we encode using an RS outer code to create ``coded-chunks'' from the message chunks. Next, we encode each chunk by using an independently drawn ensemble of low-weight random codes~\cite{CheBJ:13} that has the property that the expected codeword weight for each chunk is $\Theta(\log{n})$. 

With the above concatenated construction, the reliability analysis proceeds along familiar lines (with some parameter tweaks). Proving covertness, perhaps not surprisingly, turns out to be much more challenging. The first complication is imposed by the outer code -- the ensemble of codes that our construction generates has linear dependencies between the chunks. This breaks the analysis from~\cite{CheBJ:13} that critically relies on each bit of the codewords being generated independently. It is conceivable that since the code is known to Willie, he may  test for these dependencies and be able to come up with clever estimators of the transmission status. To overcome this problem, we use a systematic Reed-Solomon code. This decomposition of the chunks into systematic chunks and parity chunks is helpful in two ways. Firstly, this ensures that, at the very least, the systematic chunks are independently generated (since these correspond to independent message bits). Secondly, this also lets us show that, from Willie's perspective the conditional distribution of transmissions in the parity chunks (of the Reed-Solomon outer code) is essentially statistically independent of Willie's observations of transmissions in the systematic chunks, thus preventing him from gaining any advantage in estimating Alice's transmission status by using the dependencies. 

A second, and more technical, challenge is to prove that with high probability, the code for each chunk is covert. In prior works such as~\cite{CheBJ:13}, this is proved by first showing that under the ensemble-averaged distribution, the codebook is covert and then using a concentration argument over to show that with high probability over the codebook generation, the distribution imposed by the actual codebook is close to the ensemble average. Our concatenated code, however, only contains a polynomially small number of codewords in each chunk, since the chunk length scales as $\Theta{(\sqrt{\n}\log{\n})}$. Especially when $\pb$ approaches\footnote{As $\pb$ approaches $\pw$, the chunk length grows accordingly.} $\pw$, we need to provide a more sensitive analysis to ensure polynomially many plausible codewords for Willie in each chunk, but with high probability (\emph{w.h.p.}) only one for Bob. Finally, we need to carefully combine proofs of covertness in each chunk to get covertness for the overall code.

By following this intuition, our work proves that one can communicate reliably and covertly with the best known throughput~\cite{CheBJ:13}, while requiring a computational complexity that is at most polynomial in the blocklength $\n$.

%%%%%%%%%%%%%%%%%%%%%%%%%%%%%%%%%%%%%%%%%%%%%%%%%%%%%%%%%%%%%%%%%%%%%%%%%%%%%%%%%%%%%%%%%%%%%%%%%%%%%%%%%%%%%%%%%%%%%%%%%%%%%%%%%%%%%%%%%%%%%%%%%%%%%%%%%%%%
\vspace{8pt}
\section{Model}\label{sec:model}
Throughout this paper, unless otherwise stated, we use the following conventions. We take all logarithms to be binary and use $\exp(a)$  to represent $e^a$ for $a\in\mathbb{R}$. Random variables are denoted by uppercase letters, while their realizations are denoted by lowercase letters. Sets are denoted by calligraphic letters. Vectors are denoted by underlined boldface letters. The length of each vector will be clear from the context.

\begin{figure}
	\begin{center}
	\includegraphics[scale=0.7]{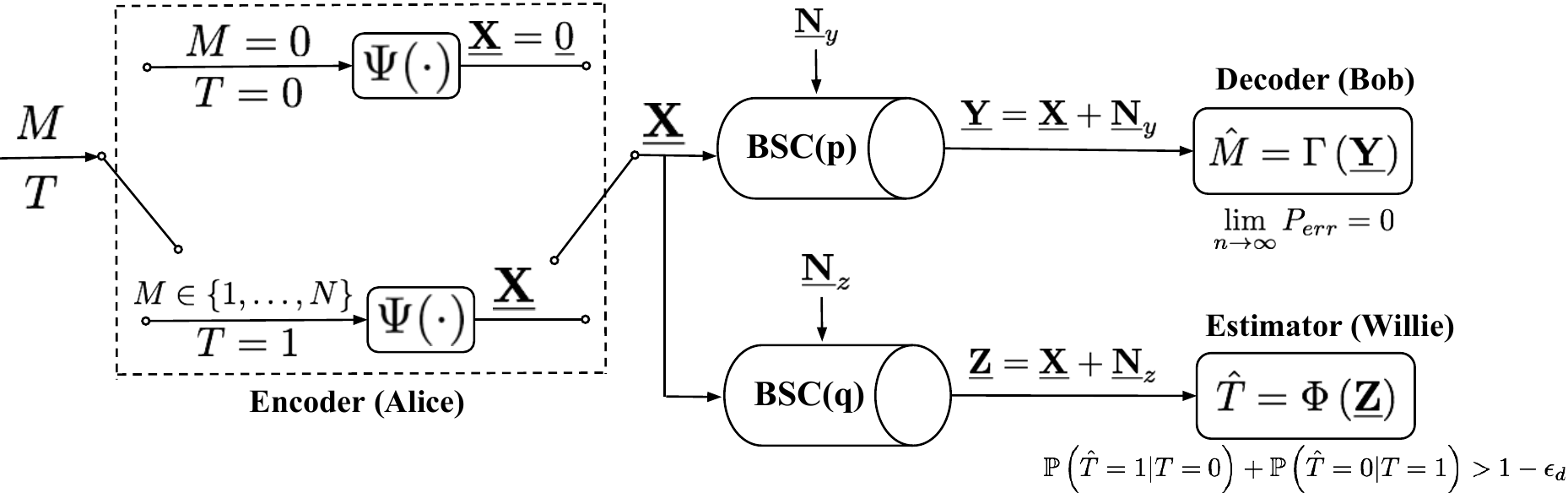}
	\caption{\scriptsize Reliable-Deniable Communication system diagram: Alice has a message $\M$ that can take $N$ values $\left\{1,\ldots,N\right\}$, and the transmission status $\T \in \left\{0,1\right\}$. If Alice's transmission status $\T = 0$, she is required to ``stay silent'' -- transmit the all zero codeword $\underline{0}$ -- this corresponds to the 0 message. On the other hand, if her transmission status $\T=1$, she uses her encoder $\Psi$ to encode her message $\M$ into a codeword $\X$. This $\X$ is broadcast to the legitimate receiver Bob, and the eavesdropper Willie, over a pair of independent Binary Symmetric Channels with respective crossover probabilities $p$ and $q$ (respectively denoted by BSC($p$) and BSC($q$)), which add Bernoulli noise vectors $\zb$ and $\zw$ respectively to $\X$, resulting in the transmissions $\Yb$ and $\Yw$ observed respectively by Bob and Willie. Bob uses a decoder $\Gamma$ to estimate Alice's transmitted message $\M$ as $\hat{\M}$, and wishes to ensure {\it reliability}, {\it i.e.} the probability (over channel noise $\zb$) that $\hat{\M} \neq \M$ is ``small''. As a by-product of his decoder, he should therefore {\it also} detect Alice's transmission status. Willie, on the other hand, {\it only} wishes to decode Alice's transmission status $\T$. A code that is {\it ($1-\epsilon_d$)-covert} ensures that, {\it regardless of Willie's estimator}, the probability (over Alice's message $\M$ and channel noise $\zb$) that $\mathbb{P}(\hat{\T} = 1 | \T = 0) + \mathbb{P}(\hat{\T} =0 | \T = 1) \ge 1-\epsilon_d$.}  \label{fig:system}
	\end{center}
\end{figure}
\noindent{\textbf{Channel model:}} The channel between the transmitter Alice and the legitimate receiver Bob is a BSC($\pb$), and the channel between Alice and the malicious eavesdropper Willie is a BSC($\pw$), where $\pw > \pb$ (note that without this asymmetry, whenever Bob can decode reliably, so can Willie). Alice's transmission status is denoted by $\T \in \left\{0, 1\right\}$ and the message is denoted by $\M \in \left\{0\right\}\cup \left\{1, 2, \ldots, \N \right\}$. When Alice communicates with Bob, her transmission status $\T = 1$ and the transmitted message $\M$ is chosen uniformly at random from $\left\{1, 2, \ldots, \N \right\}$. When Alice does not communicate with Bob, her transmission status $\T = 0$ and the default message $\M = 0$ is transmitted. All three parties know the channel parameters $\pb$ and $\pw$, but only Alice knows the transmission status $\T$ and the message $\M$ \emph{a priori}. Figure~\ref{fig:system} illustrates the system diagram of the communication model. \\
\noindent{\textbf{Encoder:}} Alice's encoder is defined through the encoding function $\Psi(\cdot): \left\{0\right\}\cup \left\{1, 2, \ldots, \N \right\} \rightarrow \left\{0, 1\right\}^{\n},$ that is applied on the message $\M$ to obtain the length-$\n$ binary codeword $\X = \Psi(\M)$. In particular, the innocent message $\M = 0$ will always be encoded to the length-$n$ zero vector, \emph{i.e.}, $\X = \underline{0}$. We define the {\it rate} of the code as $R = (\log{\N})/n$, and the {\it relative throughput} as $r = (\log{\N})/\sqrt{\n}$. It is preferable to use the relative throughput $r$ because when $\n$ goes to infinity, the relative throughput $r$ scales as a constant while the rate $R$ goes to zero.\\
\noindent{\bf Decoder:}  Bob receives the length-$\n$ binary vector $\Yb = \X \oplus \zb$, where $\zb$ is the noise vector induced by the BSC($\pb$), and applies a decoder map $\Gamma(\cdot): \left\{0, 1\right\}^{\n} \rightarrow \left\{0\right\}\cup \left\{1, 2, \ldots, \N \right\}$ to reconstruct the message $\Mhat$ from his observation $\Yb$. The goal is to guarantee the communication is reliable, \emph{i.e.}, the average probability of error satisfies $\lim_{n \to \infty} P_{err} = 0$, where $P_{err} \triangleq \max_{i \in \{0,1\}} \mathbb{P}(\Mhat \ne \M| T=i)$ \\
\noindent{\bf Estimator:} Willie aims to estimate $\T$ from his observation $\Yw = \X \oplus \zw$, where $\zw$ is the noise vector induced by the BSC($\pw$), by using an estimator $\Phi(\cdot): \left\{0, 1\right\}^{\n} \rightarrow \left\{0, 1\right\}$ that outputs the estimate $\That = \Phi(\Yw)$ of the transmission status. We use a \emph{hypothesis-testing metric} to measure the covertness of the communication. Let $\alpha(\Phi) = \mathbb{P}_{\zw}(\That = 1| \T = 0)$ be the probability of \emph{false alarm}, and $\beta(\Phi) = \mathbb{P}_{\M,\zw}(\That = 0| \T = 1)$ be the probability of \emph{missed detection}. The communication is deemed to be ($1-\epsilon_d$)-covert if there does not exist an estimator $\Phi$ such that $\alpha(\Phi) + \beta(\Phi) < 1 - \epsilon_d$. Let $P_0$ be the \emph{innocent distribution} of $\Yw$ when Alice's transmission status $\T = 0$ and $P_1$ be the \emph{active distribution} of $\Yw$ when Alice's transmission status $\T = 1$. By standard statistical arguments~\cite[Theorem~13.1.1]{lehmann2006testing}, an optimal hypothesis test $\Phi^*$ satisfies 
\begin{align}
\alpha(\Phi^*)+\beta(\Phi^*) = 1- \mathbb{V}(P_0,P_1),
\end{align} 
where $\mathbb{V}(P_0,P_1) = \frac{1}{2} \sum_{\yw \in \{0,1\}^{\n}}\left|P_0(\yw)-P_1(\yw) \right|$ denotes the \emph{variational distance}\footnote{Instead of using variational distance, other works (see, for example,~\cite{7407378, wang_limits_2015,HouK:14}) also use \emph{Kullback-Leibler (KL) divergence} to measure the covertness. In the existing literature, people usually give equal weight to the probability of false alarm and the probability of missed detection. Recently, people also propose new metrics, such as the probability of missed detection for fixed probability of false alarm~\cite{tahmasbi2017second}, to model different problem settings (for instance, the eavesdropper wishes to completely prevent missed detection, while is willing to tolerate modest false alarm in military applications). Though we focus on variational distance in this work, it is conceivable that our code construction is also applicable to other metrics. } between $P_0$ and $P_1$. Therefore, to guarantee the communication is ($1-\epsilon_d$)-covert, it suffices to show that $\mathbb{V}(P_0,P_1) \le \epsilon_d$.

%%%%%%%%%%%%%%%%%%%%%%%%%%%%%%%%%%%%%%%%%%%%%%%%%%%%%%%%%%%%%%%%%%%%%%%%%%%%%%%%%%%%%%%%%%%%%%%%%%%%%%%%%%%%%%%%%%%%%%%%%%%%%%%%%%%%%%%%%%%%%%%%%%%%%%%%%%%%
\vskip 0.3cm
\section{Main Result}
\label{sec:result}

Before stating the main theorem (Theorem 1), we need to define a variety of auxiliary functions and variables that will be useful in understanding the throughput/reliability/covertness/complexity tradeoffs in the statement of Theorem 1. We first define 
\begin{equation}
f(x) = \log{e}-(1+x)\log{\left(e/(1+x)\right)}.  \label{eq:a}
\end{equation}
Given any $0 < \pb < \pw < 1/2$ and sufficiently small $\epsilon_d > 0$, we define a \emph{code weight design parameter} 
\begin{equation}
k_2(\pw,\epsilon_d) = 2\epsilon_d \sqrt{\pw (1-\pw)}/(1-2\pw),  \label{eq:b}
\end{equation}
and a \emph{throughput parameter} 
\begin{equation}
\ru(p,q,\epsilon_d) = 2\epsilon_d\sqrt{\pw(1-\pw)}\frac{1-2\pb}{1-2\pw}\log{\left(\frac{1-\pb}{\pb}\right)}. \label{eq:c}
\end{equation}
The value of the code weight design parameter $k_2(\pw,\epsilon_d)$ is chosen to satisfy Equations~\eqref{eq:var1}-\eqref{eq:var3} in Section~\ref{sec:deniability}, and the value of the throughput parameter $\ru(p,q,\epsilon_d)$ is chosen to satisfy Claim~\ref{claim:err2} in Section~\ref{sec:reliability}. We abbreviate $k_2(\pw,\epsilon_d)$ and $r_u(p,q,\epsilon_d)$ as $k_2$ and $r_u$ respectively when the arguments are clear from the context.
Then we define four multivariable functions $g_i(u,v,w,t)$, where $1 \le i \le 4$, as 
\begin{align}
g_1(u,v,w,t) = k_2(u,v)\Bigg[u(1-w)\left(\log{\left(\frac{1-u}{u(1-w)}\right)}+\log{e}\right) + (1-u)(1+t)\left(\log{\left(\frac{u}{(1-u)(1+t)}\right)}+\log{e}\right)  - \log{e} \Bigg], \label{eq:d} \\
g_2(u,v,w,t) = k_2(u,v)\Bigg[u(1+w)\left(\log{\left(\frac{1-u}{u(1+w)}\right)}+\log{e}\right) + (1-u)(1+t)\left(\log{\left(\frac{u}{(1-u)(1+t)}\right)}+\log{e}\right)  - \log{e} \Bigg], \label{eq:d2} \\
g_3(u,v,w,t) = k_2(u,v)\Bigg[u(1-w)\left(\log{\left(\frac{1-u}{u(1-w)}\right)}+\log{e}\right) + (1-u)(1-t)\left(\log{\left(\frac{u}{(1-u)(1-t)}\right)}+\log{e}\right)  - \log{e} \Bigg], \label{eq:d3} \\
g_4(u,v,w,t) = k_2(u,v)\Bigg[u(1+w)\left(\log{\left(\frac{1-u}{u(1+w)}\right)}+\log{e}\right) + (1-u)(1-t)\left(\log{\left(\frac{u}{(1-u)(1-t)}\right)}+\log{e}\right)  - \log{e} \Bigg], \label{eq:d4}
\end{align}
The reason why we define the multivariable functions $g_i(u,v,w,t)$ will be clear in Equation~\eqref{eq:feng6}, Section~\ref{sec:deniability}.
Equipped with the auxiliary tools above, we then define the \emph{code chunk length design} parameter $k_1$
as 
\begin{align}
k_1 = \min_{\Delta_{10}^{xz}, \Delta_{11}^{xz}\in (0,1)} \max_{i \in \{1,2,3\}} \left\{ \frac{\xi_i+\delta}{\Phi_i(r_u,\pw,\epsilon_d,\Delta_{10}^{xz}, \Delta_{11}^{xz})}\right\}, \label{eq:minmax}
\end{align}
where 
\begin{align}
&\Phi_1(r_u,\pw,\epsilon_d,\Delta_{10}^{xz}, \Delta_{11}^{xz}) = \ru + \max_{j\in \{1,2,3,4\}} \left\{ g_j(\pw, \epsilon_d, \Delta_{10}^{xz}, \Delta_{11}^{xz})\right\},  \label{eq:condition1}\\
&\Phi_2(r_u,\pw,\epsilon_d,\Delta_{10}^{xz}, \Delta_{11}^{xz})=\pw \cdot k_2(\pw,\epsilon_d) \cdot f(\Delta_{10}^{xz}), \label{eq:condition2}\\
&\Phi_3(r_u,\pw,\epsilon_d,\Delta_{10}^{xz}, \Delta_{11}^{xz})=(1- \pw)\cdot k_2(\pw,\epsilon_d)\cdot f(\Delta_{11}^{xz}), \label{eq:condition3}\\
& \xi_1 = \frac{3}{2}, \ \xi_2 = \xi_3 = \frac{1}{2}, \label{eq:condition4}
\end{align}
and $\delta$ is a {\it slackness parameter} that trades off the probability that a randomly chosen code is ``good'' with the computational complexity for encoding and decoding. It can be chosen to be any value in the interval ($0,0.5$). For correctness we set $\delta = 0.01$ throughout this work. The parameters $\Delta^{xz}_{10}$ and $\Delta^{xz}_{11}$, to be formally defined in Section~\ref{sec:def-A}, play an critical role in our code design. We elaborate on the reasons why $\Delta^{xz}_{10}$ and $\Delta^{xz}_{11}$ are required to satisfy~\eqref{eq:minmax} in Equations~\eqref{eq:exp5} and~\eqref{eq:exp1}-\eqref{eq:exp4}, Section~\ref{sec:deniability}.

\begin{comment}
\ez{As $n$ grows without bound, the inequalities~\eqref{eq:k11}-\eqref{eq:k14} can be simplified to 
\begingroup\addtolength{\jot}{-0.5em}\begin{align}
%& k^{*}(\pb,\pw,\epsilon_d)&&=  \text{minimize}\ \  k \\
%& \text{subject to}
& x\left(\ru + \max_{j=1}^4 \left\{ g_i(\pw, \epsilon_d, \Delta_{10}^{xz}, \Delta_{11}^{xz})\right\}\right) \ge \frac{3}{2}+\delta \label{eq:k21}\\
& x\pw \cdot k_2(\pw,\epsilon_d) \cdot f(\Delta_{10}^{xz})  \ge \frac{1}{2}+\delta \label{eq:k22}\\
& x(1- \pw) \cdot k_2(\pw,\epsilon_d)\cdot f(\Delta_{11}^{xz}) \ge \frac{1}{2}+\delta \label{eq:k23}\\
& 0 < \Delta_{10}^{xz}, \Delta_{11}^{xz} < 1 \label{eq:k24}.
\end{align}
\endgroup 
Since the objective function is continuous with respect to $x$, the smallest positive values of $x$ satisfying~\eqref{eq:k11}-\eqref{eq:k14} and~\eqref{eq:k21}-\eqref{eq:k24} are equal for sufficiently large $n$. }
\end{comment}

The work of~\cite{CheBJ:13} shows that given $\pb,\pw$ and $\epsilon_d$, one can transmit up to $\ru \sqrt{\n}$ message bits per $\n$ channel uses covertly and reliably, but the decoding complexity as well as the space complexity for storing the codebook are exponential in $\sqrt{\n}$. Our main result, Theorem~\ref{theorem1} below, shows that it is possible to communicate reliably and covertly while reducing the complexity to be polynomial in $\n$, by using a carefully designed concatenated code $\cc_n$ chosen from the concatenated code ensemble $\cc_n^{cc}$ (for notational convenience we drop the subscript $n$ in the following) with relative throughput $\ru (1 - o(1))$. 

\begin{figure}
	\begin{center}
	\includegraphics[scale=0.65]{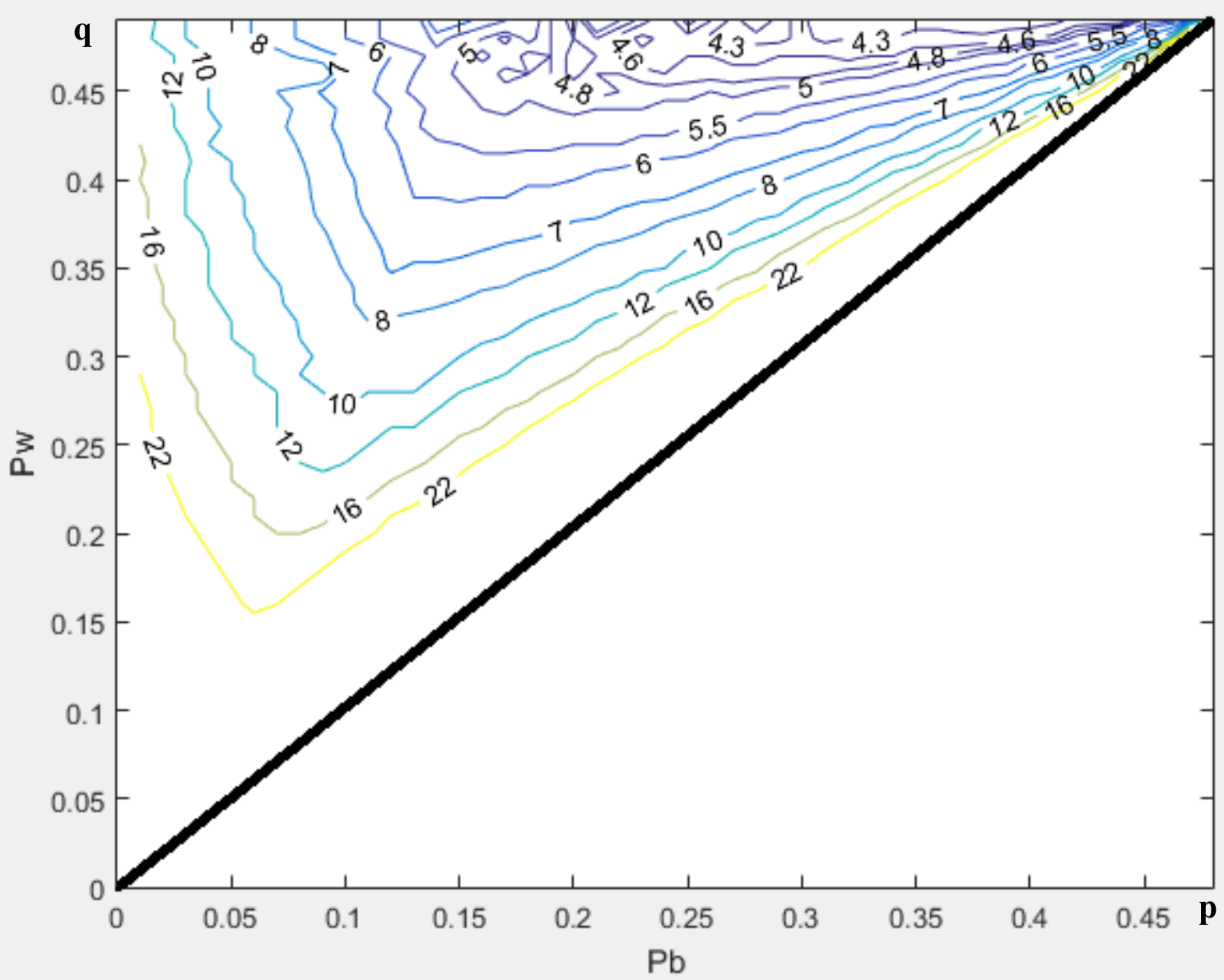}
	\caption{\scriptsize This contour plot shows the decoding complexity of our concatenated code designs as a function of the code chunk length design parameter $k_1$, as defined in Equation~\eqref{eq:minmax}, for various values of $(p,q)$. Each point on a contour labelled  $\eta$ corresponds to a $(p, q)$ value with decoding complexity ${\cal O}({\n}^{\eta})$. Our codes are only designed for the regime $p < q$ (less noisy channel to Bob than to Willie). As is to be expected, when $p$ is close to $q$, the computational complexity is high (since the channels to both parties are similar, one has to employ longer block-lengths to be able to utilize the slight asymmetries in the two channels, leading to correspondingly higher computational cost). Interestingly, even in the regime when $p$ is much smaller than $q$, the computational cost is also relatively high - in this regime the driving factor is the fact that a much higher covert throughput is possible, leading to correspondingly higher computational workload.} 
	\label{fig:contour}
	\end{center}
\end{figure}

%Let $k_1$ and $k_2$ be parameters as defined in Equations~\eqref{eq:minmax} and~\eqref{eq:b} respectively.
\begin{theorem}
\label{theorem1}
For any $0<\pb<\pw < 1/2$ and any sufficiently small $\epsilon_d > 0$, there exists a concatenated code ensemble $\cc^{cc}$ and a $\N_{\pb, \pw, \epsilon_d}$
such that for any $\n > \N_{\pb, \pw, \epsilon_d}$,
with probability super-polynomially close to one over the concatenated code ensemble $\cc^{cc}$, a randomly chosen code $\cc$ satisfies the following properties:
\begin{enumerate}
\item The relative throughput of the code $r$ is $\ru \left(1-(\log{n})^{-1/4}\right) = 2 \epsilon_d\sqrt{\pw(1-\pw)}\frac{1-2\pb}{1-2\pw}\left(\log{\left(\frac{1-\pb}{\pb}\right)}\right)\left(1-(\log{n})^{-1/4}\right)$.
\item There exists a decoder $\Gamma(\cdot)$ such that the probability of error of the code is at most $\exp{\left(-2\sqrt{n}/\left(k_1(\log{n})^2\right)\right)}$.
\item The code is at least ($1-\epsilon_d-4n^{-\delta/4}$)-covert with respect to Willie.
\item The computational complexity of Alice's encoding is $\mathcal{O}\left(\sqrt{n}\log(\sqrt{n})\right)$, and that of Bob's decoding is at most $n^{\ru k_1+1}$. The space complexity for storing the codebook is $n^{\ru k_1+1}$.
\end{enumerate}
\end{theorem}

\begin{table}\centering\caption{Effect of code design parameters on  properties of the code}\label{table:parameters}
\begin{tabular}{|c|l l|}
\hline
Parameter & Code Property &  Value 	\\
\hline $k_1$ & Chunk length &$k_1\sqrt{\n}\log(\n)$ \\
\hline $k_2$ & Average weight of codewords & $k_2\sqrt{n}$\\
\hline $r_u$ & Throughput & $r_u\sqrt{n}$\\ 
\hline
\end{tabular}
\end{table}

\begin{remark} 
%\noindent{\bf Remarks.}\\
{\em a)}\;The meaning of code parameters, as formalized in our proof, is summarized in Table~\ref{table:parameters}. The choice of these parameters leads to various tradeoffs in the complexity-throughput-covertness space.
\begin{enumerate}
\item The parameter $k_1$ determines the chunk length of our inner codes (which equal $k_1\sqrt{\n}\log n$) --- the smaller the $k_1$, the lower the complexity of the codes. However, making $k_1$ too small makes proving covertness and reliability challenging. Hence there's an inherent tradeoff, controlled by the parameter $k_1$, between desirable properties of the code --- indeed Equation~\eqref{eq:minmax} finds a ``sweet spot'' for $k_1$. %Hence the problem of finding low-complexity codes is posed as a (non-convex) optimization problem.
\item The parameter $k_2$ determines the covertness of our code, and the codewords in our codebook have average Hamming weight $k_2\sqrt{n}$ --- the specific choice of $k_2$ matches that in the (computationally inefficient) code design in~\cite{CheBJ:13}.
\item Parameters $\Delta^{xz}_{10}$ and $\Delta^{xz}_{11}$, roughly speaking, quantify the type-classes of codeword-noise pairs likeliest to cause problems for our code design.
\item The relative throughput of our codes equals $r_u(1-o(1))$, which asymptotically matches that in~\cite{CheBJ:13}. 
\item The function $f(\cdot)$ helps analyze the atypicality of codewords, while the functions $g_i(\cdot,\cdot,\cdot,\cdot)$ help analyze the covertness of our coding scheme.
\end{enumerate}
{\em b)}\; The encoding complexity is dominated by the complexity of Reed-Solomon encoding. The decoding complexity is dominated by the random inner code and is an increasing function of $k_1$. For a given value of $\pb$ and $\pw$, the choice of parameters that minimizes the overall decoding complexity is found by Equation~\eqref{eq:minmax}. In Figure~\ref{fig:contour}, we  plot the optimal value of complexities for $0 < p<q < 1/2$.\\ 
 {\em c)}\; For a specific choice of $(\pb,\pw)$, the decoding complexity is independent of the covertness parameter $\epsilon_d$, while the relative throughput scales linearly with $\epsilon_d$.  \\
 {\em d)}\; Our code is proved to be $(1-\epsilon_d-4n^{-\delta/4})$-covert. Note that $1-\epsilon_d-4n^{-\delta/4}$ converges to $1-\epsilon_d$ as $n$ grows without bound.
\end{remark}

\begin{remark} \label{remark1}
	As noted in the recent work by Tahmasbi and Bloch~\cite{tahmasbi2017second}, the optimal values of the code weight design parameter and the relative throughput that guarantees $(1-\epsilon_d)$-covertness respectively equal
	\begin{align}
	&k_2^{\ast}(q,\epsilon_d) = \frac{2\sqrt{q(1-q)}}{1-2q}\cdot Q^{-1}\left(\frac{1-\epsilon_d}{2}\right), \\
	&r_u^{\ast}(p,q,\epsilon_d) = \frac{2\sqrt{q(1-q)}\cdot Q^{-1}\left(\frac{1-\epsilon_d}{2}\right) \cdot (1-2p)}{1-2q} \log \left(\frac{1-p}{p}\right),
	\end{align} 
	where the $Q$-function is defined as $Q(x) = \frac{1}{2\pi} \int_x^{\infty}\exp\left(-\frac{u^2}{2}\right) du$. However, throughout this paper we follow the parameter settings in the preliminary version of this work~\cite{zhang2016computationally}, and stick to the definitions of $k_2(q,\epsilon_d)$ and $r_u(p,q,\epsilon_d)$ in~\eqref{eq:b} and~\eqref{eq:c} respectively. 
\end{remark}

%%%%%%%%%%%%%%%%%%%%%%%%%%%%%%%%%%%%%%%%%%%%%%%%%%%%%%%%%%%%%%%%%%%%%%%%%%%%%%%%%%%%%%%%%%%%%%%%%%%%%%%%%%%%%%%%%%%%%%%%%%%%%%%%%%%%%%%%%%%%%%%%%%%%%%%%%%%%

\section{Code Design and Computational Complexity}\label{sec:code}

In this section, we elaborate on the construction of our concatenated code. Our key technique is to use a ``low-weight'' random code to guarantee covertness. To reduce the computational cost, we divide the message of length $\Theta{(\sqrt{\n})}$ message into $\Theta (\sqrt{\n}/\log{\n})$ chunks, with each chunk containing $\Theta(\log{\n})$ message bits, and apply random inner codes to each of the chunks. In addition, we use a Reed-Solomon code as an outer code to ensure the probability of error decays with the blocklength $\n$. 

\subsection{Outer encoder and Inner encoders}\label{sec:code-A}

\begin{figure}
	\begin{center}
		\includegraphics[scale=0.8]{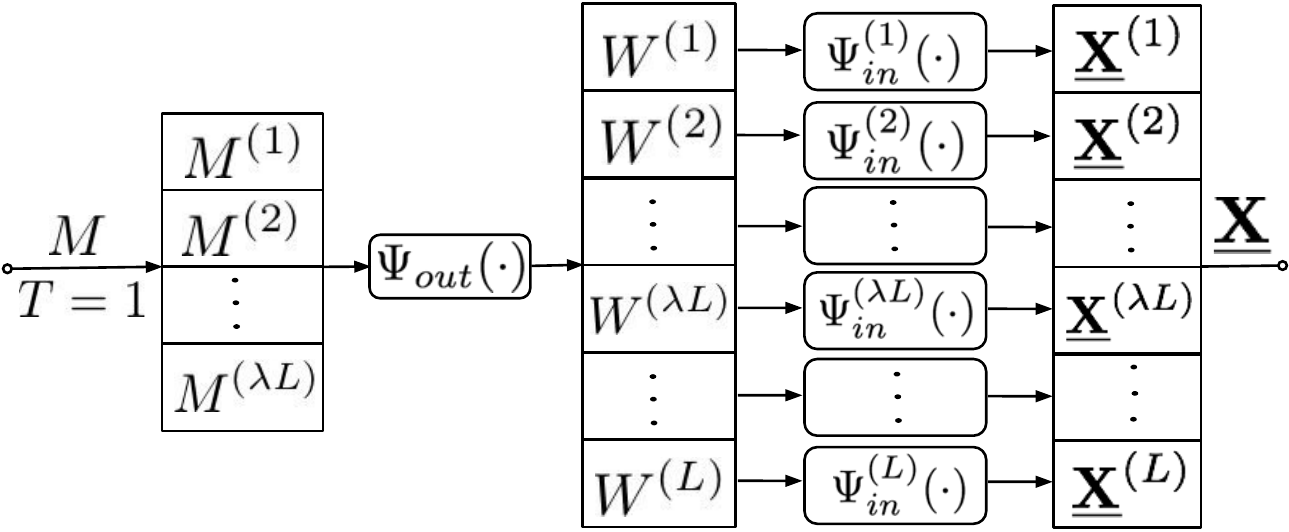}
		\caption{\scriptsize The encoder of the concatenated code: Alice first divides the message $\M$ into $\lambda L$ chunks $\M^{(1)}, \M^{(2)}, \ldots, \M^{(\lambda L)}$. The outer encoder (corresponding to Reed-Solomon code) $\Psi_{out}$ takes the $\lambda L$ chunks as input and outputs $L$ chunks $\W^{(1)}, \W^{(2)}, \ldots, \W^{(L)}$, where $\W^{(i)} = \M^{(i)}$ for $1 \le i \le \lambda L$. For each chunk $\W^{(i)}$, the inner encoder $\Psi_{in}^{(i)}$ (corresponding to the randomly-generated ``low-weight'' inner code $\Ci$) takes $\W^{(i)}$ as input and outputs an inner-codeword $\X^{(i)}$ of this inner code. The codeword $\X$ of the concatenated code is obtained by collecting all the $L$ inner-codewords.} \label{fig:encoder}
	\end{center}
\end{figure}

Figure~\ref{fig:encoder} illustrates the structure of the outer encoder and the inner encoders. Let $L \triangleq \sqrt{\n}/(k_1\log{\n})$ be the number of chunks, and $\lambda$ be the rate of the outer code, with value specified below\footnote{While a detailed discussion for this precise choice of the parameter $\lambda$ is best left to Section~\ref{sec:reliability}, where the effect of the choice of the parameter is more apparent, for now it suffices to think of each systematic chunk as having a vanishing probability of decoding error, and hence a vanishingly small fraction of parity chunks sufficing to aid Bob's decoder.}. For the outer RS code, we divide the length-$(r \sqrt{\n})$ binary vector corresponding to the message $\M$ into $\lambda L$ chunks $\M^{(1)}, \M^{(2)}, \ldots, \M^{(\lambda L)}$. Therefore, each chunk contains $rk_1\log(n)/\lambda$ message bits. Let $\rin = rk_1/\lambda$, and we regard each chunk as a symbol over finite field $\mathbb{F}$ where $|\mathbb{F}| = 2^{\rin \log{\n}}$. The encoding function of the outer code $\Psi_{out}$ takes the form 
$\Psi_{out}(\cdot): \mathbb{F}^{\lambda L} \rightarrow \mathbb{F}^{L},$
and we have $\Psi_{out}\left([\M^{(1)}, \M^{(2)}, \ldots, \M^{(\lambda L)}]\right) = [\W^{(1)}, \W^{(2)}, \ldots, \W^{(L)}]$. The first $\lambda L$ chunks $\W^{(1)}, \W^{(2)}, \ldots, \W^{(\lambda L)}$ are \emph{systematic chunks} while the last $(1-\lambda)L$ chunks $\W^{(\lambda L+1)}, \ldots, \W^{(L)}$ are \emph{parity chunks}, since we use a systematic RS code as the outer code. Note that $\W^{(i)} = \M^{(i)}$ for the systematic chunks. In this work, we set the number of parity chunks to equal $28L/(\log{n})$, and hence the rate of the outer code $\lambda = 1 - 28/(\log{n})$ approaches $1$ as $n$ grows without bound. In the following, we refer to $\W^{(i)}$ and $\X^{(i)}$ as \emph{inner-message} and \emph{inner-codeword} respectively, since they serve as the roles of ``message'' and ``codeword'' of each inner code. The length of each inner-codeword $\X^{(i)}$ is denoted by $\B \triangleq k_1\sqrt{\n}\log{\n}$, since we have $L = \sqrt{\n}/(k_1\log{\n})$ chunks in total. 

For the $i$-th chunk, we use a randomly generated ``low-weight'' inner codes $\Ci$, with distribution $P(\ci)$, to encode the inner-message $\W^{(i)}$. Note that here we use $\Ci$ to denote the random variable, while using $\ci$ to denote its realization. The inner code $\Ci$ contains $2^{\rin \log{\n}} = n^{\rin}$ inner-codewords of length-$\B$, with each bit of these inner-codewords chosen independently and identically distributed (\emph{i.i.d.}) according to Bernoulli$(\rho)$, where\footnote{Note that in this work, the parameter $\rho$ scales as $\Theta(n^{-1/2})$, while the code weight design parameter $k_2$ scales as a constant.} $\rho \triangleq k_2/\sqrt{\n}$. For each inner-message $w^{(i)}$, the corresponding inner-codeword is denoted by $\X^{(i)}_w$, with distribution
\begin{align}
P_{\X}(\x^{(i)}_w ) = \rho^{wt_H(\x^{(i)}_w)}\cdot (1-\rho)^{\npri - wt_H(\x^{(i)}_w)}, \ \forall \x^{(i)}_w \in \{0,1\}^B, \label{eq:fadai}
\end{align}
where $wt_H(\x^{(i)}_w)$ denotes the Hamming weight of $\x^{(i)}_w$.
The encoder of $\C^{(i)}$ takes the form $\Psi_{in}^{(i)}(\cdot): \left\{0, 1\right\}^{\rin \log{\n}} \rightarrow \left\{0, 1\right\}^{B}$, and outputs $\X^{(i)}_w$ for every inner-message $w^{(i)}$. 

For all $i \in \left\{1, 2, \ldots, L\right\}$, the codebooks $\C^{(i)}$ are independently and identically distributed, and hence different chunks are encoded by different inner codes. The probability distribution induced over concatenated codebooks generated via this process will be denoted by $P(C)$. By collecting all the $L$ inner-codewords, we obtain the codeword $\X = [\X^{(1)}, \X^{(2)}, \ldots, \X^{(L)}]$. 

\subsection{Outer decoder and Inner decoder}\label{sec:code-B}

Bob first partitions the channel outputs $\Yb$ into $L$ vectors $[\Yb^{(1)}, \Yb^{(2)}, \ldots, \Yb^{(L)}]$, where for each $i$ the length-$\npri$ vector $\Yb^{(i)}$ corresponds to the set of channel outputs $Y_{(i-1)\B+1},\ldots,Y_{i\B}$. The $i$-th inner decoder takes $\Yb^{(i)}$ as input and reconstructs $\What^{(i)}$ by using the decoding function $\Gamma_{in}^{(i)}(\cdot): \left\{0, 1\right\}^{B} \rightarrow \left\{0, 1\right\}^{\rin \log{\n}}$. Bob then treats each reconstructed inner-message $\What^{(i)}$ as a symbol over finite field $\mathbb{F}$, and reconstructs $\hat{\M}$ using the decoder for a systematic RS code.

\subsection{Space complexity and Computational complexity}
\subsubsection{Space complexity}
We need to store all the inner codebooks since random codes serve as the inner codes. 
An inner code $\Ci$ contains $n^{\rin}$ inner-codewords of length-$\B$, hence the space complexity for storing a single inner codebook is $\npri n^{\rin}$ bits. Note that the concatenated code $\C$ contains $L$ inner codes. Therefore, for sufficiently large $n$, the total space complexity is bounded from above by
\begin{align}
\B n^{\rin} \cdot L = n^{\rin + 1} \le n^{k_1\ru +1}, \label{eq:space}
\end{align}
where inequality~\eqref{eq:space} follows since $\B L = n$ and 
\begin{align}
\rin = \frac{rk_1}{\lambda} = \frac{r_u\left(1-(\log{n})^{-\frac{1}{4}}\right)k_1}{\lambda} = k_1r_u\frac{1-(\log n)^{-\frac{1}{4}}}{1-(28/\log n)} \le k_1\ru.
\end{align}

\subsubsection{Computational complexity for encoding}
We first consider the computational complexity of the outer RS encoder. A clever way to implement the RS encoder is to perform a \emph{Fast Fourier Transform} over the finite field $\mathbb{F}$~\cite{preparata1977computational}. Such encoding process requires $\mathcal{O}(L \log L \log |\mathbb{F}|) = \mathcal{O}(\sqrt{n}\log(\sqrt{n}))$ binary operations, since $L = \sqrt{n}/(k_1\log{n})$, $|\mathbb{F}|=2^{\rin \log n}$, and $\rin \le k_1\ru$. Next, it is worth noting that the computational complexity of the inner encoders is negligible since the inner codebooks are stored in the storage.

\subsubsection{Computational complexity for decoding}
As usual in information theory, our decoding rule for each inner code follows from the {\it typicality decoding}. In the worst case, each inner decoder needs to look through the whole inner codebook, and hence the computational complexity of the inner decoders equals the total space complexity. Moreover, the complexity of the best known RS decoder is given by $\mathcal{O}(L^2\log L \log |\mathbb{F}|)$~\cite{wicker1995error}, which is negligible compared with that of the inner decoders. Therefore, the overall computational complexity for decoding is at most $n^{k_1\ru +1}$.

%%%%%%%%%%%%%%%%%%%%%%%%%%%%%%%%%%%%%%%%%%%%%%%%%%%%%%%%%%%%%%%%%%%%%%%%%%%%%%%%%%%%%%%%%%%%%%%%%%%%%%%%%%%%%%%%%%%%%%%%%%%%%%%%%%%%%%%%%%%%%%%%%%%%%%%%%%%%

\section{Definitions and Probability Distributions of Interest}
\label{sec:def}
Since much of the analysis in this work is based on a ``chunk-wise'' manner, most of the notations and definitions introduced in this section correspond to a single chunk $i$, for $i \in \{1,\ldots,L\}$.
\subsection{Definitions used for covertness}\label{sec:def-A}
\begin{itemize}
\item For any inner-codeword $\x^{(i)}$ and Willie's channel outputs $\yw^{(i)}$, the fraction of $(0,0), \ (0,1), \ (1,0)$ and $(1,1)$ pairs in $(\x^{(i)}, \yw^{(i)})$ are respectively denoted by $f^{xz}_{00}(\x^{(i)},\yw^{(i)})$, $f^{xz}_{01}(\x^{(i)},\yw^{(i)})$, $f^{xz}_{10}(\x^{(i)},\yw^{(i)})$ and $f^{xz}_{11}(\x^{(i)},\yw^{(i)})$, with
\begin{align}
f^{xz}_{ab}(\x^{(i)},\yw^{(i)}) \triangleq \frac{\left|j \in \{1,\ldots,\npri\}:(x_j^{(i)},z_{j}^{(i)})=(a,b)\right|}{\B}, \ \ \forall a \in \{0,1\}, b \in \{0,1\},
\end{align}
where $x_j^{(i)}$ and $z_{j}^{(i)}$ are the $j$-th elements of $\x^{(i)}$ and $\yw^{(i)}$, and $\npri$ is the length of $\x^{(i)}$ and $\yw^{(i)}$.
\item The \emph{fractional Hamming weight} of the inner-codeword $\x^{(i)}$ is denoted by 
\begin{align}
f^x_{1}(\x^{(i)}) \triangleq \frac{wt_H(\x^{(i)})}{\npri}.
\end{align}
Note that $f^x_{1}(\x^{(i)}) = f^{xz}_{10}(\x^{(i)},\yw^{(i)})+f^{xz}_{11}(\x^{(i)},\yw^{(i)})$ by definition. If, as will be the case in this work, each bit of each inner-codeword is chosen to equal $1$ with probability $\rho$, then the expected value of $f^x_{1}(\x^{(i)})$ equals $\rho$.
\item The fractional Hamming weight of Willie's channel outputs $\yw^{(i)}$ is denoted by 
\begin{align}
f^z_{1}(\yw^{(i)}) \triangleq \frac{wt_H(\yw^{(i)})}{\npri}.
\end{align}
Note that $f^z_{1}(\yw^{(i)}) = f^{xz}_{01}(\x^{(i)},\yw^{(i)})+f^{xz}_{11}(\x^{(i)},\yw^{(i)})$ by definition. The expected value of $f^z_{1}(\yw^{(i)})$ when $T=1$ equals $\rho * q = \rho(1-\pw)+\pw(1-\rho)$, since $f^x_{1}(\x^{(i)})$ and $f^x_{0}(\x^{(i)})$ equal $\rho$ and ($1-\rho$) respectively, and the channel between Alice and Willie is a BSC($\pw$). 
\end{itemize}
For notational convenience we henceforth abbreviate $f^{xz}_{ab}(\x^{(i)},\yw^{(i)})$, $f^x_{1}(\x^{(i)})$, $f^z_{1}(\yw^{(i)})$ as $f^z_{ab}$, $f^x_{1}$, $f^z_{1}$ respectively (for $a,b \in \{0,1\}$), when the arguments are clear from the context.

When Alice is transmitting ($T = 1$), the random variable $Z$ is drawn from a Bernoulli($\rho *\pw$) distribution. We then define the $\B$\emph{-letter typical set} of $Z$ when $\T=1$ as\footnote{For notational convenience, we use $[a\pm b]$ to denote an interval $[a-b,a+b] \subset \mathbb{R}$.}
\begin{align}
\aywi \triangleq \left\{\yw^{(i)}\in \{0,1\}^{\npri}: f^z_{1}(\yw^{(i)}) \in \left[(\rho *\pw) \cdot (1 \pm \Delta^z_{1})\right] \right\}. \label{hexi1}
\end{align}
\begin{remark}
	Even though the elements of $\aywi$ defined in Equation~\eqref{hexi1} are labelled as $\yw^{(i)}$, the definition of this set does not depend on the $i$-th chunk. In fact, $\aywi$ can be used to classify not only Willie's channel outputs of the $i$-th chunk $\yw^{(i)}$, but also any length-$\npri$ vector with ``typical'' fractional Hamming weight. Similar remarks also apply to the definitions of other (conditional) typical sets in this Section.
\end{remark}

By choosing $\Delta^z_{1}$ carefully, we ensure that such a narrow typical set is a high probability set (as is usually the case in information-theoretic proofs), and is also as ``narrow'' as possible (includes as few type-classes as possible --- this turns out to be important since extremal type-classes in the narrow typical set dominate the performance of our codes). It can be seen via standard arguments that if $\Delta^z_{1}$ were to decay as $o(n^{-1/4})$, then the corresponding set $\mathcal{A}_n^1(Z)$ would have a vanishing probability mass --- scaling $\Delta^z_{1}$ as $O(n^{-1/4})$ results in the ``narrowest'' possible typical set. In this work, we choose $\Delta^z_{1}$ to scale as $n^{-1/4+\delta/2}$, where the slackness parameter $\delta$ (chosen in the range ($0,0.5$)) allows one to show sufficiently tight concentration of probability. In addition, we also partition the typical set $\aywi$ into many type classes. The $\npri$\emph{-letter type class} of $Z$ (of fractional Hamming weight $f_1^z$) is defined as
\begin{align}
\tywi(f^z_1) \triangleq \left\{\yw^{(i)}\in \{0,1\}^{\npri}: f^z_{1}(\yw^{(i)}) = f^z_{1} \right\}.
\end{align}
We define the set of typical fractional Hamming weight of $Z$ as
\begin{align}
\mathcal{F}^{z}_{\B}  \triangleq \left\{ f^{z}_{1} :
\begin{array}{ll}
f^z_1 \in \left[(\rho * q) (1 \pm \Delta^z_1)\right]\\
\B f^z_1 \in \mathbb{Z}
\end{array}
\right\},
\end{align}
hence the $\npri$-letter typical set of $Z$ can be represented as the union of ``typical'' type classes, \emph{i.e.}, $\aywi = \cup_{f^z_1 \in \mathcal{F}^{z}_{\B}} \ \tywi(f^z_1)$.
Moreover, for a given $\yw^{(i)}$, we define the $\B$\emph{-letter conditionally typical set} of $X$ as
\begin{align}
\axywi \triangleq \left\{ \x^{(i)}\in \{0,1\}^{\npri}: 
 	\begin{array}{ll}
	f^{xz}_{10}(\x^{(i)},\yw^{(i)}) \in \left[\rho \pw(1 \pm \Delta^{xz}_{10})\right] \\
	f^{xz}_{11}(\x^{(i)},\yw^{(i)}) \in \left[\rho (1-\pw)(1 \pm \Delta^{xz}_{11})\right]	
	\end{array}
	\right\}, \label{hexi2}
\end{align} 
where $\Delta^{xz}_{10}$ and $\Delta^{xz}_{11}$ scale as constants in the interval $(0,1)$ (with values to be specified later, in Section~\ref{sec:deniability} --- indeed, careful choice of these two parameters turns out to be critical for our code design). The $\B$-letter conditionally typical set can further be decomposed to many conditional type classes. Given $\yw^{(i)}$, the $\B$\emph{-letter conditional type class} of $X$ is defined as
\begin{align}
\txywi(f^{xz}_{10},f^{xz}_{11}) \triangleq \left\{ \x^{(i)}\in \{0,1\}^{\npri}: 
 	\begin{array}{ll}
	\big|j:(x_j^{(i)},z_{j}^{(i)})=(1,0)\big| = \B f^{xz}_{10} \\
	\big|j:(x_j^{(i)},z_{j}^{(i)})=(1,0)\big| = \B f^{xz}_{10}
	\end{array}
	\right\}, \label{hexi3}
\end{align} 
Let the set of typical fractional Hamming weight with respect to Willie be
\begin{align}
    \mathcal{F}^{xz}_{\B}  \triangleq \left\{ (f^{xz}_{10}, f^{xz}_{11}) :
                \begin{array}{ll}
                  f^{xz}_{10} \in \left[\rho \pw(1 \pm \Delta^{xz}_{10})\right]\\
                  f^{xz}_{11} \in \left[\rho (1-\pw)(1 \pm \Delta^{xz}_{11})\right]\\
                  \B f^{xz}_{10} \in \mathbb{Z} \\
		  \B f^{xz}_{11} \in \mathbb{Z}
                \end{array}
              \right\}, \label{yinghua1}
  \end{align}
and we have $\axywi = \cup_{(f^{xz}_{10}, f^{xz}_{11})\in \mathcal{F}^{xz}_{\B}} \ \txywi(f^{xz}_{10},f^{xz}_{11})$.

\subsection{Definitions used for reliability}\label{sec:def-B}
\begin{itemize}
\item For any inner-codeword $\x^{(i)}$ and Bob's channel outputs $\yb^{(i)}$, the fraction of $(0,0), \ (0,1), \ (1,0)$ and $(1,1)$ pairs in $(\x^{(i)}, \yb^{(i)})$ are respectively denoted by $f^{xy}_{00}(\x^{(i)},\yb^{(i)})$, $f^{xy}_{01}(\x^{(i)},\yb^{(i)})$, $f^{xy}_{10}(\x^{(i)},\yb^{(i)})$ and $f^{xy}_{11}(\x^{(i)},\yb^{(i)})$, with
\begin{align}
f^{xy}_{ab}(\x^{(i)},\yb^{(i)}) \triangleq \frac{\left|j \in \{1,\ldots,\npri\}:(x_j^{(i)},y_{j}^{(i)})=(a,b)\right|}{\B}, \ \ \forall a \in \{0,1\}, b \in \{0,1\}.
\end{align}
\item The fractional Hamming weight of Bob's channel outputs $\yb^{(i)}$ is denoted by 
\begin{align}
f^y_{1}(\yb^{(i)}) \triangleq \frac{wt_H(\yb^{(i)})}{\npri}.
\end{align}
\end{itemize}
The expected value of $f^y_{1}(\yb^{(i)})$ equals $\pb$ when $\T = 0$, and equals $\rho *\pb$ when $\T = 1$. We abbreviate $f^{xy}_{ab}(\x^{(i)},\yb^{(i)})$ and $f^y_{1}(\yb^{(i)})$ as $f^{xy}_{ab}$ and $f^y_{1}$ respectively (for $a,b \in \{0,1\}$), when the arguments are clear from the context.

When Alice is silent $(T = 0)$, the random variable $Y$ is drawn from a Bernoulli($\pb$) distribution. Hence we define the $\B$\emph{-letter typical set} of $Y$ when $\T = 0$ as
\begin{align}
\aaybi \triangleq \left\{\yb^{(i)}\in \{0,1\}^{\npri}: f^y_{1}(\yb^{(i)}) \in \left[\pb(1 \pm \Delta^{y}_{1})\right] \right\}. \label{hexi4}
\end{align} 
When Alice is transmitting $(T = 1)$, the random variable $Y$ is drawn from a Bernoulli($\rho *\pb$) distribution. We then define the $\B$\emph{-letter typical set} of $Y$ when $\T = 1$ as
\begin{align}
\aybi \triangleq \left\{\yb^{(i)}\in \{0,1\}^{\npri}: f^y_{1}(\yb^{(i)}) \in \left[(\rho *\pb) (1 \pm \Delta^{y}_{1})\right] \right\}. \label{hexi5}
\end{align}
The parameter $\Delta^{y}_{1}$ is set to be ${\n}^{-1/4+\delta/2}$ in the following proof\footnote{The reason for this scaling is as in Section~\ref{sec:deniability-A} for $\Delta^z_{1}$.}. For a given $\yb^{(i)}$, the $\B$\emph{-letter conditionally typical set} of $X$ when $\T=1$ is defined as 
\begin{align}
\axybi \triangleq \left\{ \x^{(i)}\in \{0,1\}^{\npri}: 
	\begin{array}{ll}
	f^{xy}_{10}(\x^{(i)}, \yb^{(i)}) \in \left[\rho \pb(1 \pm \Delta^{xy}_{10})\right] \\
	f^{xy}_{11}(\x^{(i)},\yb^{(i)}) \in \left[\rho (1-\pb)(1 \pm \Delta^{xy}_{11})\right]
	\end{array}
	\right\}, \label{hexi6}
\end{align}
where $\Delta^{xy}_{10}$ and $\Delta^{xy}_{11}$ scale as $(\log n)^{-1/3}$. The scalings of $\Delta^{xy}_{10}$ and $\Delta^{xy}_{11}$, which are analyzed in Claim~\ref{claim:err3}, guarantee simultaneously that the conditionally typical set $\axybi$ is a high probability set, and yet is also as ``narrow'' as possible. We then define the $\B$\emph{-letter conditional type class} of $X$ given $\yb^{(i)}$ as  
\begin{align}
\txybi(f^{xy}_{10},f^{xy}_{11}) \triangleq \left\{ \x^{(i)}\in \{0,1\}^{\npri}: 
 	\begin{array}{ll}
	\big|j:(x_j^{(i)},y_{j}^{(i)})=(1,0)\big| = \B f^{xy}_{10} \\
	\big|j:(x_j^{(i)},y_{j}^{(i)})=(1,0)\big| = \B f^{xy}_{10}
	\end{array}
	\right\}, \label{hexi7}
\end{align} 
and the set of typical fractional Hamming weight with respect to Bob as
\begin{align}
    \mathcal{F}^{xy}_{\B}  \triangleq \left\{ (f^{xy}_{10}, f^{xy}_{11}) :
                \begin{array}{ll}
                  f^{xy}_{10} \in \left[\rho \pb(1 \pm \Delta^{xy}_{10})\right]\\
                  f^{xy}_{11} \in \left[\rho (1-\pb)(1 \pm \Delta^{xy}_{11})\right]\\
                  \B f^{xy}_{10} \in \mathbb{Z} \\
		  \B f^{xy}_{11} \in \mathbb{Z}
                \end{array}
              \right\}. \label{yinghua2}
  \end{align}
Therefore, the conditionally typical set $\axybi$ can be represented as $\axybi = \cup_{(f^{xy}_{10},f^{xy}_{11}) \in \mathcal{F}^{xy}_{\B}} \  \txybi(f^{xy}_{10},f^{xy}_{11})$.

\subsection{Probability distributions of interest} \label{sec:distribution}
The proof of covertness essentially connects to the analysis of the distributions of Willie's channel outputs. We now introduce related distributions that are used in the proof. As noted in Section~\ref{sec:code}, each inner code comprises of $n^{\rin}$ inner-codewords, each of length $\B$. The probability that an inner-message $\wi$ is transmitted equals $ 1/n^{\rin}$.

\begin{remark}
	We follow the convention that the message $M$ is uniformly distributed, and this directly implies the inner-message $\Wi$ for systematic chunks is also uniformly distributed. Moreover, in Appendix~\ref{app:matrix} we show that Reed-Solomon codes also ensure the uniformity of the inner-message $\Wi$ for parity chunks.
\end{remark}
The probability $P(\yw^{(i)}|\x^{(i)})$ that a transmitted inner-codeword $\x^{(i)}$ gets pushed by the Bernoulli($\pw$) noise on the channel to Willie to the channel outputs $\yw^{(i)}$, at Hamming distance $d_H$($\x^{(i)},\yw^{(i)}$) from $\x^{(i)}$, equals $q^{d_H(\x^{(i)},\yw^{(i)})} (1-q)^{\B-d_H(\x^{(i)},\yw^{(i)})}$.
Hence, if $T=1$, the $\npri$\emph{-letter active distribution} $P_1^{(i)}$ of Willie's channel outputs $\Yw^{(i)}$ on chunk $i$, which depends on the particular inner code $\ci$, is given as 
\begin{align}
P_1^{(i)}(\yw^{(i)}) = \sum_{\wi} \frac{1}{n^{\rin}} P(\yw^{(i)}|\x^{(i)}_w) =\frac{1}{n^{\rin}} \sum_{\wi} q^{d_H(\x^{(i)}_w,\yw^{(i)})} (1-q)^{\B-d_H(\x^{(i)}_w,\yw^{(i)})}. \label{eq:particular_code}
\end{align}
Next, we consider the active distribution of Willie's channel outputs $\Yw^{(i)}$ averaged over the inner code design. The $\npri${\it-letter ensemble-averaged active distribution} $\mathbb{E}_{\C^{(i)}}\left(P_1^{(i)}\right)$ is given as
\begin{align}
\mathbb{E}_{\C^{(i)}}\left(P_1^{(i)}(\yw^{(i)})\right) = \sum_{\ci}P(\ci)\sum_{\wi} \frac{1}{n^{\rin}} P(\yw^{(i)}|\x^{(i)}_w) &= \frac{1}{n^{\rin}} \sum_{\wi} \sum_{\ci}P(\ci) P(\yw^{(i)}|\x^{(i)}_w) \label{eq:far} \\
&= \frac{1}{n^{\rin}} \sum_{\wi} \sum_{\x^{(i)}_w \in \{0,1\}^B}P_{\X}(\x^{(i)}_w) P(\yw^{(i)}|\x^{(i)}_w) \label{eq:chai1} \\
&= \sum_{\x^{(i)} \in \{0,1\}^B}P_{\X}(\x^{(i)}) P(\yw^{(i)}|\x^{(i)}), \label{eq:chai2}
\end{align}
where~\eqref{eq:chai1} holds since $P(\yw^{(i)}|\x^{(i)}_w)$ only depends on the inner-codeword $\x^{(i)}_w$, and~\eqref{eq:chai2} is obtained by noting that $\sum_{\x^{(i)}_w}P_{\X}(\x^{(i)}_w) P(\yw^{(i)}|\x^{(i)}_w)$ are the same for different $\wi$ (hence we use a generic symbol $\x^{(i)}$ in~\eqref{eq:chai2}). 
Using the definitions of $P_{\X}(\x^{(i)})$ and $P(\yw^{(i)}|\x^{(i)})$ above, it can be seen that this corresponds to a Binomial($n,n(\rho * \pw)$) distribution, with
\begin{align}
\mathbb{E}_{\C^{(i)}}\left(P_1^{(i)}(\yw^{(i)})\right) = (\rho * \pw)^{wt_H(\yw^{(i)})} (1-(\rho * \pw))^{\B-wt_H(\yw^{(i)})}. \label{eq:qixin}
\end{align}
Note that this can be viewed as passing the all-zero codeword through two successive BSCs, with crossover probabilities respectively $\rho$ and $q$. The ensemble-averaged distribution $\mathbb{E}_{\C^{(i)}}\left(P_1^{(i)}\right)$ itself has a relatively simple description, even though for specific codes $P_1^{(i)}$ has a complicated dependence on the inner codebook $\ci$. Indeed, this distribution plays a critical role in the following proof.   
The $\npri$\emph{-letter innocent distribution} $P_0^{(i)}$ on Willie's channel outputs is, in contrast, a Binomial($n,n\pw$) distribution, with 
\begin{align}
P_0^{(i)}(\yw^{(i)}) = q^{wt_H(\yw^{(i)})} (1-q)^{\B-wt_H(\yw^{(i)})}. 
\end{align}
%The above discussion focused on the relevant probability distributions of Willie's channel outputs $\Yw^{(i)}$ -- notice that the probability distributions of Bob's channel outputs $\Yb^{(i)}$ can also be denoted/analyzed in a similar manner.

\begin{table}[]
	\centering
	\caption{Table of Parameters}
	\label{my-label}
	\begin{tabular}{|l|l|l|l|}
		\hline
		\textbf{Symbol} & \textbf{Description} & \textbf{Equality/Range} & \textbf{Section}                   \\ \hline
		$M$      & Message     & $M \in \left\{1, \ldots, N\right\}$ & Section~\ref{sec:model}   \\ \hline
		$\T$       & Transmission status   & $\T \in \left\{0,1\right\}$ & Section~\ref{sec:model}  \\ \hline
		$\X_m$    &  Codeword of message $m$     & $\X_m \in \left\{0,1\right\}^n$  & Section~\ref{sec:model}    \\ \hline
		$\mathbf{C}$      &  Concatenated code     &  $\mathbf{C} = \{\X_m \}_{m=1}^N$    & Section~\ref{sec:result}                              \\ \hline
		$\Yb/\Yw $ &  Bob's/Willie's channel outputs         & $\Yb,\Yw \in \left\{0,1\right\}^n$       & Section~\ref{sec:model}                           \\ \hline
		$\underline{\mathbf{N}}_y/\underline{\mathbf{N}}_z$  &  Noise vector from Alice to Bob/Willie          & $\underline{\mathbf{N}}_y, \underline{\mathbf{N}}_z \in \left\{0,1\right\}^n$    & Section~\ref{sec:model}       \\ \hline
		$\pb$       &   Crossover probability of BSC (Alice to Bob)    &  $\pb \in [0, 0.5] $   & Section~\ref{sec:intro}       \\ \hline
		$\pw$       &   Crossover probability of BSC (Alice to Willie)    &  $\pw \in [0, 0.5] $  & Section~\ref{sec:intro}      \\ \hline
		$\epsilon_d$     &   Parameter of covertness   &   $0 < \epsilon_d < 1$ & Section~\ref{sec:model} \\ \hline
		$P_0(\Yw)$      &  Innocent distribution of $\Yw$ ($\T = 0$)    &  & Section~\ref{sec:model}        \\ \hline
		$P_1(\Yw)$      &  Active distribution of $\Yw$ under code $\cc$ ($\T = 1$)    &     & Section~\ref{sec:model}          \\ \hline
		$\mathbb{E}_{\C}(P_1(\Yw))$      &  Ensemble-averaged active distribution of $\Yw$ ($\T = 1$)    &   & Term~\eqref{eq:jun1}, Section~\ref{sec:deniability}   \\ \hline
		$\Psi(\cdot)$       &  Encoder       &    $\Psi(\cdot): \left\{0\right\}\cup \left\{1, 2, \ldots, \N \right\} \rightarrow \left\{0, 1\right\}^{\n}$ & Section~\ref{sec:model} \\ \hline
		$\Gamma (\cdot)$       &  Bob's decoder      &  $\Gamma(\cdot): \left\{0, 1\right\}^{\n} \rightarrow \left\{0\right\}\cup \left\{1, 2, \ldots, \N \right\}$  & Section~\ref{sec:model}    \\ \hline
		$\Phi(\cdot)$       &  Willie's estimator    &   $\Phi(\cdot): \left\{0, 1\right\}^{\n} \rightarrow \left\{0, 1\right\}$ & Section~\ref{sec:model} \\ \hline
		$\alpha(\Phi)$       &  Probability of false alarm           &    $\alpha(\Phi) = \mathbb{P}_{\zw}(\That = 1| \T = 0)$   & Section~\ref{sec:model}   \\ \hline
		$\beta(\Phi)$       &  Probability of missed detection      &  $\beta(\Phi) = \mathbb{P}_{\M,\zw}(\That = 0| \T = 1)$  & Section~\ref{sec:model}    \\ \hline
		$k_1$      &   Code chunk length design parameter  &  & Term~\eqref{eq:minmax}, Section~\ref{sec:result} \\ \hline
		$\B$     &  Chunk length         &  $\B = k_1\sqrt{n}\log{n}$           & Section~\ref{sec:code-A}                      \\ \hline
		$k_2$      &  Code weight design parameter   &  $k_2 = 2\epsilon_d \sqrt{\pw (1-\pw)}/(1-2\pw)$    & Term~\eqref{eq:b}, Section~\ref{sec:result}                            \\ \hline
		$\rho$      &  Average fraction of $1$'s in codewords  &   $\rho = k_2/\sqrt{n}$   & Section~\ref{sec:code-A}        \\ \hline
		$r_u$      &   Maximal relative throughput      &  $\ru = 2\epsilon_d\sqrt{\pw(1-\pw)}\frac{1-2\pb}{1-2\pw}\log{\left(\frac{1-\pb}{\pb}\right)}$   & Term~\eqref{eq:c}, Section~\ref{sec:result}   \\ \hline
		$r$      &  Relative throughput of the concatenated code   & $r = (\log N)/\sqrt{n} = r_u (1-(\log n)^{-1/3})$    & Section~\ref{sec:model}                               \\ \hline
		$L$      &  Number of chunks    &  $L = \sqrt{n}/(k_1\log{n})$    & Section~\ref{sec:code-A}                            \\ \hline
		$\lambda$     &   Rate of the outer code   &   $\lambda = 1 - 28/(\log n)$   & Section~\ref{sec:code-A}          \\ \hline
		$\delta$     &   Slackness parameter   &   $\delta = 0.01$   & Section~\ref{sec:result}          \\ \hline
		$\rin$    &   ``Relative throughput'' of an inner code     &  $\rin = rk_1/\lambda$   & Section~\ref{sec:code-A}                              \\ \hline
		$\mathbb{F}$     &   Finite field of the outer RS code          &   $|\mathbb{F}| = 2^{\rin \log{n}}$  & Section~\ref{sec:code-A}        \\ \hline
		$\W^{(i)}$       &  Inner-message of the $i$-th chunk     &  $\W^{(i)} \in \left\{0,1\right\}^{\rin \log{n}}$ & Section~\ref{sec:code-A}   \\ \hline
		$\X^{(i)}_w$       & Inner-codeword of $\wi$   &    $\X^{(i)}_w \in \left\{0,1\right\}^{\B}$ & Section~\ref{sec:code-A} \\ \hline
		$\Ci$      &   Inner code for the $i$-th chunk    &  $\Ci = \{\X^{(i)}_w \}_{w=1}^{n^{\rin}}$  & Section~\ref{sec:code-A}        \\ \hline
		$\Yb^{(i)}/\Yw^{(i)}$       & Bob's/Willie's channel outputs of the $i$-th chunk   &    $\Yb^{(i)} \in \left\{0,1\right\}^{B}/\Yw^{(i)} \in \left\{0,1\right\}^{B}$ & Section~\ref{sec:code-A}  \\ \hline
		$P^{(i)}_1(\Yw^{(i)})$       & $\B$-letter active distribution of $\Yw^{(i)}$ under $\ci$   &     & Term~\eqref{eq:particular_code}, Section~\ref{sec:distribution} \\ \hline
		$\mathbb{E}_{\C^{(i)}}\left(P_1^{(i)}(\Yw^{(i)})\right)$       & $\B$-letter ensemble-averaged active distribution of $\Yw^{(i)}$   &     & Term~\eqref{eq:qixin}, Section~\ref{sec:distribution} \\ \hline
		$f_{ab}^{xz}(\x^{(i)},\yw^{(i)})$/ $f_{ab}^{xy}(\x^{(i)},\yb^{(i)})$       &  Fraction of pair-($a,b$) in ($\x^{(i)},\yw^{(i)}$)/($\x^{(i)},\yb^{(i)}$), $a,b \in \left\{0,1\right\}$           &     & Sections~\ref{sec:def-A},~\ref{sec:def-B}     \\ \hline
		$f^x_{1}(\x^{(i)})/f^y_{1}(\yb^{(i)})/f^z_{1}(\yw^{(i)})$ & Fractional Hamming weight of $\x^{(i)}/\yb^{(i)}/\yw^{(i)}$   &   & Sections~\ref{sec:def-A},~\ref{sec:def-B}  \\ \hline
		$\aywi$/$\aybi$       & $\B$-letter typical set of $Z$/$Y$ ($T=1$)  &  & Sections~\ref{sec:def-A},~\ref{sec:def-B}     \\ \hline
		$\aaybi$       &  $\B$-letter typical set of $Y$ ($\T = 0$)   & & Term~\eqref{hexi4}, Section~\ref{sec:def-B}  \\ \hline
		$\axywi$/$\axybi$       &  $\B$-letter conditionally typical set of $X$ given $\yw^{(i)}/\yb^{(i)}$     & & Sections~\ref{sec:def-A},~\ref{sec:def-B}                     \\ \hline
		$\mathcal{F}^{xz}_{\B}$/$\mathcal{F}^{xy}_{\B}$    & Set of typical fractional Hamming weight  &  & Sections~\ref{sec:def-A},~\ref{sec:def-B}               \\ \hline
		$\txywi(f^{xz}_{10},f^{xz}_{11})$/$\txybi(f^{xy}_{10},f^{xy}_{11})$    & $\B$-letter conditional type class of $X$ given $\yw^{(i)}$/$\yb^{(i)}$  &        & Sections~\ref{sec:def-A},~\ref{sec:def-B}                          \\ \hline
		$\tau^{(i)}$    & Oracle revealed information of the $i$-th chunk &     & Section~\ref{sec:deniability-B}    \\ \hline
		$l_1$      &  Number of systematic chunks    &  $l_1 = \lambda L $         & Section~\ref{sec:deniability-B}                        \\ \hline
		$l_2$      &  Number of parity chunks    &  $l_2 = L(1-\lambda)$     & Section~\ref{sec:deniability-B}                            \\ \hline
		$(W^{(1)}, \ldots, W^{(l_1)})$    & Systematic inner-message vector &     & Section~\ref{sec:deniability-B}    \\ \hline
		$(W^{(l_1+1)}, \ldots, W^{(L)})$    & Parity inner-message vector &     & Section~\ref{sec:deniability-B}    \\ \hline
	\end{tabular}
\end{table}

\section{Proof of Covertness}
\label{sec:deniability}
Theorem~\ref{theorem1} states that for any sufficiently small $\epsilon_d > 0$, the code we construct is ($1-\epsilon_d - 4n^{-\delta/4}$)-covert with high probability. As discussed in Section~\ref{sec:model}, the code is deemed to be ($1-\epsilon_d - 4n^{-\delta/4}$)-covert if the variational distance between the innocent distribution $P_0$ and the active distribution $P_1$ is bounded from above as $\mathbb{V}(P_0, P_1) \le \epsilon_d + 4n^{-\delta/4}$. Note that $\forall \yw \in \{0,1\}^n$, 
\begin{align}
&P_0(\yw) = q^{wt_H(\yw)}(1-q)^{n-wt_H(\yw)}, \\
&P_1(\yw) = \sum_{m \in \{1,\ldots,N\}} \frac{1}{N} P(\yw|\x_m), 
\end{align}
where $\x_m$ is the length-$n$ codeword corresponding to the message $m$, and $1/N$ is the probability that a message $m$ is transmitted. 

\begin{figure}
	\begin{center}
	\includegraphics[scale=0.24]{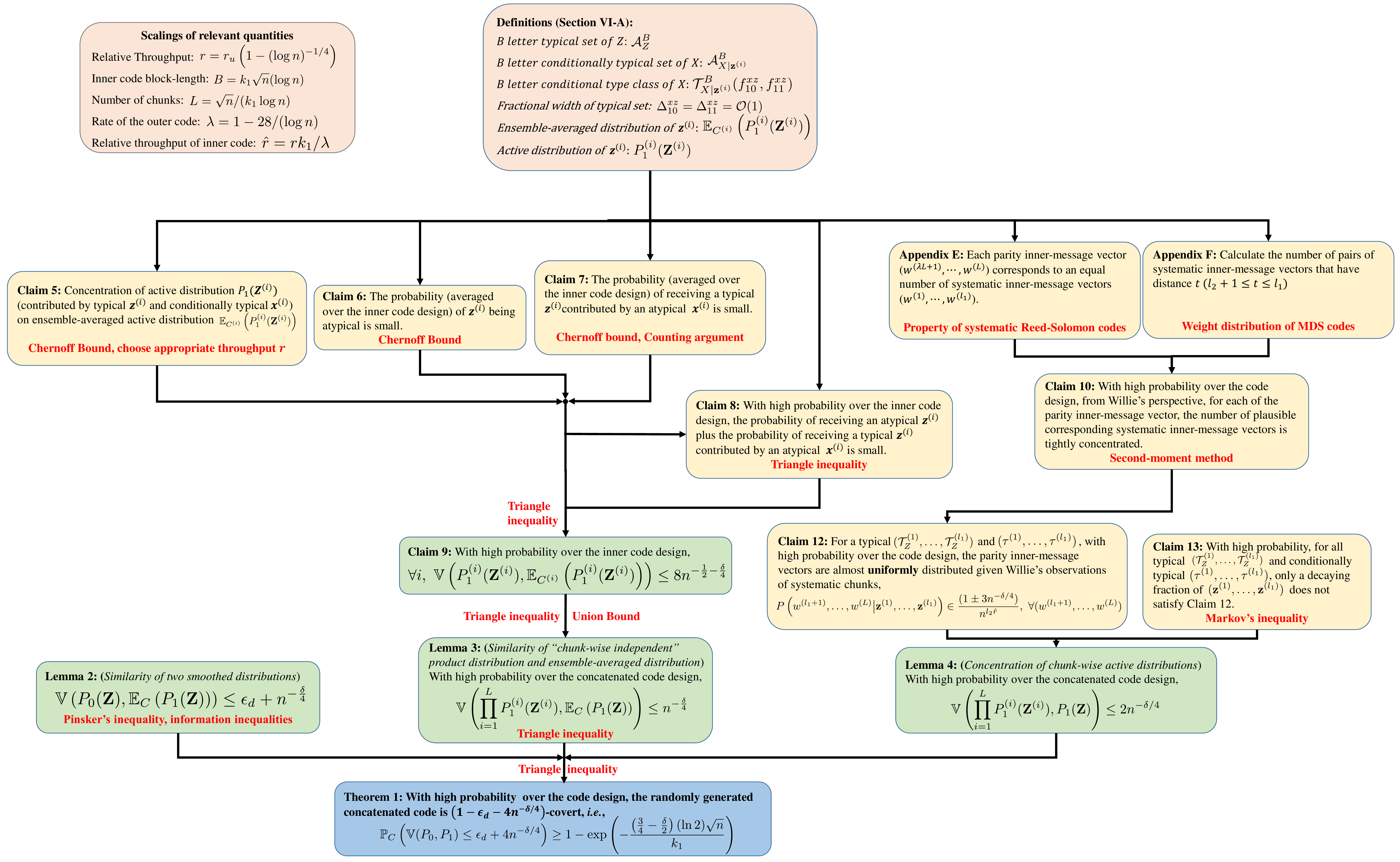}
	\caption{A road-map of our proof that our codes are covert with high probability.} \label{fig:flow}
	\end{center}
\end{figure}

The $n$\emph{-letter ensemble-averaged active distribution} of $\Yw$ is denoted by $\mathbb{E}_{\C}\left(P_1\right)$, such that the following holds for all $\yw \in \{0,1\}^n$, 
\begin{align}
\mathbb{E}_{\C}(P_1)(\yw) = \sum_{\cc}P(\cc)\sum_{m \in \{1,\ldots,N\}} \frac{1}{N}P(\yw|\x_m) &= \frac{1}{N}\sum_{m \in \{1,\ldots,N\}}\sum_{\C^{(1)}}P(C^{(1)})P(\yw^{(1)}|\x^{(1)}_{w_m})\cdots \sum_{\C^{(L)}}P(C^{(L)})P(\yw^{(L)}|\x^{(L)}_{w_m}) \label{eq:jun1} \\
&= \frac{1}{N}\sum_{m \in \{1,\ldots,N\}}\sum_{\x^{(1)}_{w_m} \in \{0,1\}^B}P_{\X}(\x^{(1)}_{w_m})P(\yw^{(1)}|\x^{(1)}_{w_m})\cdots \sum_{\x^{(L)}_{w_m} \in \{0,1\}^B}P_{\X}(\x^{(L)}_{w_m})P(\yw^{(L)}|\x^{(L)}_{w_m}) \label{eq:connie1} \\ 
&= \sum_{\x^{(1)} \in \{0,1\}^B}P_{\X}(\x^{(1)})P(\yw^{(1)}|\x^{(1)})\cdots \sum_{\x^{(L)} \in \{0,1\}^B}P_{\X}(\x^{(L)})P(\yw^{(L)}|\x^{(L)}) \label{eq:connie2} \\ 
& = \mathbb{E}_{\C^{(1)}}\left(P_1^{(1)}(\yw^{(1)})\right) \cdots \mathbb{E}_{\C^{(L)}}\left(P_1^{(L)}(\yw^{(L)})\right) \label{eq:40} \\
&= (\rho * \pw)^{wt_H(\yw)}(1-(\rho * \pw))^{n-wt_H(\yw)}. \label{eq:41}
\end{align}
Equation~\eqref{eq:jun1} is obtained by assuming the message $m$ is encoded to $(w_m^{(1)}, \ldots, w_m^{(L)})$, and the channel from Alice to Willie is memoryless. Equation~\eqref{eq:connie1} holds since for a fixed $m$, the only random variables of interest are $(\X^{(1)}_{w_m}, \ldots, \X^{(L)}_{w_m})$ --- the inner-codewords of the message $m$. Equation~\eqref{eq:connie2} is due to the fact that the term 
\begin{align}
\sum_{\x^{(1)}_{w_m} \in \{0,1\}^B}P_{\X}(\x^{(1)}_{w_m})P(\yw^{(1)}|\x^{(1)}_{w_m})\cdots \sum_{\x^{(L)}_{w_m} \in \{0,1\}^B}P_{\X}(\x^{(L)}_{w_m})P(\yw^{(L)}|\x^{(L)}_{w_m})
\end{align}  
is the same for every $m$, hence we use generic symbols $\x^{(1)}, \ldots, \x^{(L)}$ in~\eqref{eq:connie2}. Equation~\eqref{eq:40} follows from the definition of the $B$-letter ensemble-averaged active distribution $\mathbb{E}_{\C^{(i)}}\left(P_1^{(i)}(\yw^{(i)})\right)$ in~\eqref{eq:chai2}, and Equation~\eqref{eq:41} follows from~\eqref{eq:qixin}.
Since variational distance satisfies the triangle inequality, we have 
\begin{align}
\mathbb{V}(P_0, P_1) \le \mathbb{V}(P_0, \mathbb{E}_{\C}(P_1)) + \mathbb{V}(\mathbb{E}_{\C}(P_1),  P_1),
\end{align}
Following the approach in~\cite{CheBJ:13}, to prove that the proposed code is covert, it suffices to show that 
\begin{itemize}
\item (i) $\mathbb{V}(P_0, \mathbb{E}_{\C}(P_1)) \le \epsilon_d + n^{-\delta/4}$,
\item (ii) With high probability over the concatenated code design, $\mathbb{V}(\mathbb{E}_{\C}(P_1),  P_1) \le 3n^{-\delta/4}$. 
\end{itemize}
A flow-chart of the proof of covertness can be found in Figure~\ref{fig:flow}. As in~\cite{CheBJ:13}, the proof of (i) follows fairly directly from relatively standard information-theoretic inequalities. For completeness, we repeat the proof here.

\begin{lemma}\label{lemma:1} \cite{CheBJ:13}
Let the code weight design parameter $k_2 = \frac{2\epsilon_d \sqrt{\pw(1-\pw)}}{1-2\pw}$, as $n$ grows without bound, we have 
\begin{align}
\mathbb{V}\left(P_0, \mathbb{E}_\C\left(P_1\right)\right) \le \epsilon_d + n^{-\delta/4}.
\end{align}
\end{lemma}
\noindent{\emph{Proof:}} 
\begin{align}
\mathbb{V}\left(P_0, \mathbb{E}_\C\left(P_1\right)\right) & \le \sqrt{\frac{\ln 2}{2}\mathbb{D}\left(P_0 \parallel \mathbb{E}_\C\left(P_1\right)\right)} \label{eq:var1} \\
& = \sqrt{\frac{\n \ln 2}{2}\mathbb{D}\left(\pw \parallel \rho* \pw \right)}\label{eq:var2} \\
& \le \sqrt{\frac{\n \ln 2}{2}\left(\frac{\rho^2 \left(1-2\pw \right)^2}{2\pw \left(1-\pw \right) \ln 2}+\mathcal{O}(\rho^3)\right)}. \label{eq:var3}
\end{align}
In Equation~\eqref{eq:var1}, we use {\it Pinsker's inequality} to bound the variational distance in terms of Kullback-Leibler (KL) Divergence, where $\mathbb{D}(P_0 \parallel \mathbb{E}_{\C}(P_1)) \triangleq \sum_{\yw \in \{0,1\}^n}P_0(\yw)\log \frac{P_0(\yw)}{\mathbb{E}_{\C}(P_1)(\yw)}.$
Equation~\eqref{eq:var2} follows from the chain rule, since both $P_0$ and $\mathbb{E}_{\C}(P_1)$ correspond to $n$-letter sequences drawn \emph{i.i.d.} from Bernoulli($q$) and Bernoulli($\rho * q$) distributions respectively. Equation~\eqref{eq:var3} follows by taking the Taylor series expansion for KL Divergence, as in~\cite[Claim~13]{CheBJ:13}, resulting in 
\begin{align}
\mathbb{D}(\pw \parallel \rho *\pw) \le \frac{\rho^2(1-2\pw)^2}{2\pw(1-\pw)\ln{2}} + \mathcal{O}(\rho^3).
\end{align} 
By choosing $\rho = \frac{k_2}{\sqrt{\n}} = \frac{2\epsilon_d \sqrt{\pw(1-\pw)}}{(1-2\pw)\sqrt{\n}}$, as $n$ grows without bound, we have $\mathbb{V}\left(P_0, \mathbb{E}_\C\left(P_1\right)\right) \le \epsilon_d + \mathcal{O}(n^{-1/4}) \le \epsilon_d + n^{-\delta/4}.$
 \qed

\vspace{5pt}
It is worth noting that the Pinsker's inequality used in the proof of Lemma~\ref{lemma:1} is not tight. As discussed in Remark~\ref{remark1}, choosing the code weight design parameter $k_2$ to be $\frac{2\sqrt{q(1-q)}}{1-2q}\cdot Q^{-1}\left(\frac{1-\epsilon_d}{2}\right)$ still guarantees $\mathbb{V}\left(P_0, \mathbb{E}_\C\left(P_1\right)\right) \le \epsilon_d + n^{-\delta/4}$.

We now proceed to one of the major parts of our proof (proof of (ii)) --- showing that with high probability over the choice of the inner codes, the variational distance between the active distribution $P_1$ (which depends on the specific inner codes chosen) and the ensemble-averaged active distribution $\mathbb{E}_{\C}(P_1)$ is small. As mentioned in Section~\ref{sec:intuition}, this is considerably more challenging in our setting of concatenated codes comprising of multiple chunks, than in the setting of~\cite{CheBJ:13} and other works wherein a single $n$-letter code is used. By the definition of variational distance, we have 
\begin{align}
\mathbb{V}(\mathbb{E}_{\C}(P_1), P_1) & = \frac{1}{2} \sum_{\yw \in \left\{0,1\right\}^{\n}}\left|\mathbb{E}_{\C}(P_1(\yw))-P_1(\yw) \right|   \\
&  = \frac{1}{2} \sum_{\yw^{(1)} \in \{0,1\}^B \ldots \yw^{(L)} \in \{0,1\}^B } \left| \mathbb{E}_{\C}(P_1(\yw^{(1)}, \ldots, \yw^{(L)})) - P_1(\yw^{(1)}, \ldots, \yw^{(L)})\right|  \\
&  \le \frac{1}{2} \sum_{\yw^{(1)}\in \{0,1\}^B \ldots \yw^{(L)}\in \{0,1\}^B} \left| P^{(1)}_1(\yw^{(1)}) \cdots P^{(L)}_1(\yw^{(L)}) - P_1(\yw^{(1)}, \ldots, \yw^{(L)}) \right|   \\
&  + \label{eq:1} \frac{1}{2} \sum_{\yw^{(1)}\in \{0,1\}^B \ldots \yw^{(L)}\in \{0,1\}^B} \left| \mathbb{E}_{\C}(P_1(\yw^{(1)}, \ldots, \yw^{(L)})) - P^{(1)}_1(\yw^{(1)}) \cdots P^{(L)}_1(\yw^{(L)})\right|.
\end{align}
In Equation~\eqref{eq:1} above, the first term corresponds to the variational distance between the $n$-letter active distribution $P_1$ on $\Yw$, and a corresponding \emph{``chunk-wise independent'' product distribution} denoted by $P_1^{(1)}P_1^{(2)} \cdots P_1^{(L)}$; and the second term corresponds to the variational distance between the $n$-letter ensemble-averaged distribution $\mathbb{E}_{\C}(P_1)$ on $\Yw$, and the same product distribution (the inequality follows from the triangle inequality). This product distribution corresponds to the distribution that Willie would see {\it if} he were to ``assume'' that the distribution on $\Yw$ splits as a product of independent distributions on $\Yw^{(1)},\Yw^{(2)}, \ldots,\Yw^{(L)}$. There is of course no reason for this to be the case, especially since Alice is using a code that introduces correlations between chunks, but introducing such a ``proxy'' distribution and computing variational distributions with respect to it is a useful analytical tool. Intuitively, for a highly covert concatenated code, the product distribution $P_1^{(1)}P_1^{(2)} \cdots P_1^{(L)}$ should be ``close'' to {\it both} the active distribution $P_1$, and the ensemble-averaged active distribution $\mathbb{E}_{\C}(P_1)$. Indeed, this is what we show below. We prove Lemma~\ref{lemma:(a)} and Lemma~\ref{lemma:(b)} in Section~\ref{sec:deniability-A} and Section~\ref{sec:deniability-B} respectively, and conclude the proof of covertness in Section~\ref{sec:deniability-C}. 
\begin{lemma}\label{lemma:(a)}
With probability at least $1- \sqrt{n}\exp{\left(-4 \sqrt{n}/3 \right)}$ over the concatenated code design, for the randomly chosen code $\cc$, the variational distance between the ensemble-averaged active distribution $\mathbb{E}_{\C}(P_1)$ and the ``chunk-wise independent'' product distribution $P_1^{(1)}P_1^{(2)} \cdots P_1^{(L)}$ is bounded from above as
$$\frac{1}{2} \sum_{\yw^{(1)} \in \{0,1\}^B \ldots \yw^{(L)}\in \{0,1\}^B} \left| \mathbb{E}_{\C}(P_1(\yw^{(1)}, \ldots, \yw^{(L)})) - P^{(1)}_1(\yw^{(1)}) \cdots P^{(L)}_1(\yw^{(L)})\right| \le n^{-\frac{\delta}{4}}.$$
\end{lemma}

\begin{lemma}\label{lemma:(b)}
With probability at least $1 - \exp\left(-\frac{\left(\frac{3}{4}-\frac{\delta}{2}\right)(\ln 2)\sqrt{n}}{k_1}\right)$ over the concatenated code design and the channel noise to Willie, for the randomly chosen code $\cc$, the variational distance between the $n$-letter active distribution $P_1(\Yw^{(1)}, \ldots, \Yw^{(L)})$ and the ``chunk-wise independent'' product distribution $P_1^{(1)}(\Yw^{(1)})\cdots P_1^{(L)}(\Yw^{(L)})$ is bounded from above as 
$$\frac{1}{2} \sum_{\yw^{(1)}\in \{0,1\}^B \ldots \yw^{(L)}\in \{0,1\}^B} \left| P^{(1)}_1(\yw^{(1)}) \cdots P^{(L)}_1(\yw^{(L)}) - P_1(\yw^{(1)}, \ldots, \yw^{(L)}) \right| \le 2 n^{-\frac{\delta}{4}}.$$
\end{lemma}

%%%%%%%%%%%%%%%%%%%%%%%%%%%%%%%%%%%%%%%%%%%%%%%%%%%%%%%%%%%%%%%%%%%%%%%%%%%%%%%%%%%%%%%%%%%%%%%%%%%%%%%%
\subsection{Proof of Lemma~\ref{lemma:(a)}:} \label{sec:deniability-A}
We first observe that the variational distance between the ensemble-averaged active distribution and the ``chunk-wise independent'' product distribution is bounded from above as 
\begin{eqnarray}
\lefteqn{\frac{1}{2} \sum_{\yw^{(1)}\in \{0,1\}^B \ldots \yw^{(L)}\in \{0,1\}^B} \left| \mathbb{E}_{\C}(P_1(\yw^{(1)}, \ldots, \yw^{(L)})) - P^{(1)}_1(\yw^{(1)}) \cdots P^{(L)}_1(\yw^{(L)})\right|}  \\
& & = \frac{1}{2} \sum_{\yw^{(1)}\in \{0,1\}^B \ldots \yw^{(L)}\in \{0,1\}^B} \left| \mathbb{E}_{\C^{(1)}}(P_1^{(1)}(\yw^{(1)})) \cdots \mathbb{E}_{\C^{(L)}}(P_1^{(L)}(\yw^{(L)})) - P^{(1)}_1(\yw^{(1)}) \cdots P^{(L)}_1(\yw^{(L)})\right|  \label{eq:inter1}\\
& & \le \frac{1}{2} \sum_{i=1}^L \sum_{\yw^{(i)}\in \left\{0,1\right\}^{\npri}} \left| \mathbb{E}_{\Ci}\left(P_1^{(i)}(\yw^{(i)})\right)-P_1^{(i)}(\yw^{(i)})\right|. \label{eq:inter2}
\end{eqnarray}
Note that~\eqref{eq:inter1} has been proved in~\eqref{eq:40}, and~\eqref{eq:inter2} is obtained by using the triangle inequality repeatedly. 
We now follow the lead of the analysis in~\cite{CheBJ:13} by replicating the analysis there in a chunk-wise manner. Specifically, for each chunk $i \in \{1,\ldots,L\}$, we break up $\frac{1}{2}\sum_{\yw^{(i)}\in \left\{0,1\right\}^{\npri}} \left| \mathbb{E}_{\Ci}\left(P_1^{(i)}(\yw^{(i)})\right)-P_1^{(i)}(\yw^{(i)})\right|$, the variational distance between $P_1^{(i)}$ and $\mathbb{E}_{\Ci}(P_1^{(i)})$, as
\begin{eqnarray}
\lefteqn{ \frac{1}{2}\sum_{\yw^{(i)} \in \left\{0, 1\right\}^{\B}}\left|\mathbb{E}_{\Ci} \left(P_1^{(i)}(\yw^{(i)} ) \right) - P_1^{(i)}(\yw^{(i)} )\right|} \notag \\
& & \le \frac{1}{2}\sum_{\yw^{(i)} \in \aywi}\left|\sum_{\ci}P(\ci)\sum_{\wi:\x^{(i)}_w \in \axywi} \frac{1}{n^{\rin}} P(\yw^{(i)}|\x^{(i)}_w) - \sum_{\wi:\x^{(i)}_w \in \axywi} \frac{1}{n^{\rin}} P(\yw^{(i)}|\x^{(i)}_w) \right|  \label{eq:2} \\
& &+ \frac{1}{2} \sum_{\yw^{(i)} \in \aywi}\left|\sum_{\ci}P(\ci)\sum_{\wi:\x^{(i)}_w \notin \axywi} \frac{1}{n^{\rin}} P(\yw^{(i)}|\x^{(i)}_w) - \sum_{\wi:\x^{(i)}_w \notin \axywi} \frac{1}{n^{\rin}} P(\yw^{(i)}|\x^{(i)}_w) \right| \label{eq:3} \\
& &+ \frac{1}{2} \sum_{\yw^{(i)} \notin \aywi}\left|\sum_{\ci}P(\ci)\sum_{\wi} \frac{1}{n^{\rin}} P(\yw^{(i)}|\x^{(i)}_w) - \sum_{\wi} \frac{1}{n^{\rin}} P(\yw^{(i)}|\x^{(i)}_w) \right|  \label{eq:4} \\
& &\le \frac{1}{2}\sum_{\yw^{(i)} \in \aywi}\left|\sum_{\ci}P(\ci)\sum_{\wi:\x^{(i)}_w \in \axywi} \frac{1}{n^{\rin}} P(\yw^{(i)}|\x^{(i)}_w) - \sum_{\wi:\x^{(i)}_w \in \axywi} \frac{1}{n^{\rin}} P(\yw^{(i)}|\x^{(i)}_w) \right| \label{eq:5} \\
& &+ \frac{1}{2} \mathbb{E}_{\Ci} \left(\sum_{\yw^{(i)} \in \aywi} \sum_{\wi: \x^{(i)}_w \notin \axywi} \frac{1}{n^{\rin}} P(\yw^{(i)}|\x^{(i)}_w) \right) + \frac{1}{2}\mathbb{E}_{\Ci}\left(\sum_{\yw^{(i)} \notin \aywi} \pywi \right) \label{eq:6}\\
& &+ \frac{1}{2} \sum_{\yw^{(i)} \in \aywi} \sum_{\wi: \x^{(i)}_w \notin \axywi} \frac{1}{n^{\rin}} P(\yw^{(i)}|\x^{(i)}_w)  + \frac{1}{2} \sum_{\yw^{(i)} \notin \aywi}P_1^{(i)}(\yw^{(i)}). \label{eq:7} 
\end{eqnarray}
The calculation above partitions the variational distance between the active distribution $P_1^{(i)}$ and the ensemble-averaged active distribution $\mathbb{E}_{\Ci}(P_1^{(i)})$ into three components. The term in~\eqref{eq:2} corresponds to the variational distance between $P_1^{(i)}$ and $\mathbb{E}_{\Ci}(P_1^{(i)})$ contributed by typical $\yw^{(i)}$ and conditionally typical $\x^{(i)}$. The term in~\eqref{eq:3} corresponds to the variational distance contributed by typical $\yw^{(i)}$ and conditionally atypical $\x^{(i)}$. The term in~\eqref{eq:4} corresponds to the variational distance contributed by atypical $\yw^{(i)}$. Moreover, we bound~\eqref{eq:3} and~\eqref{eq:4} from above by the triangle inequality, and thus obtain the terms in~\eqref{eq:6} and~\eqref{eq:7}. In the following, we will show that each term in~\eqref{eq:5}, \eqref{eq:6}, \eqref{eq:7} asymptotically vanishes (each term decreases faster than $\Theta(1/\sqrt{n})$) with high probability over the inner code design. 

\begin{claim}[\bf Term in~\eqref{eq:5}]  \label{claim:1}
With probability at least $1 - 2\exp{\left(- 4\sqrt{n}/3 \right)}$ over the inner code design, the randomly chosen inner code $\ci$ satisfies 
$$\frac{1}{2}\sum_{\yw^{(i)} \in \aywi}\left|\sum_{\ci}P(\ci)\sum_{\wi:\x^{(i)}_w \in \axywi} \frac{1}{n^{\rin}} P(\yw^{(i)}|\x^{(i)}_w) - \sum_{\wi:\x^{(i)}_w \in \axywi} \frac{1}{n^{\rin}} P(\yw^{(i)}|\x^{(i)}_w) \right| \le \n^{-1/2-\delta/4},$$
for all large enough values of $n$.
\end{claim}
\noindent{\emph{Proof:}} We first formulate the probability (averaged over the inner code design) of one specific typical $\yw^{(i)}$ induced by conditionally typical $\x^{(i)}$. One key step is to decompose the conditionally typical set $\axywi$ into the typical conditional type classes $\txywi (f_{10}^{xz}, f_{11}^{xz})$ that comprise it, and then calculate the number of inner-codewords falling into each type class.
\begin{eqnarray}
\lefteqn{\sum_{\ci}P(\ci)\sum_{\wi: \x^{(i)}_w \in \axywi} \frac{1}{n^{\rin}} P(\yw^{(i)}|\x^{(i)}_w) }   \\
& &= \sum_{\ci}P(\ci) \sum_{(f_{10}^{xz},f_{11}^{xz})\in \mathcal{F}^{xz}_{\B}}\left( \sum_{\wi: \x^{(i)}_w \in \txywi (f_{10}^{xz}, f_{11}^{xz})} \frac{1}{n^{\rin}} P(\yw^{(i)}|\x^{(i)}_w) \right) \label{eq:8} \\
& &= \sum_{\ci}P(\ci) \sum_{(f_{10}^{xz},f_{11}^{xz})\in \mathcal{F}^{xz}_{\B}} \frac{1}{n^{\rin}} P(\yw^{(i)}|f_{10}^{xz},f_{11}^{xz}) \cdot \left|\wi: \x^{(i)}_w \in \txywi (f_{10}^{xz}, f_{11}^{xz})\right|  \label{eq:9} \\
& &= \sum_{(f_{10}^{xz},f_{11}^{xz})\in \mathcal{F}^{xz}_{\B}}\frac{1}{n^{\rin}} P(\yw^{(i)}|f_{10}^{xz},f_{11}^{xz}) \cdot  \mathbb{E}_{\Ci}\left(\left|\wi: \x^{(i)}_w \in \txywi (f_{10}^{xz}, f_{11}^{xz})\right| \right)  \label{eq:10} \\
& &= \sum_{(f_{10}^{xz},f_{11}^{xz})\in \mathcal{F}^{xz}_{\B}} \frac{1}{n^{\rin}} P(\yw^{(i)}|f_{10}^{xz},f_{11}^{xz}) \cdot \mathbb{P}_{\X^{(i)}} \left( \X^{(i)} \in \txywi (f_{10}^{xz}, f_{11}^{xz})\right) \cdot \left| \ci \right|.  \label{eq:11}
\end{eqnarray}
To obtain Equation~\eqref{eq:8}, we decompose the conditionally typical set into the typical conditional type classes that comprise it. Equation~\eqref{eq:9} follows since $\pywxi$ are identical for all $w^{(i)}$ such that their corresponding inner-codewords $\x^{(i)}_w$ are in the same conditional type class $(f_{10}^{xz},f_{11}^{xz})$, and we then use $P(\yw^{(i)}|f_{10}^{xz},f_{11}^{xz})$ to denote this channel transition probability, and interchange the order of summations to obtain Equation~\eqref{eq:10}. Equation~\eqref{eq:11} follows by noting that the expected number of inner-codewords $\x^{(i)}$ in chunk $i$ falling into a type class $\txywi$ equals the probability (averaged over the inner code design) of a single inner-codeword in chunk $i$ falling into the type class $\txywi$ times the size $|\ci|$ of the inner codebook for chunk $i$. We now bound from below the probability of a single inner-codeword falling into a specific type class as follows:
\begin{eqnarray}
\lefteqn{\mathbb{P}_{\X^{(i)}} \left( \X^{(i)} \in \txywi (f_{10}^{xz}, f_{11}^{xz})\right)}  \notag \\
& &= \binom{\npri \left( f_{01}^{xz} + f_{11}^{xz}\right)}{\npri f_{11}^{xz}}\ \rho^{\npri f_{11}^{xz}}(1-\rho)^{\npri f_{01}^{xz}} \binom{\npri \left( f_{00}^{xz} + f_{10}^{xz}\right)}{\npri f_{10}^{xz}}\ \rho^{\npri f_{10}^{xz}}(1-\rho)^{\npri f_{00}^{xz}} \label{eq:12} \\
& &\ge \frac{1}{2\pi k_1k_2\sqrt{\pw(1-\pw)}\log{\n}} \cdot 2^{\npri(f_{01}^{xz}+f_{11}^{xz})H\left(\frac{f_{11}^{xz}}{f_{01}^{xz}+f_{11}^{xz}} \right)+\npri(f_{00}^{xz}+f_{10}^{xz})H\left(\frac{f_{10}^{xz}}{f_{00}^{xz}+f_{10}^{xz}} \right)} \rho^{\npri \left( f_{10}^{xz}+f_{11}^{xz}\right)} (1-\rho)^{\npri \left(f_{00}^{xz}+f_{01}^{xz}\right)} \label{eq:13} \\
& &= \frac{1}{2\pi k_1k_2\sqrt{\pw(1-\pw)}\log{\n}} \cdot 2^{\npri(f_{01}^{xz}+f_{11}^{xz})H\left(\frac{f_{11}^{xz}}{f_{01}^{xz}+f_{11}^{xz}} \right)+\npri(f_{00}^{xz}+f_{10}^{xz})H\left(\frac{f_{10}^{xz}}{f_{00}^{xz}+f_{10}^{xz}} \right)+ \npri \left( f_{10}^{xz}+f_{11}^{xz}\right)\log \rho + \npri \left(f_{00}^{xz}+f_{01}^{xz}\right)\log \left(1-\rho \right)} \\
& &= \frac{1}{2\pi k_1k_2\sqrt{\pw(1-\pw)}\log{\n}} \cdot 2^{-\npri \left[\ixyw+\mathbb{D}\left(\x^{(i)} \parallel \rho \right) \right]} \label{eq:mutual} \\
& &= \frac{1}{2\pi k_1k_2\sqrt{\pw(1-\pw)}\log{\n}} \cdot n^{-k_1\sqrt{\n} \left[\ixyw+\mathbb{D}\left(\x^{(i)} \parallel \rho \right) \right]} \label{eq:22.5}
\end{eqnarray}
Equation~\eqref{eq:12} equals the probability that $\X^{(i)}$ falls into one type class $\txywi$, based on standard counting arguments. In Equation~\eqref{eq:13}, we bound the binomial coefficients by the inequality 
$\binom{n}{k} \ge \sqrt{\frac{1}{2\pi k}} 2^{n\mathbb{H}\left(\frac{k}{n}\right)}$,
which is derived from Stirling's approximation. In~\eqref{eq:mutual}, the term $\I(\x^{(i)}; \yw^{(i)}) \triangleq \sum_{(a,b) \in \left\{0,1\right\} \times \left\{0,1\right\}}f^{xz}_{ab}\log{\frac{f^{xz}_{ab}}{f^x_{a}\cdot f^z_{b}}}$ is the empirical mutual information between $\x^{(i)}$ and $\yw^{(i)}$, and $\mathbb{D}(\x^{(i)} \parallel \rho) \triangleq f^x_{0}\log{\frac{f^x_{0}}{1-\rho}}+f^x_{1}\log{\frac{f^x_{1}}{\rho}}$ is the empirical KL divergence between $\x^{(i)}$ and the code design parameter $\rho$. Note that $\ixyw$ is a function of the triplet ($f^z_{1}, \ f^{xz}_{10}, \ f^{xz}_{11}$), and $\mathbb{D}\left(\x^{(i)} \parallel \rho \right)$ is a function of the pair $(f^{xz}_{10}, \ f^{xz}_{11})$. The range of $f^z_{1}$, $f^{xz}_{10}$, and $f^{xz}_{11}$ are the intervals $\big[(\rho *\pw)(1 \pm \Delta^z_{1})\big]$, $\big[\rho \pw (1 \pm \Delta^{xz}_{10})\big]$, and $\big[\rho (1-\pw) (1 \pm \Delta^{xz}_{11})\big]$ respectively since we only consider typical $\yw^{(i)}$ and the conditionally typical inner-codewords $\x^{(i)}$ here. In Equation~\eqref{eq:22.5}, we substitute the value of $\npri$ as $k_1\sqrt{n}(\log{n})$. 

To figure out the value of ($f^z_{1}, \ f^{xz}_{10}, \ f^{xz}_{11}$) that maximizes $\I(\x^{(i)}; \yw^{(i)}) + \mathbb{D}(\x^{(i)} \parallel\rho)$, we take partial derivatives of $\I(\x^{(i)}; \yw^{(i)})$ and $\mathbb{D}(\x^{(i)} \parallel\rho)$ with respect to $f^z_{1}$, $f^{xz}_{10}$ and $f^{xz}_{11}$ in Appendix~\ref{app:max}. It turns out that for different value of $\pw$, the maximal value of $\I(\x^{(i)}; \yw^{(i)}) + \mathbb{D}(\x^{(i)} \parallel\rho)$ is attained at different points. Though we do not derive the specific value of ($f^z_{1}, \ f^{xz}_{10}, \ f^{xz}_{11}$) that maximizes $\I(\x^{(i)}; \yw^{(i)}) + \mathbb{D}(\x^{(i)} \parallel\rho)$, in Appendix~\ref{app:max} we can still make sure that the maximum is attained at one of the four ``corner points'' given by $f^z_{1} = \rho *\pw, f^{xz}_{10} = \rho \pw (1 \pm \Delta^{xz}_{10})$ and $f^{xz}_{11} = \rho (1-\pw) (1 \pm \Delta^{xz}_{11})$. In Appendix~\ref{app:max2}, we prove that there exists an explicitly computable constant $c_1$ such that for sufficiently large $n$,
\begin{align}
 -k_1\sqrt{n}\left[\I(\x^{(i)}; \yw^{(i)}) + \mathbb{D}(\x^{(i)} \parallel\rho)\right] \ge k_1\cdot \left( \max_{i\in \{1,2,3,4\}} \{g_i(\pw,\epsilon_d,\Delta^{xz}_{10},\Delta^{xz}_{11})\}+c_1n^{-1/2}\right),\label{eq:feng6}
\end{align}
where the auxiliary functions $g_i(\cdot,\cdot,\cdot,\cdot)$ are defined in~\eqref{eq:d}-\eqref{eq:d4}, Section~\ref{sec:result}. Recall that as specified in Section~\ref{sec:code-A}, the size of the codebook $\ci$ equals $2^{\rin \log n}$, where
\begin{align}
\rin = \frac{r_u(1-(\log{n})^{-1/4})k_1}{\lambda} = k_1r_u\frac{1-(\log n)^{-1/4}}{1-(28/\log n)} \ge k_1\left(r_u - c_2(\log n)^{-1/4}\right), \label{eq:rate_copy_2}
\end{align}
for some constant $c_2 > 0$.
Hence substituting~\eqref{eq:22.5},~\eqref{eq:feng6},~\eqref{eq:rate_copy_2} into~\eqref{eq:11} yields that the expected number of inner-codewords $\x^{(i)}$ falling into the type-class $\txywi$ equals
\begin{align}
\mathbb{E}_{\Ci}\left(\left|\wi: \x^{(i)}_w \in \txywi (f_{10}^{xz}, f_{11}^{xz})\right| \right) &= \mathbb{P}_{\X^{(i)}} \left( \X^{(i)} \in \txywi (f_{10}^{xz}, f_{11}^{xz})\right) \cdot \left| \ci \right|   \\
&\ge \frac{1}{2\pi k_1k_2\sqrt{\pw(1-\pw)}\log{\n}} \cdot n^{-k_1\sqrt{\n} \left[\ixyw+\mathbb{D}\left(\x^{(i)} \parallel \rho \right) \right]} \cdot n^{k_1\left(r_u - c_2(\log n)^{-1/4}\right)}  \\
&\ge \frac{1}{2\pi k_1k_2\sqrt{\pw(1-\pw)}\log{\n}} \cdot \n^{k_1\left[\ru+ \max_{i\in \{1,2,3,4\}} \{g_i(\pw,\epsilon_d,\Delta^{xz}_{10},\Delta^{xz}_{11})\}-c_2(\log n)^{-1/4}+c_1n^{-1/2}\right]}. \label{eq:exp6} \\
& \ge \n^{k_1\left[\ru+ \max_{i\in \{1,2,3,4\}} \{g_i(\pw,\epsilon_d,\Delta^{xz}_{10},\Delta^{xz}_{11})\}\right]-\frac{\delta}{2}},\label{eq:exp5}
\end{align}
where~\eqref{eq:exp5} is true for sufficiently large $n$, and $\delta=0.01$ is the slackness parameter first defined in Section~\ref{sec:result}.
The code chunk length design parameter $k_1$ is chosen to satisfy Equations~\eqref{eq:minmax},~\eqref{eq:condition1} and~\eqref{eq:condition4}, and in turn guarantees that $k_1\ru+ k_1 \max_{j\in \{1,2,3,4\}} g_j(\pw,\epsilon_d,\Delta^{xz}_{10},\Delta^{xz}_{11}) \ge 3/2+\delta$. Therefore, we have 
\begin{align}
\mathbb{E}_{\Ci}\left(\left|\wi: \x^{(i)}_w \in \txywi (f_{10}^{xz}, f_{11}^{xz})\right| \right) \ge n^{\frac{3}{2}+\frac{\delta}{2}}.\label{eq:claim}
\end{align}
By the Chernoff bound\footnote{We state the version of the Chernoff bound we used here (and throughout this paper) in Appendix~\ref{app:chernoff}, since there are many different versions of the Chernoff bound in the literature.}~\cite{chernoff1952measure}, the \emph{actual} number of inner-codewords falling into one type class is tightly concentrated around its expectation, \emph{i.e.},
\begin{align}
\mathbb{P}_{\Ci}\left[ \left|\wi: \x^{(i)}_w \in \txywi (f_{10}^{xz}, f_{11}^{xz})\right| \in \left(1 \pm 2\n^{-\frac{1}{2}-\frac{\delta}{4}}\right) \mathbb{E}_{\Ci}\left( \left|\wi: \x^{(i)}_w \in \txywi (f_{10}^{xz}, f_{11}^{xz})\right| \right) \right] = 1 - 2\exp{\left(- 4\sqrt{n}/3 \right)}.  \label{eq:concentration}
\end{align}
Returning now to estimating the term in~\eqref{eq:5}, we thus conclude that with probability (over the inner code design) at least $1- 2\exp{\left(- 4\sqrt{n}/3 \right)}$,
\begin{align}
&\frac{1}{2}\sum_{\yw^{(i)} \in \aywi}\left|\sum_{\ci}P(\ci)\sum_{\wi: \x^{(i)}_w \in \axywi} \frac{1}{n^{\rin}} \pywxi - \sum_{\wi: \x^{(i)}_w \in \axywi} \frac{1}{n^{\rin}} \pywxi \right|  \\
&\le \frac{1}{2} \sum_{\yw^{(i)} \in \aywi} \sum_{(f_{10}^{xz},f_{11}^{xz})\in \mathcal{F}^{xz}_{\B}} \frac{1}{n^{\rin}} P(\yw^{(i)}|f_{10}^{xz},f_{11}^{xz}) \cdot \left| \mathbb{E}_{\Ci} \left( \left|\wi: \x^{(i)}_w \in \txywi (f_{10}^{xz}, f_{11}^{xz})\right| \right) - \left|\wi: \x^{(i)}_w \in \txywi (f_{10}^{xz}, f_{11}^{xz})\right| \right|  \label{eq:14} \\
&\le \frac{1}{2} \sum_{\yw^{(i)} \in \aywi} \sum_{(f_{10}^{xz},f_{11}^{xz})\in \mathcal{F}^{xz}_{\B}} \frac{1}{n^{\rin}} P(\yw^{(i)}|f_{10}^{xz},f_{11}^{xz}) \cdot 2\n^{-\frac{1}{2}-\frac{\delta}{4}} \mathbb{E}_{\Ci}\left(\left|\wi: \x^{(i)}_w \in \txywi (f_{10}^{xz}, f_{11}^{xz})\right| \right)   \label{eq:15} \\
&\le \n^{-\frac{1}{2}-\frac{\delta}{4}}.  \label{eq:16}
\end{align}  
The reasons for inequalities~\eqref{eq:14}-\eqref{eq:16} are as follows. Analogously to the decomposition in Equations~\eqref{eq:8}-\eqref{eq:11}, to obtain Equation~\eqref{eq:14}, we decompose the conditionally typical set $\axywi$ into the summation over all the conditional type class $\txywi (f_{10}^{xz}, f_{11}^{xz})$. Equation~\eqref{eq:15} follows from the fact, stated in~\eqref{eq:concentration}, that the number of inner-codewords falling into one conditional type class is tightly concentrated around its expectation. Equation~\eqref{eq:16} holds since 
\begin{eqnarray}
\lefteqn{\sum_{\yw^{(i)} \in \aywi} \sum_{(f_{10}^{xz},f_{11}^{xz})\in \mathcal{F}^{xz}_{\B}} \frac{1}{n^{\rin}} P(\yw^{(i)}|f_{10}^{xz},f_{11}^{xz}) \cdot \mathbb{E}_{\Ci}\left( \left|\wi: \x^{(i)}_w \in \txywi (f_{10}^{xz}, f_{11}^{xz})\right| \right) }  \\ 
& & = \sum_{\ci} P(\ci) \sum_{\yw^{(i)} \in \aywi} \sum_{(f_{10}^{xz},f_{11}^{xz})\in \mathcal{F}^{xz}_{\B}} \frac{1}{n^{\rin}} P(\yw^{(i)}|f_{10}^{xz},f_{11}^{xz}) \cdot  \left|\wi: \x^{(i)}_w \in \txywi (f_{10}^{xz}, f_{11}^{xz})\right| \label{xinyuan1} \\
& & = \sum_{\ci} P(\ci) \sum_{\yw^{(i)} \in \aywi} \sum_{(f_{10}^{xz},f_{11}^{xz})\in \mathcal{F}^{xz}_{\B}}  \sum_{\wi: \x^{(i)}_w \in \txywi(f^{xz}_{10},f^{xz}_{11})}\frac{1}{n^{\rin}} \pywxi  \\
& & \le \sum_{\ci} P(\ci) \sum_{\yw^{(i)}} \sum_{\wi}  \frac{1}{n^{\rin}} \pywxi \label{xinyuan2} \\
& & = 1.
\end{eqnarray}
We interchange the order of summations in Equation~\eqref{xinyuan1}. Inequality~\eqref{xinyuan2} is obtained by dropping the requirements that $\yw^{(i)}$ is typical and $\x^{(i)}$ is conditionally typical.
This completes the proof of Claim~\ref{claim:1}.  \qed

In the following, we show that as $n$ grows without bound, the probability (averaged over the inner code design) of receiving an atypical $\yw^{(i)}$ goes to zero, and the probability (averaged over the inner code design) that a typical $\yw^{(i)}$ is received and a conditionally atypical inner-codeword $\x^{(i)}$ is transmitted also goes to zero. We choose $\Delta_{1}^z$ as $n^{-1/4+\delta/2}$ (recall that $\Delta_{1}^z$ is the parameter, defined in Section~\ref{sec:def}, specifying the ``width'' of the narrow typical set $\mathcal{A}_{\B}^1(Z)$).

\begin{claim}[\bf Second term in~\eqref{eq:6}] \label{claim:3}
The probability (averaged over the inner code design) of receiving an atypical $\yw^{(i)}$ is bounded from above as
$$\mathbb{E}_{\Ci} \left( \sum_{\yw^{(i)} \notin \aywi} \pywi \right) \le 2\exp{\left(-\frac{k_1 (\rho * q)}{3} n^{\delta} \log n \right)}.$$
\end{claim}

\noindent{\emph{Proof:}} Note that the ensemble-averaged distribution $\mathbb{E}_{\Ci}\left(P_1^{(i)}(\Yw^{(i)})\right)$ is a Bernoulli($\rho * \pw$) distribution, since it corresponds to an inner-codeword $\x^{(i)}$ being chosen according to a Bernoulli($\rho$) distribution, and then $\X^{(i)}$ passing through a BSC($q$). The probability that a $\Yw^{(i)}$ generated in this manner is atypical, \emph{i.e.}, the type-class $f^z_{1}$ falls outside the range $[(\rho*\pw)(1 \pm \Delta^{z}_{1}) ]$, is at most $2\exp{\left(-\frac{k_1 (\rho * q)}{3} n^{\delta} \log n \right)}$ by the Chernoff bound, since the value of $\Delta_{1}^z$ is chosen as $n^{-1/4+\delta/2}$. More specifically, we have
\begin{align}
\mathbb{E}_{\Ci} \left(\sum_{\yw^{(i)} \notin \aywi} \pywi \right) &=  \mathbb{P}_{\Ci, W^{(i)}, \zw}\left(\Yw^{(i)} \notin \aywi \right)  \\
&= \mathbb{P}_{\Ci, W^{(i)}, \zw}\left(f^{z}_{1}(\Yw^{(i)}) \notin \left[(1 \pm \Delta^{z}_{1} )\rho*\pw \right] \right)  \\
&= 2\exp{\left(-\frac{k_1 (\rho * q)}{3} n^{\delta} \log n \right)} .
\end{align}
   \qed

\begin{claim}[\bf First term in~\eqref{eq:6}] \label{claim:4}
The probability (averaged over the inner code design) that a typical $\yw^{(i)}$ is received and a conditionally atypical inner-codeword $\x^{(i)}$ is transmitted is bounded from above as
$$\mathbb{E}_{\Ci} \left(\sum_{\yw^{(i)} \in \aywi} \sum_{\wi: \x^{(i)}_w \notin \axywi} \frac{1}{n^{\rin}} \pywxi \right) \le 4 n^{-\frac{1}{2}-\frac{\delta}{2}}.$$
\end{claim}
\noindent{\emph{Proof:}} We rewrite the first term in~\eqref{eq:6} as 
\begin{align}
&\mathbb{E}_{\Ci} \left(\sum_{\yw^{(i)}} \sum_{\wi} \frac{1}{n^{\rin}} \pywxi \mathbbm{1}\left\{\yw^{(i)} \in \aywi, \x^{(i)}_w \notin \axywi \right\} \right)   \\
&= \frac{1}{n^{\rin}} \sum_{\wi} \sum_{\x^{(i)}_w \in \{0,1\}^B }P_{\X}(\x^{(i)}_w) \sum_{\yw^{(i)}} \pywxi \mathbbm{1}\left\{\yw^{(i)} \in \aywi, \x^{(i)}_w \notin \axywi \right\}  \label{eq:lab1} \\
&= \sum_{\x^{(i)}\in \{0,1\}^B} \sum_{\yw^{(i)}} P_{\X}(\x^{(i)}) P(\yw^{(i)}|\x^{(i)}) \mathbbm{1}\left\{\yw^{(i)} \in \aywi, \x^{(i)} \notin \axywi \right\}  \label{eq:lab3} \\
&= \mathbb{P}_{\X^{(i)}\Yw^{(i)}}\left(\Yw^{(i)} \in \aywi, \X^{(i)} \notin \axYwi \right) \label{eq:18} \\
&\le \mathbb{P}_{\X^{(i)}\Yw^{(i)}}\left(\X^{(i)} \notin \axYwi \right)  \\
&=  \mathbb{P}_{\X^{(i)}\Yw^{(i)}}\left( \left\{f_{10}^{xz}(\X^{(i)},\Yw^{(i)}) \notin \left[(1-\Delta_{10}^{xz})\rho \pw, (1+\Delta_{10}^{xz})\rho \pw\right]\right\} \bigcup \left\{ f_{11}^{xz}(\X^{(i)},\Yw^{(i)}) \notin \left[(1-\Delta_{11}^{xz})\rho (1-\pw), (1+\Delta_{11}^{xz})\rho (1-\pw)\right] \right\} \right). 
\end{align}
Equation~\eqref{eq:lab1} follows since for each $w^{(i)}$, the only random variable of interests is $\X^{(i)}_w$. We use the generic symbol $\x^{(i)}$ starting from~\eqref{eq:lab3} since the term in~\eqref{eq:lab3} is exactly the same for each $\wi$. By the standard counting arguments, we obtain
\begin{align}
&\mathbb{P}_{\X^{(i)}\Yw^{(i)}}\left( \left\{f_{10}^{xz}(\X^{(i)},\Yw^{(i)}) \notin \left[(1-\Delta_{10}^{xz})\rho \pw, (1+\Delta_{10}^{xz})\rho \pw\right]\right\} \bigcup \left\{ f_{11}^{xz}(\X^{(i)},\Yw^{(i)}) \notin \left[(1-\Delta_{11}^{xz})\rho (1-\pw), (1+\Delta_{11}^{xz})\rho (1-\pw)\right] \right\} \right) \notag \\
& \le \sum_{i_1=k_1k_2\pw (\log{\n})\left(1+\Delta_{10}^{xz}\right)}^{k_1\sqrt{\n}\log{\n}} \binom{k_1\sqrt{\n}\log{\n}}{i_1}\left(\frac{k_2\pw}{\sqrt{\n}}\right)^{i_1}\left(1-\frac{k_2\pw}{\sqrt{\n}}\right)^{k_1\sqrt{\n}\log{\n}-i_1}  \label{eq:term1} \\
&+ \sum_{i_2=0}^{k_1k_2\pw (\log{\n})\left(1-\Delta_{10}^{xz}\right)} \binom{k_1\sqrt{\n}\log{\n}}{i_2}\left(\frac{k_2\pw}{\sqrt{\n}}\right)^{i_2}\left(1-\frac{k_2\pw}{\sqrt{\n}}\right)^{k_1\sqrt{\n}\log{\n}-i_2}  \label{eq:term2} \\
&+ \sum_{i_3=k_1k_2(1-\pw) (\log{\n})\left(1+\Delta_{11}^{xz}\right)}^{k_1\sqrt{\n}\log{\n}} \binom{k_1\sqrt{\n}\log{\n}}{i_3}\left(\frac{k_2(1-\pw)}{\sqrt{\n}}\right)^{i_3}\left(1-\frac{k_2(1-\pw)}{\sqrt{\n}}\right)^{k_1\sqrt{\n}\log{\n}-i_3} \label{eq:term3}  \\
&+ \sum_{i_4=0}^{k_1k_2(1-\pw) (\log{\n})\left(1-\Delta_{11}^{xz}\right)} \binom{k_1\sqrt{\n}\log{\n}}{i_4}\left(\frac{k_2(1-\pw)}{\sqrt{\n}}\right)^{i_4}\left(1-\frac{k_2(1-\pw)}{\sqrt{\n}}\right)^{k_1\sqrt{\n}\log{\n}-i_4}. \label{eq:term4}
\end{align} 
Here the four terms in~\eqref{eq:term1}-\eqref{eq:term4} correspond to the four possible atypical ranges for the pair ($f^{xz}_{10},f^{xz}_{11}$). For notational convenience we define an auxiliary function $h(i)$ as 
\begin{align}
h(i) = \binom{k_1\sqrt{\n}\log{\n}}{i}\left(\frac{k_2\pw}{\sqrt{\n}}\right)^{i}\left(1-\frac{k_2\pw}{\sqrt{\n}}\right)^{k_1\sqrt{\n}\log{\n}-i}. \label{eq:hi}
\end{align}
Appendix~\ref{app:series} shows that as $n$ grows without bound, the terms in~\eqref{eq:term1}-\eqref{eq:term4} respectively satisfy
\begin{align}
\sum_{i_1=k_1k_2\pw (\log{\n})\left(1+\Delta_{10}^{xz}\right)}^{k_1\sqrt{\n}\log{\n}} h(i_1) &\le n^{-k_1k_2\pw f(\Delta^{xz}_{10})+\delta/2}, \label{eq:exp1}\\
\sum_{i_2=0}^{k_1k_2\pw (\log{\n})\left(1-\Delta_{10}^{xz}\right)} h(i_2) &\le n^{-k_1k_2\pw f(\Delta^{xz}_{10})+\delta/2}, \label{eq:exp2}\\
\sum_{i_3=k_1k_2(1-\pw) (\log{\n})\left(1+\Delta_{11}^{xz}\right)}^{k_1\sqrt{\n}\log{\n}} h(i_3) &\le n^{-k_1k_2(1-\pw) f(\Delta^{xz}_{11})+\delta/2}, \label{eq:exp3} \\
\sum_{i_4 = 0}^{k_1k_2(1-\pw) (\log{\n})\left(1-\Delta_{11}^{xz}\right)} h(i_4) &\le n^{-k_1k_2(1-\pw) f(\Delta^{xz}_{11})+\delta/2}, \label{eq:exp4}
\end{align}
where the auxiliary function $f(\cdot)$ is as defined in Section~\ref{sec:result}. Recall that Equation~\eqref{eq:minmax} together with Equations~\eqref{eq:condition2}-\eqref{eq:condition4} require the code chunk length design parameter $k_1$ to satisfy the following two conditions\footnote{In order to show the probability (averaged over the inner code design) that a typical $\yw^{(i)}$ is received and a conditionally atypical inner-codeword $\x^{(i)}$ is transmitted is bounded from above by $\mathcal{O}(n^{-1/2-\delta})$, we require each of the four terms in~\eqref{eq:term1}-\eqref{eq:term4} to scale as $\mathcal{O}(n^{-1/2-\delta})$. And this requirement, in turn, forces us to scale $\Delta^{xz}_{10}$ and $\Delta^{xz}_{11}$ as constants (in the interval $[0,1]$) that satisfy the constraints in~\eqref{eq:case1} and~\eqref{eq:case2}. By contrast, the parameters $\Delta^{xy}_{10}$ and $\Delta^{xy}_{11}$, first defined in~\eqref{hexi6}, specifying the ``width'' of Bob's conditionally typical set $\axybi$, scale as $(\log n)^{-1/3}$ since the probability of error is not necessarily required to decay faster than $\mathcal{O}(n^{-1/2})$. The scalings of $\Delta^{xy}_{10}$ and $\Delta^{xy}_{11}$, which are diminishing functions of $n$, make the proof of reliability a lot easier since the Chernoff bound is applicable in this case.}:
\begin{align}
k_1k_2\pw \cdot f(\Delta_{10}^{xz}) &\ge 1/2+\delta,  \label{eq:case1}\\
k_1k_2(1- \pw) \cdot f(\Delta_{11}^{xz}) &\ge 1/2+\delta.  \label{eq:case2}
\end{align}
Therefore, we have 
\begin{align}
\mathbb{E}_{\Ci} \left(\sum_{\yw^{(i)} \in \aywi} \sum_{\wi: \x^{(i)}_w \notin \axywi} \frac{1}{n^{\rin}} \pywxi \right) \le  4n^{-\frac{1}{2}-\frac{\delta}{2}}. 
\end{align}
\qed

In Claim~\ref{claim:5}, we show with high probability over the inner code design, for the randomly chosen inner code $\ci$, the probability of receiving an atypical $\yw^{(i)}$ plus the probability of receiving a typical $\yw^{(i)}$ induced by a conditionally atypical inner-codeword $\x^{(i)}$ is polynomially small. 

\begin{claim}[\bf Terms in~\eqref{eq:7}] \label{claim:5}
With probability at least $1- 2\exp{\left(-4 \sqrt{n}/3 \right)}$ over the inner code design, the randomly chosen inner code $\ci$ satisfies
\begin{align}
\frac{1}{2} \sum_{\yw^{(i)} \in \aywi}\sum_{\wi: \x^{(i)}_w \in \axywi}\frac{1}{n^{\rin}} \pywxi + \frac{1}{2} \sum_{\yw^{(i)} \notin \aywi}P_1^{(i)}(\yw^{(i)}) \le 4 n^{-\frac{1}{2}-\frac{\delta}{4}}. \label{eq:claim9}
\end{align}
\end{claim}
\noindent{\emph{Proof:}} By combining Claim~\ref{claim:1} and Claim~\ref{claim:4}, with probability (over the inner code design) at least $1- 2\exp{\left(-4 \sqrt{n}/3 \right)}$, we have
\begin{eqnarray}
\lefteqn{n^{-\frac{1}{2}-\frac{\delta}{4}}+2 n^{-\frac{1}{2}-\frac{\delta}{2}}}  \\
& &\ge  \frac{1}{2}\sum_{\yw^{(i)} \in \aywi}\left|\sum_{\ci}P(\ci)\sum_{\wi: \x^{(i)}_w \in \axywi} \frac{1}{n^{\rin}} \pywxi - \sum_{\wi:\x^{(i)}_w \in \axywi}\frac{1}{n^{\rin}} \pywxi \right| \notag \\
& &\qquad\qquad\qquad\qquad\qquad\qquad\qquad\qquad\qquad+ \frac{1}{2}\mathbb{E}_{\Ci} \left(\sum_{\yw^{(i)} \in \aywi} \sum_{\wi: \x^{(i)}_w \notin \axywi}\frac{1}{n^{\rin}} \pywxi \right) \label{eq:combine} \\
& &\ge  \frac{1}{2}\sum_{\yw^{(i)} \in \aywi}\left(\sum_{\ci}P(\ci)\sum_{\wi:\x^{(i)}_w \in \axywi} \frac{1}{n^{\rin}} \pywxi \right) + \frac{1}{2}\mathbb{E}_{\Ci} \left(\sum_{\yw^{(i)} \in \aywi} \sum_{\wi: \x^{(i)}_w \notin \axywi}\frac{1}{n^{\rin}} \pywxi \right)  \notag \\
& &\qquad\qquad\qquad\qquad\qquad\qquad\qquad\qquad\qquad - \frac{1}{2}\sum_{\yw^{(i)} \in \aywi}\sum_{\wi: \x^{(i)}_w \in \axywi} \frac{1}{n^{\rin}} \pywxi,  \label{eq:20}
\end{eqnarray}
where Equation~\eqref{eq:combine} follows from Claim~\ref{claim:1} and Claim~\ref{claim:4}, and Equation~\eqref{eq:20} follows from the triangle inequality. Note that the summation of the first two terms of~\eqref{eq:20} equals the probability of receiving a typical $\yw^{(i)}$ under the ensemble-averaged active distribution, which, by Claim~\ref{claim:3}, equals 
\begin{align}
\frac{1}{2} \mathbb{E}_{\Ci} \left(\sum_{\yw^{(i)} \in \aywi} \pywi \right) = \frac{1}{2} - \frac{1}{2}\mathbb{E}_{\Ci} \left(\sum_{\yw^{(i)} \notin \aywi} \pywi \right) \ge \frac{1}{2} - \exp{\left(-\frac{k_1 (\rho * q)}{3} n^{\delta} \log n \right)}. \label{eq:first}
\end{align}
For the third term of Equation~\eqref{eq:20}, we have 
\begin{eqnarray}
\lefteqn{\frac{1}{2}\sum_{\yw^{(i)} \in \aywi} \sum_{\wi:\x^{(i)}_w \in \axywi}\frac{1}{n^{\rin}} \pywxi  }  \\
& & = \frac{1}{2} - \frac{1}{2} \sum_{\yw^{(i)} \in \aywi} \sum_{\wi:\x^{(i)}_w \notin \axywi}\frac{1}{n^{\rin}} \pywxi  - \frac{1}{2} \sum_{\yw^{(i)} \notin \aywi}P_1^{(i)}(\yw^{(i)}). \label{eq:third}
\end{eqnarray}
Hence, combining Equations~\eqref{eq:20},~\eqref{eq:first} and~\eqref{eq:third}, for sufficiently large $n$, with probability (over the inner code design) at least $1- 2\exp{\left(-4 \sqrt{n}/3 \right)}$, the chosen inner code $\ci$ satisfies 
\begin{align}
&\frac{1}{2} \sum_{\yw^{(i)} \in \aywi} \sum_{\wi:\x^{(i)}_w \notin \axywi}\frac{1}{n^{\rin}} \pywxi + \frac{1}{2} \sum_{\yw^{(i)} \notin \aywi}P_1^{(i)}(\yw^{(i)}) \\
&\le  n^{-\frac{1}{2}-\frac{\delta}{4}} + 2 n^{-\frac{1}{2}-\frac{\delta}{2}} + \exp{\left(-\frac{k_1 (\rho * q)}{3} n^{\delta} \log n \right)} \\
&\le  4 n^{-\frac{1}{2}-\frac{\delta}{4}}.
\end{align}  
This completes the proof of Claim~\ref{claim:5}.  \qed

\vskip 0.2cm
Equipped with Claims~\ref{claim:1}-\ref{claim:5}, we are able to show that for any chunk $i \in \left\{1,\ldots, L\right\}$, with high probability over the inner code design, the ensemble-averaged active distribution $\mathbb{E}_{\Ci}\left(P_1^{(i)}(\Yw^{(i)})\right)$ and the active distribution $P_1^{(i)}(\Yw^{(i)})$ are sufficiently close.  
\begin{claim} \label{claim:6}
For any $i \in \left\{1,\ldots, L\right\}$, with probability at least $1- 2\exp{\left(-4 \sqrt{n}/3 \right)}$ over the inner code design, for the randomly chosen inner code $\ci$, the variational distance between the ensemble-averaged active distribution $\mathbb{E}_{\Ci}\left(P_1^{(i)}(\Yw^{(i)})\right)$ and the active distribution $P_1^{(i)}(\Yw^{(i)})$ is bounded from above as 
$$\frac{1}{2}\sum_{\yw^{(i)} \in \left\{0, 1\right\}^{\B}}\left|\mathbb{E}_{\Ci} \left(P_1^{(i)}(\yw^{(i)}) \right) - P_1^{(i)}(\yw^{(i)} )\right| \le 8 n^{-\frac{1}{2}-\frac{\delta}{4}}.$$
\end{claim}
\noindent{\emph{Proof:}} For any $i \in \left\{1,\ldots, L\right\}$ and sufficiently large $n$, with probability (over the inner code design) at least $1- 2\exp{\left(-4 \sqrt{n}/3 \right)}$,  
\begin{eqnarray}
\lefteqn{\frac{1}{2}\sum_{\yw^{(i)} \in \left\{0, 1\right\}^{\B}}\left|\mathbb{E}_{\Ci} \left(P_1^{(i)}(\yw^{(i)}) \right) - P_1^{(i)}(\yw^{(i)} )\right|} \notag \\
& &\le \frac{1}{2}\sum_{\yw^{(i)} \in \aywi}\left|\sum_{\ci}P(\ci)\sum_{\wi: \x^{(i)}_w \in \axywi}\frac{1}{n^{\rin}} \pywxi - \sum_{\wi: \x^{(i)}_w \in \axywi}\frac{1}{n^{\rin}} \pywxi \right|  \label{eq:rep1}\\
& &+ \frac{1}{2} \sum_{\yw^{(i)} \in \aywi} \mathbb{E}_{\Ci} \left( \sum_{\wi: \x^{(i)}_w \notin \axywi}\frac{1}{n^{\rin}} \pywxi \right) + \frac{1}{2}\mathbb{E}_{\Ci}\left(\sum_{\yw^{(i)} \notin \aywi} \pywi \right)  \label{eq:rep2}\\
& &+ \frac{1}{2}\sum_{\yw^{(i)} \in \aywi} \sum_{\wi: \x^{(i)}_w \notin \axywi}\frac{1}{n^{\rin}} \pywxi  + \frac{1}{2} \sum_{\yw^{(i)} \notin \aywi}P_1^{(i)}(\yw^{(i)}) \label{eq:rep3} \\
& &\le n^{-\frac{1}{2}-\frac{\delta}{4}}+\exp{\left(-\frac{k_1 (\rho * q)}{3} n^{\delta} \log n \right)}+2 n^{-\frac{1}{2}-\frac{\delta}{2}}+4 n^{-\frac{1}{2}-\frac{\delta}{4}} \label{eq:rep4}\\
& &\le 8 n^{-\frac{1}{2}-\frac{\delta}{4}}, \notag
\end{eqnarray}
where inequalities~\eqref{eq:rep1}-\eqref{eq:rep3} are adapted from~\eqref{eq:2}-\eqref{eq:4}, and inequality~\eqref{eq:rep4} follows from Claims~\ref{claim:1}-\ref{claim:5}. \qed
\vskip 0.2cm
In the following, we take one more step to show that with high probability over the concatenated code design, the $n$-letter ensemble-averaged active distribution $\mathbb{E}_{\C}(P_1)$ and the ``chunk-wise independent'' distribution $P_1^{(1)}P_1^{(2)}\cdots P_1^{(L)}$ are close. 

\vspace{6pt}
\noindent{\textbf{Lemma 3}} \ (Restated). \emph{With probability at least $1- \sqrt{n}\exp{\left(-4 \sqrt{n}/3 \right)}$ over the concatenated code design, for the randomly chosen code $\cc$, the variational distance between the ensemble-averaged active distribution $\mathbb{E}_{\C}(P_1)$ and the ``chunk-wise independent'' product distribution $P_1^{(1)}P_1^{(2)} \cdots P_1^{(L)}$ is bounded from above as} 
$$\frac{1}{2} \sum_{\yw^{(1)} \in \{0,1\}^B \ldots \yw^{(L)}\in \{0,1\}^B} \left| \mathbb{E}_{\C}(P_1(\yw^{(1)}, \ldots, \yw^{(L)})) - P^{(1)}_1(\yw^{(1)}) \cdots P^{(L)}_1(\yw^{(L)})\right| \le n^{-\frac{\delta}{4}}.$$

\noindent{\emph{Proof:}} Based on Claim~\ref{claim:6} and the union bound, it is then the case that with probability at least $1- L\cdot 2\exp{\left(-4 \sqrt{n}/3 \right)}$ over the concatenated code design, the variational distance between the ensemble-averaged active distribution and ``chunk-wise independent'' product distribution is bounded from above as 
\begin{eqnarray}
\lefteqn{\frac{1}{2} \sum_{\yw^{(1)} \in \{0,1\}^B \ldots \yw^{(L)}\in \{0,1\}^B} \left| \mathbb{E}_{\C}(P_1(\yw^{(1)}, \ldots, \yw^{(L)})) - P^{(1)}_1(\yw^{(1)}) \cdots P^{(L)}_1(\yw^{(L)})\right| } \\
& &\le  \frac{1}{2} \sum_{i=1}^L \sum_{\yw^{(i)}\in \left\{0,1\right\}^{\npri}} \left| \mathbb{E}_{\Ci}\left(P_1^{(i)}(\yw^{(i)})\right)-P_1^{(i)}(\yw^{(i)})\right|  \\
& &= 8L \cdot n^{-\frac{1}{2}-\frac{\delta}{4}} \\
& &= \frac{8\sqrt{n}}{k_1\log{n}}\cdot n^{-\frac{1}{2}-\frac{\delta}{4}}  \\
& &\le  n^{-\frac{\delta}{4}},
\end{eqnarray}
for sufficiently large $n$. Note that $1- L\cdot 2\exp{\left(-4 \sqrt{n}/3 \right)} \le 1- \sqrt{n}\exp{\left(-4 \sqrt{n}/3 \right)}$, since $L = \sqrt{n}/(k_1\log{n})$.
This completes the proof of Lemma~\ref{lemma:(a)}.
\qed

%%%%%%%%%%%%%%%%%%%%%%%%%%%%%%%%%%%%%%%%%%%%%%%%%%%%%%%%%%%%%%%%%%%%%%%%%%%%%%%%%%%%%%%%%%%%%%%%%%%%%%%%

\subsection{Proof of Lemma~\ref{lemma:(b)}:} \label{sec:deniability-B}

Lemma~\ref{lemma:(b)} aims to bound the variational distance between the $n$-letter active distribution $P_1(\Yw^{(1)}, \ldots, \Yw^{(L)})$ and the ``chunk-wise independent'' product distribution $P_1^{(1)}(\Yw^{(1)}) \cdots P_1^{(L)}(\Yw^{(L)})$. Let $l_1 \triangleq \lambda L$ be the number of systematic chunks and $l_2 \triangleq L(1-\lambda)$ be the number of parity chunks. We first note that for any $\yw^{(1)}, \ldots, \yw^{(L)}$, the $n$-letter active distribution can be decomposed as
\begin{align}
P_1(\yw^{(1)}, \ldots, \yw^{(L)}) &= P(\yw^{(1)}, \ldots, \yw^{(l_1)}) \cdot P(\yw^{(l_1+1)},\ldots, \yw^{(L)}|\yw^{(1)},\ldots, \yw^{(l_1)}) \\
&= P^{(1)}_1(\yw^{(1)}) \cdots P^{(l_1)}_1(\yw^{(l_1)}) \cdot P(\yw^{(l_1+1)},\ldots, \yw^{(L)}|\yw^{(1)},\ldots, \yw^{(l_1)}),
\end{align}
since the inner codes in the first $l_1$ systematic chunks, and also the messages $\W^{(i)} = \M^{(i)}$ that are inputs to those chunks, are all independent. However, the analysis for the remaining $l_2$ parity chunks is more involved since the Reed-Solomon outer code in general introduces correlations between $\W^{(i)}$ in the $l_1$ systematic chunks (which are $l_1$-wise independent) and any $\W^{(i^\prime)}$ in a parity chunk --- in particular, any such $\W^{(i^\prime)}$ is a linear combination of $\W^{(i)}$ in the $l_1$ systematic chunks. The crux of the proof is to show that conditioned on Willie's typical observations ($\yw^{(1)}, \ldots, \yw^{(l_1)}$) on systematic chunks, the {\it parity inner-message vectors} $(W^{(l_1+1)}, \ldots, W^{(L)})$ are almost uniformly distributed from Willie's perspective (essentially statistically independent of ($\yw^{(1)}, \ldots, \yw^{(l_1)}$)). The uniformity of parity inner-message vectors further implies that $P(\yw^{(l_1+1)}\cdots \yw^{(L)}|\yw^{(1)}\cdots \yw^{(l_1)}) \approx  \prod_{i=l_1+1}^L P^{(i)}_1(\yw^{(i)})$ and $P(\yw^{(1)}\cdots \yw^{(L)}) \approx  \prod_{i=1}^L P^{(i)}_1(\yw^{(i)})$. We make it more concrete in the following.

Willie's observations $(\yw^{(1)},\ldots, \yw^{(\LL)})$ on systematic chunks is said to be typical if each of $\yw^{(i)}$ is typical, \emph{i.e.}
\begin{align}
(\yw^{(1)},\ldots, \yw^{(\LL)}) \in \aywiL \text{ if } \yw^{(i)} \in \aywi, \ \forall i \in \{1,2,\ldots,\LL \}.
\end{align}
To simplify our analysis, we assume there is an oracle revealing to Willie which conditional type class does each of the transmitted inner-codeword (on systematic chunks) $\x^{(i)}_w$ fall into. The oracle revealed information on the $i$-th chunk is denoted by $\tau^{(i)}$, which exactly equals the type class $\txywi(f^{xz}_{10}, f^{xz}_{11})$ defined in Section~\ref{sec:def-A}. Note that this extra information only strengthens Willie since it reduces his uncertainty about Alice's transmissions. Due to the fact that the channel transition probabilities $P(\yw^{(i)}|\x^{(i)}_w)$ are the same for all $\x^{(i)}_w \in \tau^{(i)}$, all the inner-messages $w^{(i)}$ with $\x^{(i)}_w \in \tau^{(i)}$ are equally likely from Willie's perspective. The collection of oracle revealed information $(\tau^{(1)}, \ldots, \tau^{(\LL)})$ is said to be typical if each of $\tau^{(i)}$ is typical, \emph{i.e.},
\begin{align}
  (\tau^{(1)},\ldots, \tau^{(\LL)}) \in \axywiL \text{ if } \tau^{(i)} \in \mathcal{F}^{xz}_{\B}, \ \forall i \in \{1,2,\ldots,\LL \},
\end{align}  
where $\mathcal{F}^{xz}_{\B}$ (defined in Section~\ref{sec:def-A}) contains all conditionally typical type class indicated by $(f^{xz}_{10}, f^{xz}_{11})$.
Note that the distribution of $(W^{(l_1+1)}, \ldots, W^{(L)})$ conditioned on $(\yw^{(1)}, \ldots, \yw^{(\LL)})$ can be expressed as
\begin{align}
&P\left(w^{(l_1+1)} \ldots w^{(L)} \big| \yw^{(1)}\ldots \yw^{(l_1)} \right) \\
&= \sum_{w^{(1)}\ldots w^{(l_1)}} P\left(w^{(l_1+1)} \ldots w^{(L)} \big| w^{(1)} \ldots w^{(l_1)}, \yw^{(1)}\ldots \yw^{(l_1)} \right) \cdot P\left(w^{(1)}\ldots w^{(l_1)} \big| \yw^{(1)} \ldots \yw^{(l_1)}\right) \\
&= \sum_{w^{(1)}\ldots w^{(l_1)}} P\left(w^{(l_1+1)} \ldots w^{(L)} \big| w^{(1)} \ldots w^{(l_1)} \right) \cdot P\left(w^{(1)}\ldots w^{(l_1)} \big| \yw^{(1)} \ldots \yw^{(l_1)}\right) \\
&= \sum_{\left(w^{(1)}\ldots w^{(l_1)}\right) \in \mathcal{S}\left(w^{(l_1+1)} \ldots w^{(L)}\right)} P\left(w^{(1)}\ldots w^{(l_1)} \big| \yw^{(1)} \ldots \yw^{(l_1)}\right)  \label{eq:wei1} \\
& = \frac{\sum_{\left(w^{(1)}\ldots w^{(l_1)}\right) \in \mathcal{S}\left(w^{(l_1+1)} \ldots w^{(L)}\right)} \mathbbm{1}\left\{\x^{(1)}_w \in \tau^{(1)}, \ldots, \x^{(l_1)}_w \in \tau^{(l_1)} \right\} }{\prod_{i=1}^{\LL}\Big|w^{(i)}: \x^{(i)}_w \in \tau^{(i)}\Big|},  \label{eq:wei2}
\end{align}
where $\mathcal{S}(w^{(l_1+1)} \ldots w^{(L)})$ is the set of {\it systematic inner-message vectors} $(w^{(1)} \ldots w^{(\LL)})$ that are encoded to $(w^{(1)} \ldots w^{(\LL)}, w^{(\LL+1)} \ldots w^{(L)})$ by the systematic outer code. Equation~\eqref{eq:wei1} holds since $P(w^{(l_1+1)} \ldots w^{(L)} | w^{(1)} \ldots w^{(l_1)} )$ equals one if $(w^{(1)}\ldots w^{(l_1)}) \in \mathcal{S}(w^{(l_1+1)} \ldots w^{(L)})$, and equals zero otherwise. Equation~\eqref{eq:wei2} is obtained by noting that 
\begin{align}
P(w^{(1)}\cdots w^{(l_1)} | \yw^{(1)} \ldots \yw^{(l_1)})=\frac{1}{\prod_{i=1}^{l_1}|w^{(i)}: \x^{(i)}_w \in \tau^{(i)}|}
\end{align}
if the corresponding inner-codewords (on systematic chunks) satisfy $\x^{(1)}_w \in \tau^{(1)}, \ldots, \x^{(l_1)}_w \in \tau^{(l_1)}$, and equals zero otherwise. 

Let $\Upsilon_i \triangleq \mathbb{P}_{\X^{(i)}}(\X^{(i)} \in \tau^{(i)})$ be the probability that a randomly generated inner-codeword falls into the oracle revealed type class $\tau^{(i)}$, and note that $\Upsilon_i$ depends on the type $\tau^{(i)}$ only. Without loss of generality, we assume $\Upsilon_1 \ge \Upsilon_2 \ge \cdots \ge \Upsilon_{l_1}$. We then concentrate the numerator of~\eqref{eq:wei2} in Claim~\ref{claim:new1}.

\begin{claim} \label{claim:new1}
	For any typical oracle revealed type classes $(\tau^{(1)},\ldots, \tau^{(\LL)}) \in \axywiL$ and {\it parity inner-message vector} $(w^{(l_1+1)} \ldots w^{(L)})$, with probability at least $1 - 2\exp\left(-\frac{3(\ln 2)\sqrt{n}}{2k_1}\right)$ over the code design , 
	\begin{align}
	 \sum_{\left(w^{(1)}\ldots w^{(l_1)}\right) \in \mathcal{S}\left(w^{(l_1+1)} \ldots w^{(L)}\right)} \mathbbm{1}\left\{\x^{(1)}_w \in \tau^{(1)}, \ldots, \x^{(l_1)}_w \in \tau^{(l_1)} \right\} \in (1\pm n^{-1})\left(\nu \cdot \prod_{i=1}^{l_1}\Upsilon_i\right) .
	\end{align}
\end{claim}
\noindent{\it Proof:} Let $\nu \triangleq n^{(l_1-l_2)\rin}$ (where $n^{\rin}$ is the size of inner code). We show in Appendix~\ref{app:matrix} that $|\mathcal{S}(w^{(l_1+1)} \ldots w^{(L)})| = \nu$ for every parity inner-message vector $(w^{(l_1+1)} \ldots w^{(L)})$. For the $j$-th element $(w^{(1)}_j \ldots w^{(\LL)}_j)$ in $\mathcal{S}(w^{(\LL+1)} \ldots w^{(L)})$, we define the corresponding random variable as 
\begin{align}
U_j \triangleq \mathbbm{1}\left\{\x^{(1)}_{w_j} \in \tau^{(1)}, \ldots, \x^{(l_1)}_{w_j} \in \tau^{(l_1)} \right\}, \forall j \in \{1,2,\ldots, \nu\}. \label{eq:james}
\end{align}
Let 
\begin{align}
U \triangleq \sum_{j=1}^{\nu}U_j = \sum_{\left(w^{(1)}\ldots w^{(l_1)}\right) \in \mathcal{S}\left(w^{(l_1+1)} \ldots w^{(L)}\right)} \mathbbm{1}\left\{\x^{(1)}_w \in \tau^{(1)}, \ldots, \x^{(l_1)}_w \in \tau^{(l_1)} \right\}.
\end{align}
It is worth noting that for $j_1 \ne j_2$, the random variables $U_{j_1}$ and $U_{j_2}$ are not necessarily independent since their corresponding systematic inner-message vectors may share some common inner-messages (\emph{i.e.}, $w^{(i)}_{j_1} = w^{(i)}_{j_2}$ for some $i$). Specifically, the probability that $U_{j_1}$ equals one will be larger if $U_{j_2}=1$ and $w^{(i)}_{j_1} = w^{(i)}_{j_2}$ for some $i$, since $\x^{(i)}_{j_1}$ (the inner-codeword for $w^{(i)}_{j_1}$) is already known to belong to $\tau^{(i)}$. In the following, we use the {\it second-moment method} to concentrate $U$. The first moment is given as
\begin{align}
\mathbb{E}(U) = \sum_{j=1}^{\nu}\mathbb{E}(U_j) = \nu \cdot \mathbb{P}\left(\X^{(1)} \in \tau^{(1)}, \ldots, \X^{(l_1)} \in \tau^{(l_1)} \right) = \nu \cdot \prod_{i=1}^{\LL}\Upsilon_i.
\end{align}
Calculating the second moment is more involved because of the dependencies between random variables. We have 
\begin{align}
\mathbb{E}(U^2) = \mathbb{E}\left(\sum_{j=1}^{\nu}U_j^2 + \sum_{j_1 \ne j_2} U_{j_1}U_{j_2} \right) &=\sum_{i=j}^{\nu} \mathbb{E}\left(U_j^2\right) + \sum_{j_1 \ne j_2}\mathbb{E} \left(U_{j_1}U_{j_2} \right) \\
&= \sum_{j=1}^{\nu} \mathbb{E}\left(U_j\right) + \sum_{j_1 \ne j_2} P\left(U_{j_1} = 1, U_{j_2} = 1\right) \\
&= \sum_{j=1}^{\nu} \mathbb{E}\left(U_j\right) + \sum_{j_1 \ne j_2} \left(\prod_{i=1}^{\LL}\Upsilon_i\right) \cdot P\left(U_{j_2} = 1|U_{j_1} = 1 \right). \label{eq:zou}
\end{align}
\begin{definition}
	For $j_1 \ne j_2$, the distance $d(U_{j_1},U_{j_2})$ between two random variables $U_{j_1}$ and $U_{j_2}$ (as introduced in~\eqref{eq:james}) is defined as 
	\begin{align}
	d(U_{j_1},U_{j_2}) \triangleq \left|i \in \{1,2,\ldots, l_1\}: w^{(i)}_{j_1} \ne w^{(i)}_{j_2} \right|.
	\end{align}  
\end{definition} 
Note that the conditional probability $P(U_{j_2} = 1|U_{j_1} = 1 )$ depends on the distance between $U_{j_1}$ and $U_{j_2}$. Specifically, if $d(U_{j_1},U_{j_2}) = t$ and the locations that they differ in are denoted by $\{i_1, \ldots, i_t\}$, then we have 
\begin{align}
P\left(U_{j_2} = 1|U_{j_1} = 1 \right) &= \mathbb{P}\left(\X^{(i_1)} \in \tau^{(i_1)}, \ldots, \X^{(i_t)} \in \tau^{(i_t)} \right)  = \prod_{i=i_1}^{i_t} \Upsilon_{i} \le \prod_{i=1}^{t} \Upsilon_{i},
\end{align}  
where the last step follows from the assumption $\Upsilon_1 \ge \Upsilon_2 \ge \cdots \ge \Upsilon_{l_1}$.
In Appendix~\ref{app:mds}, we show that the number of $j_1 \ne j_2$ such that $d(U_{j_1},U_{j_2}) = t$ equals
\begin{align} 
\nu \cdot \binom{l_1}{t}(n^{\rin}-1)\sum_{i=0}^{t-l_2-1}(-1)^i \binom{t-1}{i}\left(n^{\rin}\right)^{t-i-l_2-1} \ \text{ if } t \ge l_2 + 1, \label{eq:wdmds}
\end{align}
and equals 0 otherwise. Roughly speaking, the proof of~\eqref{eq:wdmds} is inspired by the {\it weight distribution} of {\it Maximum Distance Separable (MDS) code}.
Hence the term in~\eqref{eq:zou} can be expressed as
\begin{align}
&\sum_{j=1}^{\nu} \mathbb{E}\left(U_j\right) + \sum_{t=l_2+1}^{l_1} \sum_{j_1 \ne j_2: d(U_{j_1},U_{j_2}) = t} \left(\prod_{i=1}^{\LL}\Upsilon_i\right) \cdot P\left(U_{j_2} = 1|U_{j_1} = 1 \right) \\
& \le \sum_{j=1}^{\nu} \mathbb{E}\left(U_j\right) + \sum_{t=l_2+1}^{l_1} \left(\prod_{i=1}^{\LL}\Upsilon_i\right)\left(\prod_{i=1}^{t}\Upsilon_i\right)\cdot \nu \cdot \binom{l_1}{t}(n^{\rin}-1)\sum_{i=0}^{t-l_2-1}(-1)^i \binom{t-1}{i}\left(n^{\rin}\right)^{t-i-l_2-1}. \label{eq:zou3}
\end{align}
Let the auxiliary function $\tilde{f}(t) = \left(\prod_{i=1}^{\LL}\Upsilon_i\right)\left(\prod_{i=1}^{t}\Upsilon_i\right) \cdot \nu \cdot \binom{l_1}{t}(n^{\rin}-1)\sum_{i=0}^{t-l_2-1}(-1)^i \binom{t-1}{i}\left(n^{\rin}\right)^{t-i-l_2-1}$, we calculate the ratio between two successive terms as follows:
\begin{align}
\frac{\tilde{f}(t)}{\tilde{f}(t-1)} &= \frac{\left(\prod_{i=1}^{l_1}\Upsilon_i\right)\left(\prod_{i=1}^{t}\Upsilon_i\right) \cdot \left[\nu \cdot \binom{l_1}{t}(n^{\rin}-1)\left(\left(n^{\rin}\right)^{t-l_2-1} - (t-1)\left(n^{\rin}\right)^{t-l_2-2} + \cdots \right)\right]}{\left(\prod_{i=1}^{l_1}\Upsilon_i\right)\left(\prod_{i=1}^{t-1}\Upsilon_i\right) \cdot \left[\nu \cdot \binom{l_1}{t-1}(n^{\rin}-1)\left(\left(n^{\rin}\right)^{t-l_2-2} - (t-2)\left(n^{\rin}\right)^{t-l_2-3} + \cdots \right)\right]} \\
&= \frac{\left(\prod_{i=1}^{l_1}\Upsilon_i\right)\left(\prod_{i=1}^{t}\Upsilon_i\right) \cdot \left[\nu \cdot \binom{l_1}{t}(n^{\rin}-1)\left(n^{\rin(t-l_2-1)} - \mathcal{O}\left(n^{\rin(t-l_2-2)+1/2} \right)\right)\right]}{\left(\prod_{i=1}^{l_1}\Upsilon_i\right)\left(\prod_{i=1}^{t-1}\Upsilon_i\right) \cdot \left[\nu \cdot \binom{l_1}{t-1}(n^{\rin}-1)\left(n^{\rin(t-l_2-2)} - \mathcal{O}\left(n^{\rin(t-l_2-3)+1/2} \right) \right)\right]} \\
&\stackrel{n \to \infty}{=} \Upsilon_t \cdot n^{\rin} \cdot \frac{l_1-t+1}{t} \\
& \ge \frac{n^{\rin} \Upsilon_{l_1}}{l_1}, \label{eq:zou2}
\end{align}
where inequality~\eqref{eq:zou2} holds for all $l_2+1 \le t \le l_1$. It turns out that $\tilde{f}(t)$ is an increasing function since $n^{\rin} \Upsilon_{l_1} \ge n^{\frac{3}{2}+\delta}$ and $l_1 \le \sqrt{n}$. Substituting~\eqref{eq:zou2} into~\eqref{eq:zou3}, we have
\begin{align}
\mathbb{E}(U^2) &= \sum_{j=1}^{\nu} \mathbb{E}\left(U_j\right) + \sum_{t=l_2+1}^{l_1} \tilde{f}(t) \\
&\le \nu \cdot \prod_{i=1}^{l_1}\Upsilon_i + \tilde{f}(l_1)\left(1+ \frac{l_1}{\Upsilon_{l_1}\cdot \left(n^{\rin}\right)} + \left(\frac{l_1}{\Upsilon_{l_1}\cdot \left(n^{\rin}\right)}\right)^2 + \cdots \right) \\
&\le \nu \cdot \prod_{i=1}^{l_1}\Upsilon_i + \left(\prod_{i=1}^{l_1}\Upsilon_i \right)^2 \left(\nu\left( n^{\rin(l_1-l_2)} - l_1 n^{\rin(l_1-l_2-1)} + \cdots \right)\right)\left(1+ \left(\frac{l_2}{\Upsilon_{l_1}\cdot n^{\rin}}\right) + \left(\frac{l_1}{\Upsilon_{l_1}\cdot n^{\rin}}\right)^2 + \cdots \right) \\
&= \nu \cdot \prod_{i=1}^{l_1}\Upsilon_i + \left(\prod_{i=1}^{l_1}\Upsilon_i \right)^2 \left(\nu \left( n^{\rin(l_1-l_2)} - \mathcal{O}\left(n^{\rin(l_1-l_2-1)}\right)\right)\right) \\
&\le \nu \cdot \prod_{i=1}^{l_1}\Upsilon_i + \left(\nu \cdot \prod_{i=1}^{l_1}\Upsilon_i\right)^2.
\end{align}
For any small $\varepsilon >0$, we define an auxiliary random variable $V \triangleq U - (1-\varepsilon)\mathbb{E}(U)$, with 
\begin{align}
&\mathbb{E}(V) = \varepsilon \nu \cdot \prod_{i=1}^{l_1}\Upsilon_i, \\
&\mathbb{E}(V^2) = \mathbb{E}(U^2) + (\varepsilon^2-1)\left(\mathbb{E}(U)\right)^2 \le \nu \cdot \prod_{i=1}^{l_1}\Upsilon_i + \left(\varepsilon \nu \cdot \prod_{i=1}^{l_1}\Upsilon_i\right)^2.
\end{align}
Furthermore, let 
\begin{align}
V' = \begin{cases}
V, \ \text{ if } V > 0, \\
0, \ \text{ otherwise},
\end{cases}
\end{align}
and one can show that $\mathbb{E}(V') \ge \mathbb{E}(V)$ and $\mathbb{E}(V'^2) \le \mathbb{E}(V^2)$.
By setting $\varepsilon = n^{-1}$, as $n$ grows without bound, we have
\begin{align}
\mathbb{P}\left(U > (1-n^{-1})\mathbb{E}(U)\right) = \mathbb{P}(V > 0) &= \mathbb{P}(V' > 0) \\
& \ge \frac{\left(\mathbb{E}(V')\right)^2}{\mathbb{E}(V'^2)} \label{eq:zou4} \\
&\ge \frac{\left(\mathbb{E}(V)\right)^2}{\mathbb{E}(V^2)}\\
& \ge \frac{\left(n^{-1} \nu \cdot \prod_{i=1}^{l_1}\Upsilon_i\right)^2}{\nu \cdot \prod_{i=1}^{l_1}\Upsilon_i + \left(n^{-1} \nu \cdot \prod_{i=1}^{l_1}\Upsilon_i\right)^2} \\
&\ge 1 - n^{-\left(\left(\frac{3}{2}+\delta\right)l_1 - l_2\rin-2\right)} \label{eq:zou5}\\
&\ge 1 - \exp\left(-\frac{3(\ln 2)\sqrt{n}}{2k_1}\right). \label{eq:zou6}
\end{align}
where~\eqref{eq:zou4} follows from the second moment method, and~\eqref{eq:zou5} holds since $n^{\rin}\Upsilon_i \ge n^{\frac{3}{2}+\delta}$ for any typical $\tau^{(i)}$ (as indicated by~\eqref{eq:minmax} and~\eqref{eq:feng6}).  

Similarly, by setting $V = (1+n^{-1})\mathbb{E}(U) - U$, one can also prove that as $n$ grows without bound, 
\begin{align}
\mathbb{P}\Big(U < (1+n^{-1})\mathbb{E}(U)\Big) \ge 1 - \exp\left(-\frac{3(\ln 2)\sqrt{n}}{2k_1}\right),
\end{align}
which completes the proof.
 \qed

\begin{claim} \label{claim:new3}
	For any $(\yw^{(1)},\ldots, \yw^{(\LL)}) \in \aywiL$ and typical oracle revealed type classes $(\tau^{(1)},\ldots, \tau^{(\LL)}) \in \axywiL$, with probability at least $1- 4\exp\left(-\frac{3(\ln 2)\sqrt{n}}{2k_1}\right)$, we have 
	\begin{align}
 P\left(w^{(l_1+1)}, \ldots, w^{(L)} \big| \yw^{(1)}, \ldots, \yw^{(l_1)}\right) \in \frac{(1 \pm 3n^{-\frac{\delta}{4}})}{n^{l_2 \rin}}, \ \ \forall (w^{(\LL+1)}, \ldots, w^{(L)}). 
	\end{align}
\end{claim}

\noindent{\it Proof:}
As noted in~\eqref{eq:wei2}, 
\begin{align}
P\left(w^{(l_1+1)}, \ldots, w^{(L)} \big| \yw^{(1)}, \ldots, \yw^{(l_1)}\right) &= \frac{\sum_{\left(w^{(1)}\ldots w^{(l_1)}\right) \in \mathcal{S}\left(w^{(l_1+1)} \ldots w^{(L)}\right)} \mathbbm{1}\left\{\x^{(1)}_w \in \tau^{(1)}, \ldots, \x^{(l_1)}_w \in \tau^{(l_1)} \right\} }{\prod_{i=1}^{\LL}\Big|w^{(i)}: \x^{(i)}_w \in \tau^{(i)}\Big|}
\end{align}
Recall that in~\eqref{eq:concentration}, we have shown that for the $i$-th systematic chunk, given a typical received vector $\yw^{(i)}$ and a conditionally typical type class $\tau^{(i)}$, the number of inner-codewords falling into $\tau^{(i)}$ is tightly concentrated around its expectation, \emph{i.e.},
\begin{align}
\mathbb{P}\left(\Big|\big|w^{(i)}: \x^{(i)}_w \in \tau^{(i)}\big|-\Upsilon_in^{\rin}\Big| \le 2n^{-\frac{1}{2}-\frac{\delta}{4}}\Upsilon_i n^{\rin} \right) \ge 1 - 2\exp\left(-\frac{4}{3}\sqrt{n}\right),
\end{align}
where $\Upsilon_in^{\rin}$ is the expected number of inner-codewords falling into $\tau^{(i)}$.
By taking a union bound over all $l_1$ ($l_1< \sqrt{n}$) systematic chunk, we prove that the product of $\big|w^{(i)}: \x^{(i)}_w \in \tau^{(i)}\big|$ can also be concentrated, \emph{i.e.}, 
\begin{align}
\mathbb{P}\left(\left|\prod_{i=1}^{l_1}\Big|w^{(i)}: \x^{(i)}_w \in \tau^{(i)}\Big|-n^{l_1\rin} \prod_{i=1}^{l_1}\Upsilon_i\right| \le 2n^{-\frac{\delta}{4}}n^{l_1\rin} \prod_{i=1}^{l_1}\Upsilon_i \right) \ge 1 - 2\sqrt{n}\exp\left(-\frac{4}{3}\sqrt{n}\right). \label{eq:zou7}
\end{align}
Combining~\eqref{eq:zou7} and Claim~\ref{claim:new1}, we obtain that with probability at least\footnote{Note that $2\sqrt{n}\exp\left(-\frac{4}{3}\sqrt{n}\right)$ is decaying faster than $2\exp\left(-\frac{3(\ln 2)\sqrt{n}}{2k_1}\right)$ because of value of $k_1$.} $1- 4\exp\left(-\frac{3(\ln 2)\sqrt{n}}{2k_1}\right)$, 
\begin{align}
P\left(w^{(l_1+1)}, \ldots, w^{(L)} \big| \yw^{(1)}, \ldots, \yw^{(l_1)}\right) &\le \frac{(1+n^{-1})\nu \prod_{i=1}^{l_1}\Upsilon_i}{\left(1-2n^{-\frac{\delta}{4}}\right)n^{l_1\rin} \prod_{i=1}^{l_1}\Upsilon_i} \\
& = \frac{1}{n^{l_2\rin}} (1+n^{-1}) \left(1+2n^{-\frac{\delta}{4}} + \left(2n^{-\frac{\delta}{4}}\right)^2 + \cdots\right) \label{eq:zou8} \\
&\stackrel{n \to \infty}{\le} \frac{(1+3n^{-\frac{\delta}{4}})}{n^{l_2 \rin}}, \\
P\left(w^{(l_1+1)}, \ldots, w^{(L)} \big| \yw^{(1)}, \ldots, \yw^{(l_1)}\right) &\ge \frac{(1-n^{-1})\nu \prod_{i=1}^{l_1}\Upsilon_i}{\left(1+2n^{-\frac{\delta}{4}}\right)n^{l_1\rin} \prod_{i=1}^{l_1}\Upsilon_i} \stackrel{n \to \infty}{\ge} \frac{(1+3n^{-\frac{\delta}{4}})}{n^{l_2 \rin}},
\end{align}
where~\eqref{eq:zou8} follows since $\nu = n^{\rin(l_1-l_2)}$. Since the number of $(w^{(\LL+1)}, \ldots, w^{(L)})$ is only $\exp\left(o(\sqrt{n})\right)$, we are able to take a union bound over all $(w^{(\LL+1)}, \ldots, w^{(L)})$.
\qed

With Claim~\ref{claim:new3}, we are able to prove Lemma~\ref{lemma:(b)} --- the variational distance between the active distribution $P_1(\Yw^{(1)}, \ldots, \Yw^{(L)})$ and the ``chunk-wise independent'' distribution $P_1^{(1)}(\Yw^{(1)})\cdots P_1^{(L)}(\Yw^{(L)})$ goes to zero as $n$ goes to infinity. 

\noindent{\textbf{Lemma 4}} \ (Restated). \emph{With probability at least $1 - \exp\left(-\frac{\left(\frac{3}{4}-\frac{\delta}{2}\right)(\ln 2)\sqrt{n}}{k_1}\right)$ over the concatenated code design and the channel noise to Willie, for the randomly chosen code $\cc$, the variational distance between the $n$-letter active distribution $P_1(\Yw^{(1)}, \ldots, \Yw^{(L)})$ and the ``chunk-wise independent'' product distribution $P_1^{(1)}(\Yw^{(1)})\cdots P_1^{(L)}(\Yw^{(L)})$ is bounded from above as} 
$$\frac{1}{2} \sum_{\yw^{(1)}\in \{0,1\}^B \ldots \yw^{(L)}\in \{0,1\}^B} \left| P^{(1)}_1(\yw^{(1)}) \cdots P^{(L)}_1(\yw^{(L)}) - P_1(\yw^{(1)}, \ldots, \yw^{(L)}) \right| \le 2 n^{-\frac{\delta}{4}}.$$

\noindent{\it Proof:} We first partition $(\yw^{(1)}\ldots \yw^{(l_1)})$ into $(\yw^{(1)}\ldots \yw^{(l_1)}) \in \aywiL$ and $(\yw^{(1)}\ldots \yw^{(l_1)}) \notin \aywiL$ as follows.
\begin{align}
\sum_{\yw^{(1)}\in \{0,1\}^B \ldots \yw^{(L)}\in \{0,1\}^B} \left| P_1^{(1)}(\yw^{(1)})\cdots P_1^{(L)}(\yw^{(L)}) - P_1(\yw^{(1)}, \ldots, \yw^{(L)}) \right|  &= \sum_{\left(\yw^{(1)}\ldots \yw^{(l_1)}\right) \in \aywiL} \sum_{\yw^{(l_1+1)}\ldots \yw^{(L)}} \left| P_1^{(1)}(\yw^{(1)})\cdots P_1^{(L)}(\yw^{(L)}) - P_1(\yw^{(1)}, \ldots, \yw^{(L)}) \right|  \label{eq:May9_1}\\
&+\sum_{\left(\yw^{(1)}\ldots \yw^{(l_1)}\right) \notin \aywiL} \sum_{\yw^{(l_1+1)}\ldots \yw^{(L)}} \left| P_1^{(1)}(\yw^{(1)})\cdots P_1^{(L)}(\yw^{(L)}) - P_1(\yw^{(1)}, \ldots, \yw^{(L)}) \right|  \label{eq:May9_2}.
\end{align}
The right-hand side (RHS) of~\eqref{eq:May9_1} can further be decomposed into two parts --- (i) all $\yw^{(i)}$ ($i \in \{1,\ldots, \LL\}$) are contributed by typical inner-codewords, and (ii) there exists at least one $\yw^{(i)}$ ($i \in \{1,\ldots, \LL\}$) is contributed by atypical inner-codewords.     
\begin{align}
&\sum_{\left(\yw^{(1)}\ldots \yw^{(l_1)}\right) \in \aywiL} \sum_{\yw^{(l_1+1)}\ldots \yw^{(L)}} \left| P_1^{(1)}(\yw^{(1)})\cdots P_1^{(L)}(\yw^{(L)}) - P_1(\yw^{(1)}, \ldots, \yw^{(L)}) \right| \\
&= \sum_{\left(\yw^{(1)}\ldots \yw^{(l_1)}\right) \in \aywiL} P_1^{(1)}(\yw^{(1)})\cdots P_1^{(\LL)}(\yw^{(\LL)}) \sum_{\yw^{(\LL+1)}\ldots \yw^{(L)}} \left| P_1^{(\LL+1)}(\yw^{(\LL+1)})\cdots P_1^{(L)}(\yw^{(L)}) - P\left(\yw^{(\LL+1)} \cdots \yw^{(L)} \big| \yw^{(1)} \cdots \yw^{(\LL)} \right) \right| \\
&\le \sum_{\left(\yw^{(1)}\ldots \yw^{(l_1)}\right) \in \aywiL} \sum_{\left(\tau^{(1)} \ldots \tau^{(\LL)}\right) \in \axywiL} \ \sum_{w^{(1)}: \x^{(1)}_w \in \tau^{(1)}}\cdots \sum_{ w^{(\LL)}: \x^{(\LL)}_w \in \tau^{(\LL)}} \left(\prod_{i=1}^{\LL} \frac{1}{n^{\rin}} P(\yw^{(i)}|\x^{(i)}_w)\right) \notag \\
&\qquad\qquad\qquad\qquad\qquad\qquad\qquad\qquad\qquad\sum_{\yw^{(\LL+1)}\ldots \yw^{(L)}} \left|P_1^{(\LL+1)}(\yw^{(\LL+1)})\cdots P_1^{(L)}(\yw^{(L)}) - P\left(\yw^{(l_1+1)}\ldots \yw^{(L)} \big|\yw^{(1)}\ldots \yw^{(l_1)}\right) \right|  \label{eq:May9_3} \\
&+ \sum_{\left(\yw^{(1)}\ldots \yw^{(l_1)}\right) \in \aywiL} \sum_{\left(\tau^{(1)} \ldots \tau^{(\LL)}\right) \notin \axywiL} \ \sum_{w^{(1)}: \x^{(1)}_w \in \tau^{(1)}}\cdots \sum_{ w^{(\LL)}: \x^{(\LL)}_w \in \tau^{(\LL)}} \left(\prod_{i=1}^{\LL} \frac{1}{n^{\rin}} P(\yw^{(i)}|\x^{(i)}_w)\right) \notag \\
&\qquad\qquad\qquad\qquad\qquad\qquad\qquad\qquad\qquad\sum_{\yw^{(\LL+1)}\ldots \yw^{(L)}} \left|P_1^{(\LL+1)}(\yw^{(\LL+1)})\cdots P_1^{(L)}(\yw^{(L)}) - P\left(\yw^{(l_1+1)}\ldots \yw^{(L)} \big|\yw^{(1)}\ldots \yw^{(l_1)}\right) \right|,  \label{eq:May9_4}
\end{align}
where~\eqref{eq:May9_3} and~\eqref{eq:May9_4} follow from the total probability theorem. 
To prove Lemma~\ref{lemma:(b)}, we need to show that with high probability over the code design, terms~\eqref{eq:May9_2},~\eqref{eq:May9_3}, and~\eqref{eq:May9_4} all go to zero as $n$ grows without bound. 
\subsubsection{Bounding the term in~\eqref{eq:May9_3}}
We first consider the typical event --- the term in~\eqref{eq:May9_3}. Note that
\begin{align}
&P_1^{(\LL+1)}(\yw^{(\LL+1)})\cdots P_1^{(L)}(\yw^{(L)}) =  \sum_{w^{(l_1+1)} \ldots w^{(L)}}\left(\prod_{i=l_1+1}^{L}\frac{1}{n^{\rin}} P(\yw^{(i)}|w^{(i)}) \right), \\
&P\left(\yw^{(l_1+1)}\ldots \yw^{(L)} \big|\yw^{(1)}\ldots \yw^{(l_1)}\right) \\
&=  \sum_{w^{(l_1+1)} \ldots w^{(L)}} P\left(w^{(l_1+1)} \ldots w^{(L)} \big|\yw^{(1)}\ldots \yw^{(l_1)}\right) P\left(\yw^{(l_1+1)}\cdots \yw^{(L)} \big|w^{(l_1+1)} \ldots w^{(L)},\yw^{(1)}\ldots \yw^{(l_1)}\right) \\
&= \sum_{w^{(l_1+1)} \ldots w^{(L)}} P\left(w^{(l_1+1)} \ldots w^{(L)} \big|\yw^{(1)}\ldots \yw^{(l_1)}\right) P\left(\yw^{(l_1+1)}\cdots \yw^{(L)} \big|w^{(l_1+1)} \ldots w^{(L)}\right) \label{eq:home1} \\
&= \sum_{w^{(l_1+1)} \ldots w^{(L)}} P\left(w^{(l_1+1)} \ldots w^{(L)} \big|\yw^{(1)}\ldots \yw^{(l_1)}\right) \prod_{i=\LL+1}^L P\left(\yw^{(i)} \big|\x^{(i)}_w\right), \label{eq:home2}
\end{align}
where~\eqref{eq:home1} holds since the received vectors of the systematic chunks $(\yw^{(1)}\ldots \yw^{(l_1)})$, the parity inner-message vector $(w^{(l_1+1)} \ldots w^{(L)})$, and the received vectors of the parity chunks $(\yw^{(l_1+1)}\cdots \yw^{(L)})$, form a Markov chain. Equation~\eqref{eq:home2} follows from the memoryless property of the channel.
By the triangle inequality, the term in~\eqref{eq:May9_3} can further be bounded from above as
\begin{align}
&\sum_{\left(\yw^{(1)}\ldots \yw^{(l_1)}\right) \in \aywiL} \sum_{\left(\tau^{(1)} \ldots \tau^{(\LL)}\right) \in \axywiL} \ \sum_{w^{(1)}: \x^{(1)}_w \in \tau^{(1)}}\cdots \sum_{ w^{(\LL)}: \x^{(\LL)}_w \in \tau^{(\LL)}} \left(\prod_{i=1}^{\LL} \frac{1}{n^{\rin}} P(\yw^{(i)}|\x^{(i)}_w)\right) \notag \\
&\qquad\qquad\qquad\qquad\qquad \sum_{\yw^{(\LL+1)}\ldots \yw^{(L)}} \sum_{w^{(l_1+1)}\ldots w^{(L)}} \left|\left(\prod_{i=l_1+1}^L \frac{1}{n^{\rin}} \right) - P\left(w^{(l_1+1)} \ldots w^{(L)} \big|\yw^{(1)}\ldots \yw^{(l_1)}\right) \right| \cdot \left(\prod_{i=l_1+1}^L P(\yw^{(i)}|\x^{(i)}_w)\right)\\
&= \sum_{\TT_Z^{(1)} \in \F_B^z, \ldots, \TT_Z^{(l_1)} \in \F_B^z} \ \sum_{\yw^{(1)} \in \TT_Z^{(1)}, \ldots, \yw^{(l_1)} \in \TT_Z^{(l_1)}} \ \sum_{\left(\tau^{(1)} \ldots \tau^{(\LL)}\right) \in \axywiL} \ \ \sum_{w^{(1)}: \x^{(1)}_w \in \tau^{(1)}}\cdots \sum_{ w^{(\LL)}: \x^{(\LL)}_w \in \tau^{(\LL)}}  \left(\prod_{i=1}^{\LL} \frac{1}{n^{\rin}} P(\yw^{(i)}|\x_w^{(i)})\right)  \notag\\
&\qquad\qquad\qquad\qquad\qquad \sum_{w^{(l_1+1)}\ldots w^{(L)}} \left|\left(\prod_{i=l_1+1}^L \frac{1}{n^{\rin}} \right) - P\left(w^{(l_1+1)} \ldots w^{(L)}\big|\yw^{(1)}\ldots \yw^{(l_1)}\right) \right| \sum_{\yw^{(l_1+1)}\ldots \yw^{(L)}} \left(\prod_{i=l_1+1}^L P(\yw^{(i)}|\x^{(i)}_w)\right) \label{eq:jj} \\
&= \sum_{\TT_Z^{(1)} \in \F_B^z, \ldots, \TT_Z^{(l_1)} \in \F_B^z} \ \sum_{\left(\tau^{(1)} \ldots \tau^{(\LL)}\right) \in \axywiL} \ \left(\prod_{i=1}^{\LL} \frac{1}{n^{\rin}}P\left(\TT_Z^{(i)} \big|\tau^{(i)}\right)\right) \sum_{\yw^{(1)} \in \TT_Z^{(1)}, \ldots, \yw^{(l_1)} \in \TT_Z^{(l_1)}} \sum_{w^{(1)}: \x^{(1)}_w \in \tau^{(1)}} \cdots \sum_{ w^{(\LL)}: \x^{(\LL)}_w \in \tau^{(\LL)}} \notag\\
&\qquad\qquad\qquad\qquad\qquad \sum_{w^{(l_1+1)}\ldots w^{(L)}} \left|\frac{1}{n^{l_2\rin}} - P\left(w^{(l_1+1)} \ldots w^{(L)}\big|\yw^{(1)}\ldots \yw^{(l_1)}\right) \right| \sum_{\yw^{(l_1+1)}\ldots \yw^{(L)}} \left(\prod_{i=l_1+1}^L P(\yw^{(i)}|\x^{(i)}_w)\right). \label{eq:jj2}
\end{align}
In~\eqref{eq:jj}, for each of the systematic chunk, we decompose the typical set $\aywi$ into typical type classes $\TT_Z^{(i)} \in \mathcal{F}_B^z$ that comprise it. Equation~\eqref{eq:jj2} is obtained by noting that the values of $\prod_{i=1}^{\LL} \frac{1}{n^{\rin}}P(\yw^{(i)}|\x_w^{(i)})$ are the same for all $(\yw^{(1)}, \ldots, \yw^{(\LL)}) \in (\TT_Z^{(1)}, \ldots, \TT_Z^{(\LL)})$ and $(\x_w^{(1)}, \ldots, \x_w^{(\LL)}) \in (\tau^{(1)}, \ldots, \tau^{(\LL)})$, hence we use $\prod_{i=1}^{\LL} \frac{1}{n^{\rin}}P(\TT_Z^{(i)}|\tau^{(i)})$ to denote this value. Recall that Claim~\ref{claim:new3} shows that for a typical $(\yw^{(1)},\ldots, \yw^{(\LL)})$ and a conditionally typical $(\tau^{(1)},\ldots, \tau^{(\LL)})$, with probability at least $1- 4\exp\left(-\frac{3(\ln 2)\sqrt{n}}{2k_1}\right)$, 
\begin{align}
P\left(w^{(l_1+1)} \ldots w^{(L)}\big|\yw^{(1)}\ldots \yw^{(l_1)}\right) \in \frac{(1 \pm 3n^{-\frac{\delta}{4}})}{n^{l_2\rin}}, \ \forall \left(w^{(l_1+1)} \ldots w^{(L)}\right). \label{eq:hom}
\end{align}
We wish to take a union bound to show that with high probability, the concentration inequality~\eqref{eq:hom} holds for all typical $(\yw^{(1)},\ldots, \yw^{(\LL)})$ and conditionally typical type class $(\tau^{(1)},\ldots, \tau^{(\LL)})$, however, it is not valid since the number of $(\yw^{(1)},\ldots, \yw^{(\LL)})$ is exponential in $n$, and the probability that~\eqref{eq:hom} holds is only sub-exponentially close to one. Instead, we circumvent this issue by a less ambitious approach. For each typical type class $(\TT_Z^{(1)}, \ldots, \TT_Z^{(l_1)})$ and conditionally typical type class $(\tau^{(1)},\ldots, \tau^{(\LL)})$, we define 
\begin{align}
&\Omega\left(\TT_Z^{(1)} \ldots \TT_Z^{(l_1)}\right) \triangleq \left\{(\yw^{(1)}, \ldots, \yw^{(\LL)}): \left(\yw^{(1)} \in \TT_Z^{(1)}, \ldots, \yw^{(l_1)} \in \TT_Z^{(\LL)}\right)\right\}, \\
&\Omega_{\checkmark}\left(\TT_Z^{(1)} \ldots \TT_Z^{(l_1)}, \tau^{(1)}\ldots \tau^{(\LL)}\right) \triangleq \left\{(\yw^{(1)}, \ldots, \yw^{(\LL)}) \in \Omega: P\left(w^{(l_1+1)} \ldots w^{(L)}\big|\yw^{(1)}\ldots \yw^{(l_1)}\right) \in \frac{(1 \pm 3n^{-\frac{\delta}{4}})}{n^{l_2\rin}}, \ \forall \left(w^{(l_1+1)} \ldots w^{(L)}\right) \right\}, \\
& \Omega_{\times}\left(\TT_Z^{(1)} \ldots \TT_Z^{(l_1)}, \tau^{(1)}\ldots \tau^{(\LL)}\right) \triangleq \Omega\left(\TT_Z^{(1)} \ldots \TT_Z^{(l_1)}\right) \setminus \Omega_{\checkmark}\left(\TT_Z^{(1)} \ldots \TT_Z^{(l_1)}, \tau^{(1)}\ldots \tau^{(\LL)}\right).
\end{align}
For notational convenience we ignore the arguments inside $\Omega$, $\Omega_{\checkmark}$, and $\Omega_{\times}$ in the following. 
Note that $\Omega_{\checkmark}$ contains all $(\yw^{(1)}, \yw^{(2)}, \ldots, \yw^{(\LL)}) \in \Omega$ that satisfies~\eqref{eq:hom}. In Claim~\ref{claim:pa1} we use Markov's inequality to show that with probability at least $1 - \exp\left(-\frac{\left(\frac{3}{4}-\frac{\delta}{2}\right)(\ln 2)\sqrt{n}}{k_1}\right)$, for all typical type class $(\TT_Z^{(1)}, \ldots, \TT_Z^{(l_1)})$ and conditionally typical type class $(\tau^{(1)},\ldots, \tau^{(\LL)})$, only a small fraction of $(\yw^{(1)}, \yw^{(2)}, \ldots, \yw^{(\LL)}) \in \Omega$ belongs to $\Omega_{\times}$.
\begin{claim} \label{claim:pa1}
	With probability at least $1 - \exp\left(-\frac{\left(\frac{3}{4}-\frac{\delta}{2}\right)(\ln 2)\sqrt{n}}{k_1}\right)$ over the code design, for all typical type class $(\TT_Z^{(1)}, \ldots, \TT_Z^{(l_1)})$ and conditionally typical type class $(\tau^{(1)},\ldots, \tau^{(\LL)})$, we have
	\begin{align}
	\left| \Omega_{\times} \right| \le |\Omega|\cdot 4\exp\left(-\frac{(\ln 2)\sqrt{n}}{2k_1}\right).
	\end{align}
\end{claim}
\noindent{\it Proof:} For a fixed typical type class $(\TT_Z^{(1)}, \ldots, \TT_Z^{(l_1)})$ and a conditionally typical type class $(\tau^{(1)},\ldots, \tau^{(\LL)})$, by the fact that  $\mathbb{E}\left(\left|\Omega_{\times}\right|\right) \le \left|\Omega\right| \cdot 4\exp\left(-\frac{3(\ln 2)\sqrt{n}}{2k_1}\right)$ and the Markov's inequality, we have 
\begin{align} 
\mathbb{P}\left(\left|\Omega_{\times}\right| \ge  |\Omega| \cdot 4\exp\left(-\frac{(\ln 2)\sqrt{n}}{2k_1}\right) \right) \le \frac{|\Omega| \cdot 4\exp\left(-\frac{3(\ln 2)\sqrt{n}}{2k_1}\right)}{|\Omega| \cdot 4\exp\left(-\frac{(\ln 2)\sqrt{n}}{2k_1}\right)} = \exp\left(-\frac{(\ln 2)\sqrt{n}}{k_1}\right). 
\end{align}
We then take a union bound over all typical $(\TT_Z^{(1)}, \ldots, \TT_Z^{(l_1)})$ and all conditionally typical $(\tau^{(1)},\ldots, \tau^{(\LL)})$. The union bound is valid since 
\begin{align}
& \left|\left(\TT_Z^{(1)}, \ldots, \TT_Z^{(l_1)}\right): \TT_Z^{(1)} \in \F_B^z, \ldots, \TT_Z^{(l_1)} \in \F_B^z \right| = \left(2(\rho * q) \Delta^z_1 B\right)^{\LL} = 2^{\left(\frac{1}{4}+\frac{\delta}{2}\right)\frac{\sqrt{n}}{k_1}\cdot \left(1 + \mathcal{O}\left(\frac{\log\log n}{\log n}\right)\right)}, \\
& \left|\left(\tau^{(1)} \ldots \tau^{(\LL)}\right): \left(\tau^{(1)} \ldots \tau^{(\LL)}\right) \in \axywiL\right| = \left((2\rho q \Delta^{xz}_{10}B)\cdot (2\rho (1-q) \Delta^{xz}_{11}B)\right)^{\LL} = 2^{\mathcal{O}\left(\frac{\sqrt{n} (\log\log n)}{\log n}\right)}.
\end{align}
Therefore, with probability at least $1 - \exp\left(-\frac{\left(\frac{3}{4}-\frac{\delta}{2}\right)(\ln 2)\sqrt{n}}{k_1}\right)$ over the code design, $\left| \Omega_{\times} \right| \le |\Omega|\cdot 4\exp\left(-\frac{(\ln 2)\sqrt{n}}{2k_1}\right)$ holds for all typical type class $(\TT_Z^{(1)}, \ldots, \TT_Z^{(l_1)})$ and conditionally typical type class $(\tau^{(1)},\ldots, \tau^{(\LL)})$.
\qed

Roughly speaking, with the help of Claim~\ref{claim:pa1}, we are able to show that the $n$-letter active distribution $P_1(\Yw^{(1)},\ldots, \Yw^{(\LL)})$ and the ``chunk-wise independent'' distribution $P_1^{(1)}(\Yw^{(1)}) \cdots P_1^{(L)}(\Yw^{(L)})$ are sufficiently close under the ``good'' event $(\yw^{(1)},\ldots, \yw^{(\LL)}) \in \Omega_{\checkmark}$, since $(\yw^{(1)},\ldots, \yw^{(\LL)}) \in \Omega_{\checkmark}$ guarantees
\begin{align}
\left|\frac{1}{n^{l_2\rin}} - P\left(w^{(l_1+1)} \ldots w^{(L)}\big|\yw^{(1)}\ldots \yw^{(l_1)}\right) \right| \le \frac{3n^{-\frac{\delta}{4}}}{n^{l_2\rin}}, \ \forall \left(w^{(l_1+1)}\ldots w^{(L)}\right),
\end{align}
while the ``bad'' event $(\yw^{(1)},\ldots, \yw^{(\LL)}) \in \Omega_{\times}$ occurs with decaying probability.
Returning back to~\eqref{eq:jj2}, for each each typical type class $(\TT_Z^{(1)}, \ldots, \TT_Z^{(l_1)})$ and conditionally typical type class $(\tau^{(1)},\ldots, \tau^{(\LL)})$, we further decompose $(\yw^{(1)},\ldots, \yw^{(\LL)})$ into $(\yw^{(1)},\ldots, \yw^{(\LL)}) \in \Omega_{\checkmark}$ and $(\yw^{(1)},\ldots, \yw^{(\LL)}) \in \Omega_{\times}$. We first bound the term corresponding to $\Omega_{\checkmark}$ as follows. 
\begin{align}
&\sum_{\TT_Z^{(1)} \in \F_B^z, \ldots, \TT_Z^{(l_1)} \in \F_B^z} \ \sum_{\left(\tau^{(1)} \ldots \tau^{(\LL)}\right) \in \axywiL} \ \left(\prod_{i=1}^{\LL} \frac{1}{n^{\rin}}P\left(\TT_Z^{(i)} \big|\tau^{(i)}\right)\right) \sum_{(\yw^{(1)} \ldots \yw^{(l_1)}) \in \Omega_{\checkmark}} \sum_{w^{(1)}: \x^{(1)}_w \in \tau^{(1)}} \cdots \sum_{ w^{(\LL)}: \x^{(\LL)}_w \in \tau^{(\LL)}} \notag\\
&\qquad\qquad\qquad\qquad\qquad \sum_{w^{(l_1+1)}\ldots w^{(L)}} \left| \frac{1}{n^{l_2\rin}} - P\left(w^{(l_1+1)} \ldots w^{(L)}\big|\yw^{(1)}\ldots \yw^{(l_1)}\right) \right| \sum_{\yw^{(l_1+1)}\ldots \yw^{(L)}} \left(\prod_{i=l_1+1}^L P(\yw^{(i)}|\x^{(i)}_w)\right) \label{eq:teng1} \\
&\le \sum_{\TT_Z^{(1)} \in \F_B^z, \ldots, \TT_Z^{(l_1)} \in \F_B^z} \ \sum_{\left(\tau^{(1)} \ldots \tau^{(\LL)}\right) \in \axywiL} \ \left(\prod_{i=1}^{\LL} \frac{1}{n^{\rin}}P\left(\TT_Z^{(i)} \big|\tau^{(i)}\right)\right) \sum_{(\yw^{(1)} \ldots \yw^{(l_1)}) \in \Omega_{\checkmark}} \sum_{w^{(1)}: \x^{(1)}_w \in \tau^{(1)}} \cdots \sum_{ w^{(\LL)}: \x^{(\LL)}_w \in \tau^{(\LL)}} \notag\\
&\qquad\qquad\qquad\qquad\qquad\qquad\qquad\qquad\qquad\qquad \cdot 3n^{-\frac{\delta}{4}} \sum_{w^{(l_1+1)}\ldots w^{(L)}} \frac{1}{n^{l_2\rin}} \sum_{\yw^{(l_1+1)}\ldots \yw^{(L)}} \left(\prod_{i=l_1+1}^L P(\yw^{(i)}|\x^{(i)}_w)\right) \label{eq:teng4} \\
&\le 3n^{-\frac{\delta}{4}} \sum_{\yw^{(1)}} \cdots \sum_{\yw^{(L)}} \sum_{w^{(1)}} \cdots \sum_{ w^{(L)}} \left(\prod_{i=1}^L \frac{1}{n^{\rin}} P(\yw^{(i)}|\x^{(i)}_w)\right)\label{eq:huiji} \\
& = 3n^{-\frac{\delta}{4}}, \label{eq:teng3}
\end{align} 
where the inequality~\eqref{eq:huiji} is obtained by changing order of summations and dropping the requirements that $\yw^{(i)}$ is typical and $\x_w^{(i)}$ is conditionally typical. We then bound the term corresponding to $\Omega_{\times}$.
\begin{align}
&\sum_{\TT_Z^{(1)} \in \F_B^z, \ldots, \TT_Z^{(l_1)} \in \F_B^z} \ \sum_{\left(\tau^{(1)} \ldots \tau^{(\LL)}\right) \in \axywiL} \ \left(\prod_{i=1}^{\LL} \frac{1}{n^{\rin}}P\left(\TT_Z^{(i)} \big|\tau^{(i)}\right)\right) \sum_{(\yw^{(1)} \ldots \yw^{(l_1)}) \in \Omega_{\times}} \sum_{w^{(1)}: \x^{(1)}_w \in \tau^{(1)}} \cdots \sum_{ w^{(\LL)}: \x^{(\LL)}_w \in \tau^{(\LL)}} \notag\\
&\qquad\qquad\qquad\qquad\qquad \sum_{w^{(l_1+1)}\ldots w^{(L)}} \left|\frac{1}{n^{l_2\rin}} - P\left(w^{(l_1+1)} \ldots w^{(L)}\big|\yw^{(1)}\ldots \yw^{(l_1)}\right) \right| \sum_{\yw^{(l_1+1)}\ldots \yw^{(L)}} \left(\prod_{i=l_1+1}^L P(\yw^{(i)}|\x^{(i)}_w)\right), \label{eq:teng2} \\
&\stackrel{\text{w.h.p.}}{\le} 4\exp\left(-\frac{(\ln 2)\sqrt{n}}{2k_1}\right) \cdot \sum_{\TT_Z^{(1)} \in \F_B^z, \ldots, \TT_Z^{(l_1)} \in \F_B^z} \ \sum_{\left(\tau^{(1)} \ldots \tau^{(\LL)}\right) \in \axywiL} \ \left(\prod_{i=1}^{\LL} \frac{1}{n^{\rin}}P\left(\TT_Z^{(i)} \big|\tau^{(i)}\right)\right) \sum_{(\yw^{(1)} \ldots \yw^{(l_1)}) \in \Omega} \sum_{w^{(1)}: \x^{(1)}_w \in \tau^{(1)}} \cdots \sum_{ w^{(\LL)}: \x^{(\LL)}_w \in \tau^{(\LL)}} \notag\\
&\qquad\qquad\qquad\qquad\qquad \sum_{w^{(l_1+1)}\ldots w^{(L)}} \left|\frac{1}{n^{l_2\rin}} - P\left(w^{(l_1+1)} \ldots w^{(L)}\big|\yw^{(1)}\ldots \yw^{(l_1)}\right) \right| \sum_{\yw^{(l_1+1)}\ldots \yw^{(L)}} \left(\prod_{i=l_1+1}^L P(\yw^{(i)}|\x^{(i)}_w)\right) \label{eq:teng5} \\
& \le 4\exp\left(-\frac{(\ln 2)\sqrt{n}}{2k_1}\right) \cdot \sum_{\yw^{(1)}} \cdots \sum_{\yw^{(\LL)}} \sum_{w^{(1)}} \cdots \sum_{ w^{(\LL)}} \left(\prod_{i=1}^{\LL} P(\yw^{(i)}|\x^{(i)}_w)\right) \notag \\
&\qquad\qquad\qquad\qquad\qquad \sum_{w^{(l_1+1)}\ldots w^{(L)}} \left(\frac{1}{n^{l_2\rin}} + P\left(w^{(l_1+1)} \ldots w^{(L)}\big|\yw^{(1)}\ldots \yw^{(l_1)}\right) \right) \sum_{\yw^{(l_1+1)}\ldots \yw^{(L)}} \left(\prod_{i=l_1+1}^L P(\yw^{(i)}|\x^{(i)}_w)\right) \label{eq:teng6} \\
& \le 8\exp\left(-\frac{(\ln 2)\sqrt{n}}{2k_1}\right). \label{eq:teng7}
\end{align}
Note that inequality~\eqref{eq:teng5} holds with probability at least $1 - \exp\left(-\frac{\left(\frac{3}{4}-\frac{\delta}{2}\right)(\ln 2)\sqrt{n}}{k_1}\right)$ by Claim~\ref{claim:pa1}, and inequality~\eqref{eq:teng6} follows from the triangle inequality. By combining~\eqref{eq:jj2},~\eqref{eq:teng3} and~\eqref{eq:teng7}, we prove that with probability at least $1 - \exp\left(-\frac{\left(\frac{3}{4}-\frac{\delta}{2}\right)(\ln 2)\sqrt{n}}{k_1}\right)$, the term in~\eqref{eq:May9_3} (corresponding to the typical event) is bounded from above by $3n^{-\frac{\delta}{4}} + 8\exp\left(-\frac{(\ln 2)\sqrt{n}}{2k_1}\right)$. 

\subsubsection{Bounding the terms in~\eqref{eq:May9_2} and~\eqref{eq:May9_4}}
We now consider the atypical events expressed in~\eqref{eq:May9_2} and~\eqref{eq:May9_4}. Recall that in~\eqref{eq:May9_2}, $(\yw^{(1)}\ldots \yw^{(l_1)}) \notin \aywiL$ implies there exists at least one $i \in \{1,2,\ldots, \LL\}$ such that $\yw^{(i)} \notin \aywi$, hence 
\begin{align}
&\sum_{\left(\yw^{(1)}\ldots \yw^{(l_1)}\right) \notin \aywiL} \sum_{\yw^{(l_1+1)}\ldots \yw^{(L)}} \left| P_1^{(1)}(\yw^{(1)})\cdots P_1^{(L)}(\yw^{(L)}) - P_1(\yw^{(1)}, \ldots, \yw^{(L)}) \right| \\
& \le \sum_{i=1}^{\LL} \sum_{\yw^{(i)} \notin \aywi} \sum_{\left(\yw^{(1)}\ldots \yw^{(i-1)}, \yw^{(i+1)} \ldots \yw^{(l_1)}\right)}\sum_{\yw^{(l_1+1)}\ldots \yw^{(L)}} \left( P_1^{(1)}(\yw^{(1)})\cdots P_1^{(L)}(\yw^{(L)}) + P_1(\yw^{(1)}, \ldots, \yw^{(L)}) \right) \\
& \le 2 \LL \sum_{\yw^{(i)} \notin \aywi} P_1^{(i)}(\yw^{(i)}).
\end{align}
Similarly, the term~\eqref{eq:May9_4} implies there exists at least one $i \in \{1,2,\ldots, \LL\}$ such that $\yw^{(i)} \in \aywi$ and $\tau^{(i)} \notin \mathcal{F}_B^{xz}$, hence
\begin{align}
&\sum_{\left(\yw^{(1)}\ldots \yw^{(l_1)}\right) \in \aywiL} \sum_{\left(\tau^{(1)} \ldots \tau^{(\LL)}\right) \notin \axywiL} \ \sum_{w^{(1)}: \x^{(1)}_w \in \tau^{(1)}}\cdots \sum_{ w^{(\LL)}: \x^{(\LL)}_w \in \tau^{(\LL)}} \left(\prod_{i=1}^{\LL} \frac{1}{n^{\rin}} P(\yw^{(i)}|\x^{(i)}_w)\right) \notag \\
&\qquad\qquad\qquad\qquad\qquad\qquad\qquad\qquad\qquad\sum_{\yw^{(\LL+1)}\ldots \yw^{(L)}} \left|P_1^{(\LL+1)}(\yw^{(\LL+1)})\cdots P_1^{(L)}(\yw^{(L)}) - P\left(\yw^{(l_1+1)}\ldots \yw^{(L)} \big|\yw^{(1)}\ldots \yw^{(l_1)}\right) \right| \\
&\le \sum_{i=1}^{\LL} \sum_{\left(\yw^{(1)}\ldots \yw^{(l_1)}\right) \in \aywiL} \sum_{\tau^{(i)} \notin \mathcal{F}_B^{xz}} \sum_{\left(\tau^{(1)} \ldots \tau^{(i-1)}, \tau^{(i+1)} \ldots \tau^{(\LL)}\right)} \ \sum_{w^{(1)}: \x^{(1)}_w \in \tau^{(1)}}\cdots \sum_{ w^{(\LL)}: \x^{(\LL)}_w \in \tau^{(\LL)}} \left(\prod_{i=1}^{\LL} \frac{1}{n^{\rin}} P(\yw^{(i)}|\x^{(i)}_w)\right) \notag \\
&\qquad\qquad\qquad\qquad\qquad\qquad\qquad\qquad\qquad\sum_{\yw^{(\LL+1)}\ldots \yw^{(L)}} \left(P_1^{(\LL+1)}(\yw^{(\LL+1)})\cdots P_1^{(L)}(\yw^{(L)}) + P\left(\yw^{(l_1+1)}\ldots \yw^{(L)} \big|\yw^{(1)}\ldots \yw^{(l_1)}\right) \right) \\
& \le 2 \LL \sum_{\yw^{(i)} \in \aywi} \sum_{\tau^{(i)} \notin \mathcal{F}_B^{xz}} \sum_{w^{(i)}: \x^{(i)}_w \in \tau^{(i)}} \frac{1}{n^{\rin}} P(\yw^{(i)}|\x^{(i)}_w) \\
&= 2 \LL \sum_{\yw^{(i)} \in \aywi} \sum_{w^{(i)}: \x^{(i)}_w \notin \axywi} \frac{1}{n^{\rin}} P(\yw^{(i)}|\x^{(i)}_w).
\end{align}
By Claim~\ref{claim:5}, we are able to show that with probability at least $1- \exp\left(-4\sqrt{n}/3\right)$ over the code design, the terms in~\eqref{eq:May9_2} and~\eqref{eq:May9_4} can be bounded from above as
\begin{align}
2 \LL \sum_{\yw^{(i)} \notin \aywi} P_1(\yw^{(i)}) + 2 \LL \sum_{\yw^{(i)} \in \aywi} \sum_{w^{(i)}: \x^{(i)}_w \notin \axywi} \frac{1}{n^{\rin}} P(\yw^{(i)}|\x^{(i)}_w) \le 2 \LL \cdot \left(8n^{-\frac{1}{2}-\frac{\delta}{4}}\right) \le \frac{16n^{-\frac{\delta}{4}}}{k_1 \log n}.
\end{align}

Combining the proofs for~\eqref{eq:May9_2},~\eqref{eq:May9_3}, and~\eqref{eq:May9_4}, we finally obtain that 
\begin{align}
\frac{1}{2} \sum_{\yw^{(1)} \in \{0,1\}^B \ldots \yw^{(L)}\in \{0,1\}^B} \left| P^{(1)}_1(\yw^{(1)}) \cdots P^{(L)}_1(\yw^{(L)}) - P_1(\yw^{(1)}, \ldots, \yw^{(L)}) \right| \le \frac{3}{2} n^{-\frac{\delta}{4}} + 4\exp\left(-\frac{(\ln 2)\sqrt{n}}{2k_1}\right) +  \frac{8n^{-\frac{\delta}{4}}}{k_1 \log n} \stackrel{n \to \infty}{\le} 2n^{-\frac{\delta}{4}},
\end{align}
with probability at least $1 - \exp\left(-\frac{\left(\frac{3}{4}-\frac{\delta}{2}\right)(\ln 2)\sqrt{n}}{k_1}\right)$ over the code design. This completes the proof of Lemma~\ref{lemma:(b)}. \qed

\subsection{Concluding remarks for the proof of covertness} \label{sec:deniability-C}

To bound the variational distance between the innocent distribution $P_0$ and the active distribution $P_1$, we repeatedly use the triangle inequality to obtain 
\begin{align}
\mathbb{V}\left(P_0, P_1\right) &\le \mathbb{V}\left(P_0, \mathbb{E}_{\C}(P_1)\right) + \mathbb{V}\left( \mathbb{E}_{\C}(P_1), P_1 \right) \\
&\le \mathbb{V}\left(P_0, \mathbb{E}_{\C}(P_1)\right) + \mathbb{V}\left(\mathbb{E}_{\C}(P_1), P_1^{(1)}(\yw^{(1)}) \cdots P_1^{(L)}(\yw^{(L)})\right) + \mathbb{V}\left(P_1, P_1^{(1)}(\yw^{(1)}) \cdots P_1^{(L)}(\yw^{(L)})\right).
\end{align}
By Lemmas~\ref{lemma:1},~\ref{lemma:(a)} and~\ref{lemma:(b)}, we obtain that with probability at least $1 - \exp\left(-\frac{\left(\frac{3}{4}-\frac{\delta}{2}\right)(\ln 2)\sqrt{n}}{k_1}\right)$ over the concatenated code design and the channel noise to Willie, for the randomly chosen code $\cc$, 
\begin{align}
\mathbb{V}(P_0, P_1) &\le \epsilon_d + n^{-\delta/4}+n^{-\delta/4}+2n^{-\delta/4}  \\
&\le \epsilon_d + 4n^{-\delta/4}, \label{eq:approx}
\end{align}
This completes the proof of covertness of our proposed codes, as in Property 3) in Theorem 1.

%%%%%%%%%%%%%%%%%%%%%%%%%%%%%%%%%%%%%%%%%%%%%%%%%%%%%%%%%%%%%%%%%%%%%%%%%%%%%%%%%%%%%%%%%%%%%%%%%%%%%%%%%%%%%%%%%%%%%%%%%%%%%%%%%%%%%%%%%%%%%%%%%%%%%%%%%%%%

\section{Proof of Reliability}
\label{sec:reliability}

\begin{figure}
	\begin{center}
	\includegraphics[scale=0.28]{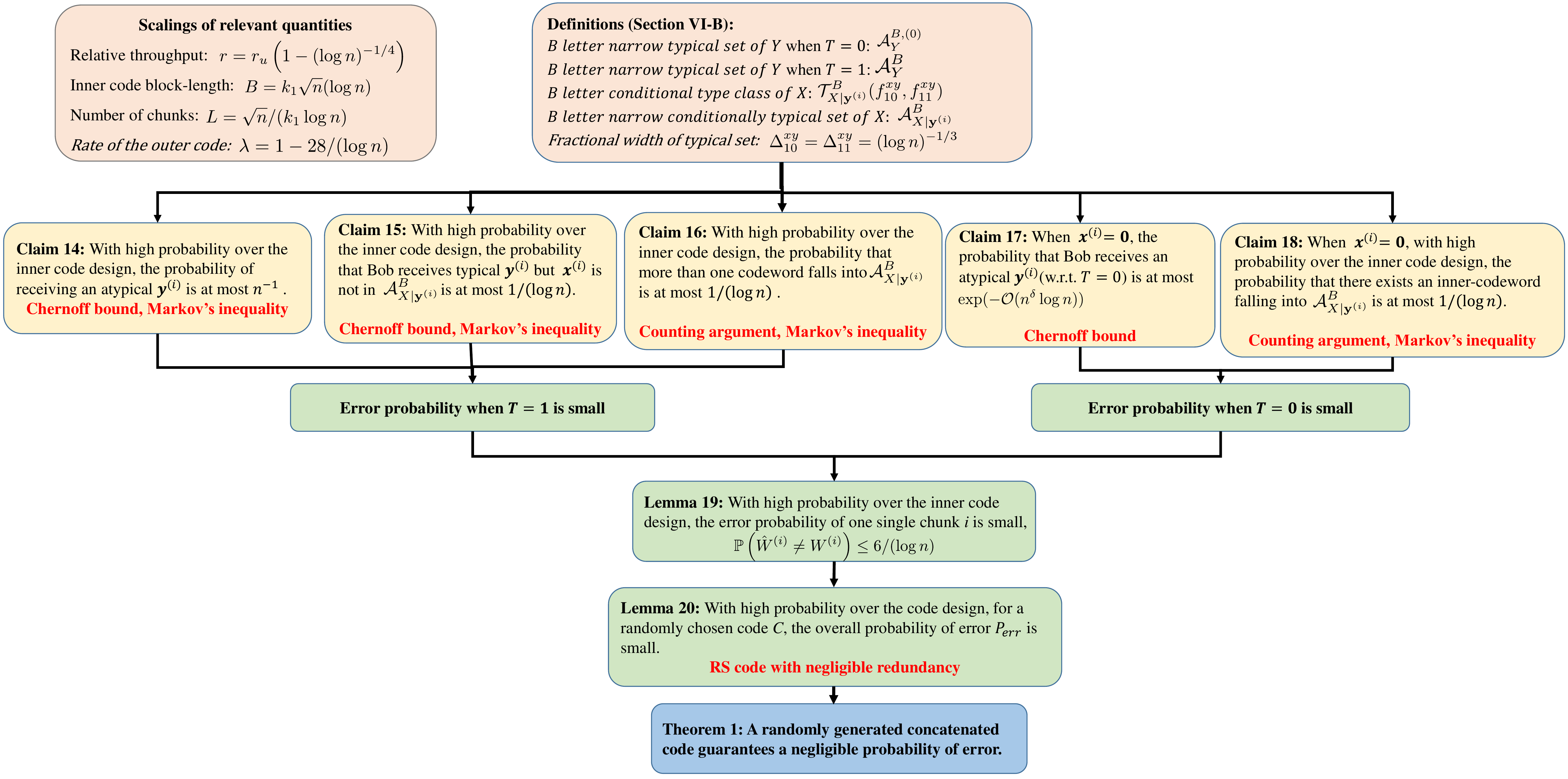}
	\caption{A road-map of our proof that our codes are highly reliable with high probability.} \label{fig:flow_rel}
	\end{center}
\end{figure}

In this section we show that with high probability over the concatenated code design, the probability of error of a randomly chosen code $\cc$ is at most $\exp{\left(-2\sqrt{n}/(k_1(\log{n})^2)\right)}$. Figure~\ref{fig:flow_rel} is a road-map summarizing our proof of reliability. Upon receiving $\yb$, Bob first partitions $\yb$ into $L$ chunks $(\yb^{(1)}, \yb^{(2)}, \ldots, \yb^{(L)})$. For each chunk $i$, Bob decodes the inner-message $\What^{(i)} = \Gamma^{(i)}_{in}(\yb^{(i)})$ by using the inner decoder $\Gamma^{(i)}_{in}(\cdot)$, and then reconstructs $\Mhat$ from Reed-Solomon code. We now elaborate on the \textbf{decoding rule} of Bob's inner decoder $\Gamma^{(i)}_{in}(\cdot)$, for reconstructing the inner-message $\hat{\W}^{(i)}$ as follows.
\vskip 0.1cm
\begin{tcolorbox}
\textbf{Decoding Rule $\What^{(i)}(\yb^{(i)})$ for reconstructing the inner-message on chunk $i$:}
\begin{enumerate}
\item If $\yb^{(i)} \in \aaybi \setminus \aybi$, then Bob decodes $\What^{(i)} = 0$. 
\vskip 0.2cm 
\item If $\yb^{(i)} \in \aybi$, then 
\begin{enumerate}
    \item If there is exactly one inner-message $w^{(i)}$ such that $\x^{(i)}_w \in \axybi$, then Bob decodes $\What^{(i)} = w^{(i)}$. 
    \item If there exists two inner-messages $w^{(i)}$ and $\tilde{w}^{(i)}$ such that $w^{(i)} \ne \tilde{w}^{(i)}$,  $\x^{(i)}_w \in \axybi$ and $\x^{(i)}_{\tilde{w}} \in \axybi$, then Bob declares an error.
    \item If there does not exist an inner-message $w^{(i)}$ such that $\x^{(i)}_w \in \axybi$, then Bob decodes $\What^{(i)} = 0$.  
  \end{enumerate}
\vskip 0.2cm
\item If $\yb^{(i)}$ is neither in $\aaybi$ nor in $\aybi$, then Bob declares an error. 
\end{enumerate} 
\end{tcolorbox}

\begin{figure}
	\begin{center}
	\includegraphics[scale=0.7]{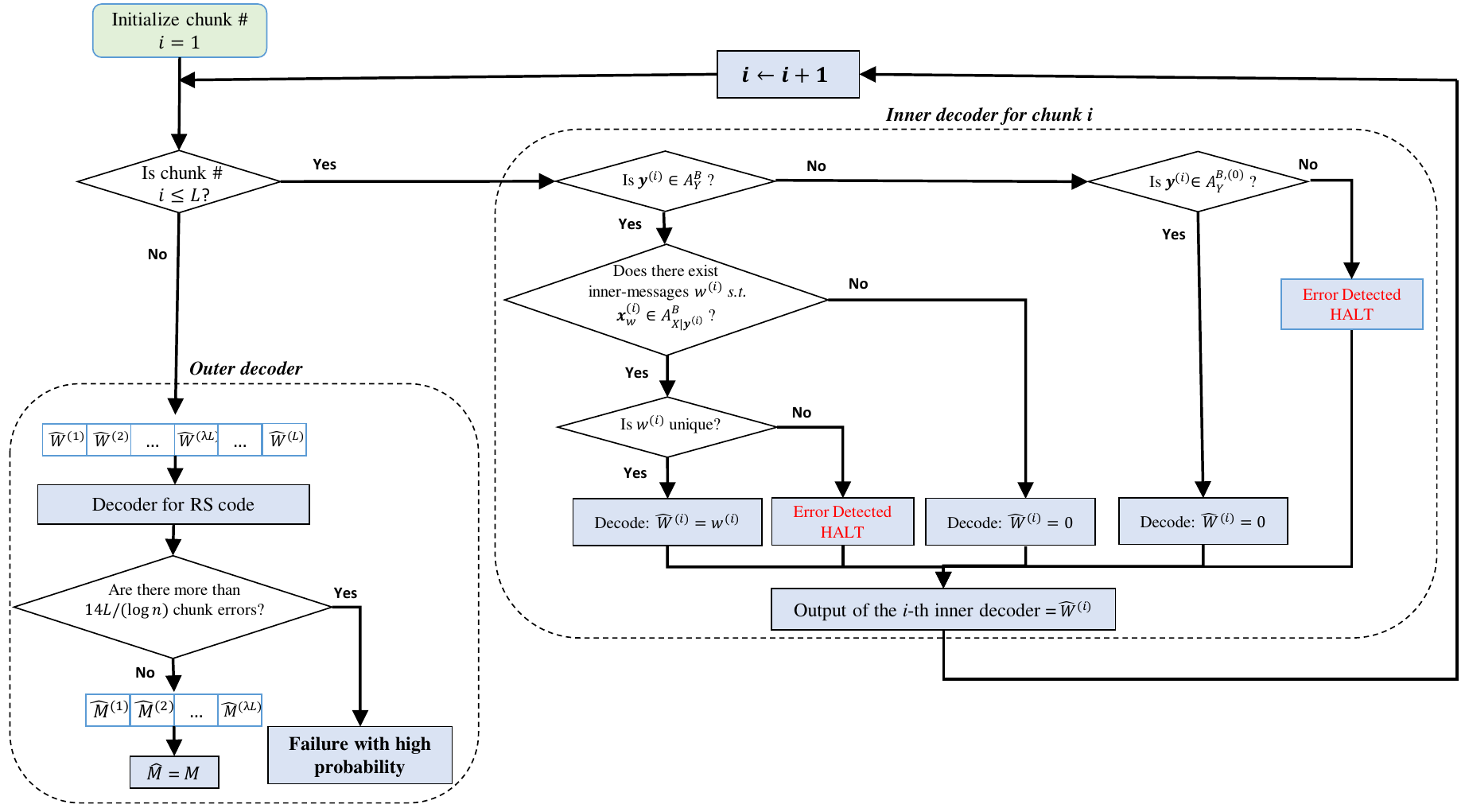}
	\caption{A flow-chart describing the decoding procedure and the potential error events.} \label{fig:decoding-flow_updated}
	\end{center}
\end{figure}

Figure~\ref{fig:decoding-flow_updated} is a flow-chart describing the decoding procedure and the potential error events of Bob's decoder $\Gamma(\cdot)$.   
In the following, we first consider the probability of decoding error of one single chunk, $\mathbb{P} (\What^{(i)} \ne \W^{(i)} )$, under the decoder $\Gamma^{(i)}_{in}(\cdot)$ in Claims~\ref{claim:err1}-\ref{claim:err5} and Lemma~\ref{lem:err}, and then analyze the probability of error $P_{err}$ of the outer RS code in Lemma~\ref{lemma:last}.

\subsection{Probability of decoding error of one single chunk}
\begin{figure}
	\begin{center}
		\includegraphics[scale=0.4]{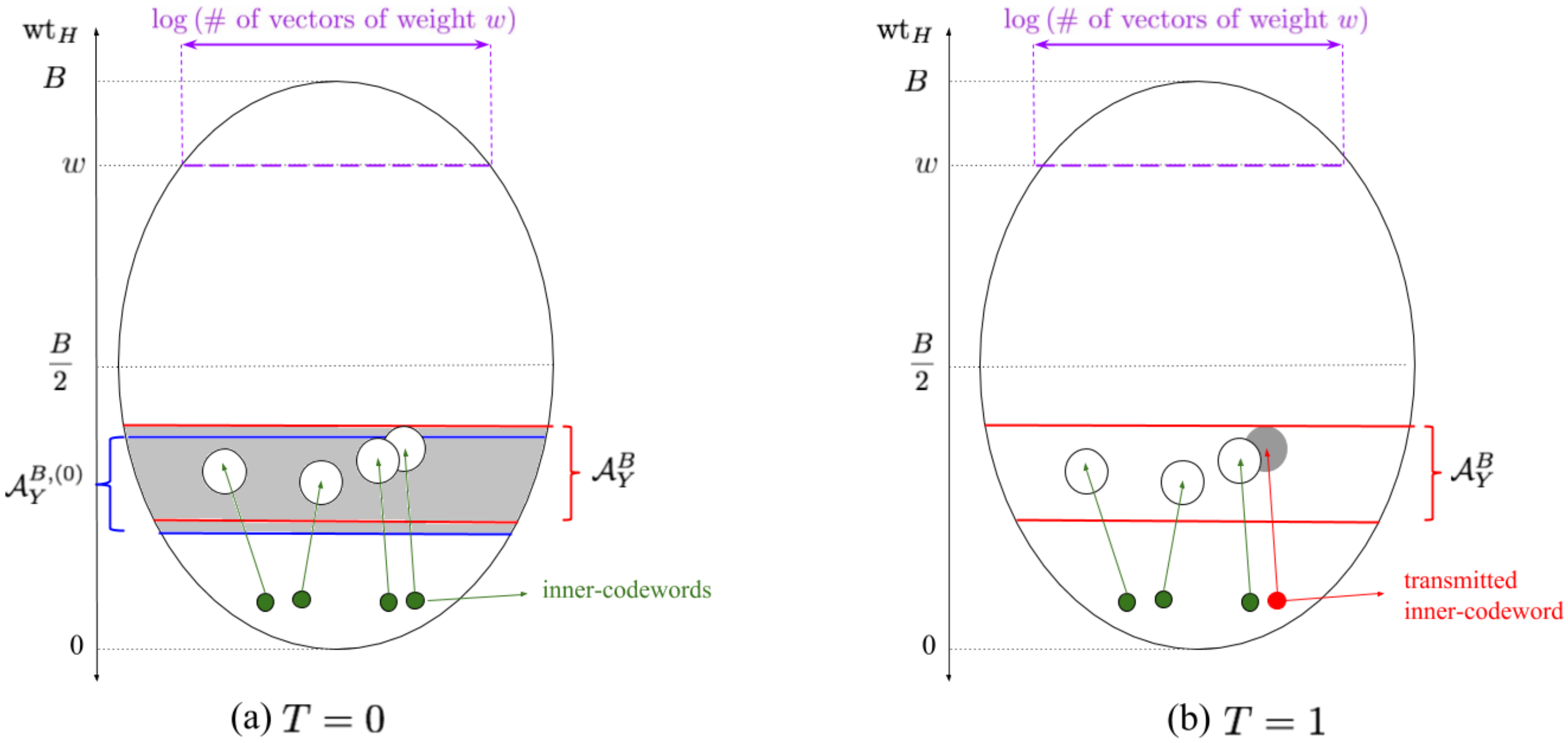}
		\caption{(a) The inner decoder outputs $\What^{(i)} = 0$ if $\yb^{(i)}$ falls into the gray region. Therefore, the blank region represents the possible error events when $\T = 0$, i.e., (i) $\yb^{(i)} \notin \left(\aaybi \cup \aybi\right)$ or (ii) $\yb^{(i)} \in \aybi$, but there exists an inner-codeword falling into the conditionally typical set. \\
			(b) The inner decoder outputs the transmitted inner-codeword if $\yb^{(i)}$ falls into the gray region, i.e., (i) $\yb^{(i)} \in \aybi$, (ii) the transmitted inner-codeword is conditionally typical with $\yb^{(i)}$, and (iii) there does not exist another inner-codeword that is conditionally typical with $\yb^{(i)}$. Therefore, the blank region represents the error events when $\T = 1$ and the ``red'' inner-codeword is transmitted.}
		\label{fig:decode_error}
	\end{center}
\end{figure}

When Alice's transmission status $\T = 1$, without loss of generality, we assume the inner-message $\wi$ is transmitted. Since the inner-messages (for the $i$-th chunk) are equiprobable and each inner-codeword is generated \emph{i.i.d.}, the analysis of error probability is the same no matter which inner-message is transmitted. Therefore, the probability of error $P_{e,(1)}$ when Alice is transmitting is defined as 
\begin{align}
P_{e,(1)} = \mathbb{P} \left(\What^{(i)} \ne W^{(i)} \big| \T = 1\right) = \mathbb{P} \left(\What^{(i)} \ne \wi \big| \W^{(i)} = \wi, \T = 1\right).
\end{align}
When Alice's transmission status $\T = 0$, the probability of error $P_{e,(2)}$ is defined as
\begin{align}
P_{e,(2)} = \mathbb{P} \left(\What^{(i)} \ne 0 \big| \T = 0\right).
\end{align}
The probability of decoding error of one single chunk is given as 
\begin{align}
\mathbb{P} (\What^{(i)} \ne \W^{(i)}) = P_{e,(1)} + P_{e,(2)}.
\end{align}
In the following, we show that both the probability of error when $\T = 0$ and the probability of error when $\T=1$ go to zero asymptotically. Figure~\ref{fig:decode_error} depicts the region of various error events in greater detail.

\vskip 0.1cm
\noindent{\underline{\textbf{Upper bound on $P_{e,(1)}$:}}} From Bob's decoding rule, the probability of error when transmitting ($\T = 1$) can be expanded as follows.
\begin{align}
&\mathbb{P}{\left(\What^{(i)} \ne \wi \big| \W^{(i)}=\wi, \T = 1\right)} \\
&\le \sum_{\yb^{(i)} \in \{0,1\}^B }P(\yb^{(i)}|\x_w^{(i)})\cdot  \mathbbm{1}\left\{\yb^{(i)} \notin \aybi\right\} \label{eq:err1}\\ 
&+ \sum_{\yb^{(i)}\in \{0,1\}^B}P(\yb^{(i)}|\x_w^{(i)})\cdot  \mathbbm{1}\left\{\yb^{(i)} \in \aybi,\x_w^{(i)} \notin \axybi \right\} \label{eq:err2} \\
&+ \sum_{\yb^{(i)}\in \{0,1\}^B}P(\yb^{(i)}|\x_w^{(i)})\cdot  \mathbbm{1}\left\{\exists \tilde{w}^{(i)}\ne w^{(i)} \mbox{\ s.t.\ } \yb^{(i)} \in \aybi, \x^{(i)}_{\tilde{w}} \in \axybi\right\}. \label{eq:err3}
\end{align}
The term in~\eqref{eq:err1} corresponds to the probability of receiving an atypical $\yb^{(i)}$. The term in~\eqref{eq:err2} corresponds to the probability that Bob receives a typical $\yb^{(i)}$, but the true inner-codeword $\x_w^{(i)}$ does not belong to the conditionally typical set $\axybi$. The term in~\eqref{eq:err3} corresponds to the probability that Bob receives a typical $\yb^{(i)}$, but there exists another inner-codeword $\x^{(i)}_{\tilde{w}}$ ($\tilde{w} \ne w$) falling into the conditionally typical set $\axybi$. 

In Claims~\ref{claim:err1}-\ref{claim:err3}, we present that the probabilities of the three error components, presented in~\eqref{eq:err1},~\eqref{eq:err2} and~\eqref{eq:err3}, respectively go to zero as $n$ goes to infinity. In Claim~\ref{claim:err1}, we set $\Delta^{y}_{1} = n^{-1/4+\delta/2}$ (recall that $\Delta^{y}_{1}$ is the parameter, defined in Section~\ref{sec:def}, specifying the ``width'' of the typical set $\mathcal{A}_{\B}^1(Y)$).

\begin{claim}[\bf Term in~\eqref{eq:err1}]\label{claim:err1}
With probability at least $1-2\exp{\left(-\left(\frac{k_1(\rho *\pb)}{3} n^{\delta}+ (\ln 2)\right)(\log n) \right)}$ over the inner code design, for the randomly chosen inner code $\ci$, the probability that Bob receives an atypical $\yb^{(i)}$ is bounded from above as
\begin{align*}
 \sum_{\yb^{(i)}\in \{0,1\}^B}P(\yb^{(i)}|\x_w^{(i)})\cdot  \mathbbm{1}\left\{\yb^{(i)} \notin \aybi\right\} \le {\n}^{-1}.
\end{align*}
\end{claim}
\noindent{\emph{Proof:}} The probability (averaged over the inner code design) that Bob receives an atypical $\yb^{(i)}$ equals
\begin{align}
\mathbb{E}_{\Ci}\left( \sum_{\yb^{(i)}\in \{0,1\}^B}P(\yb^{(i)}|\x_w^{(i)})\cdot  \mathbbm{1}\left\{\yb^{(i)} \notin \aybi\right\} \right) &= \sum_{\x^{(i)}_w \in \{0,1\}^B} \sum_{\yb^{(i)} \in \{0,1\}^B}P_{\X}(\x^{(i)}_w) P(\yb^{(i)}|\x_w^{(i)})\cdot  \mathbbm{1}\left\{\yb^{(i)} \notin \aybi\right\}\\ 
& = \mathbb{P}_{\X^{(i)}_w \Yb^{(i)}} {\left(\Yb^{(i)} \notin \aybi \right)}  \\
&\le 2\exp{\left(-\frac{k_1(\rho *\pb)}{3} n^{\delta}(\log n) \right)} \label{eq:ch1}.
\end{align}
Equation~\eqref{eq:ch1} follows from the Chernoff bound, since the narrow typical set $\aybi$ is centered at $\rho *\pb$ and with width $\Delta_{1}^{y}=n^{-1/4+\delta/2}$. By the Markov's inequality, we have 
\begin{align}
\mathbb{P}_{\Ci}\left(\sum_{\yb^{(i)}\in \{0,1\}^B}P(\yb^{(i)}|\x_w^{(i)})\cdot  \mathbbm{1}\left\{\yb^{(i)} \notin \aybi\right\} \ge {\n}^{-1} \right) \le 2\exp{\left(-\left(\frac{k_1(\rho *\pb)}{3} n^{\delta}+ (\ln 2)\right)(\log n) \right)}.
\end{align}
    \qed

\begin{claim}[\bf Term in~\eqref{eq:err2}] \label{claim:err3}
	With probability at least $1 - 4 (\log{n})\cdot \exp \left(-\frac{1}{3} k_1k_2\pb (\log n)^{1/3}\right)$ over the inner code design, for the randomly chosen inner code $\ci$, the probability that Bob receives a typical $\yb^{(i)}$ and a conditionally atypical inner-codeword $\x^{(i)}_{w}$ is transmitted is bounded from above as
	$$\sum_{\yb^{(i)}\in \{0,1\}^B}P(\yb^{(i)}|\x_w^{(i)})\cdot  \mathbbm{1}\left\{\yb^{(i)} \in \aybi,\x_w^{(i)} \notin \axybi \right\} \le 1/(\log{n}).$$ 
\end{claim}

\noindent{\emph{Proof:}} 
Note that 
\begin{align}
&\mathbb{E}_{\Ci}\left(\sum_{\yb^{(i)}\in \{0,1\}^B}P(\yb^{(i)}|\x_w^{(i)})\cdot  \mathbbm{1}\left\{\yb^{(i)} \in \aybi,\x_w^{(i)} \notin \axybi \right\}\right) \\
&= \sum_{\x^{(i)}_w \in \{0,1\}^B} \sum_{\yb^{(i)} \in \{0,1\}^B}P_{\X}(\x^{(i)}_w)P(\yb^{(i)}|\x_w^{(i)})\cdot  \mathbbm{1}\left\{\yb^{(i)} \in \aybi,\x_w^{(i)} \notin \axybi \right\}\\
&= \mathbb{P}_{\X^{(i)}_w,\Yb^{(i)}}\left(\Yb^{(i)} \in \aybi, \X^{(i)}_w \notin \axYbi \right) \\
&\le \mathbb{P}_{\X^{(i)}_w,\Yb^{(i)}}\left(\X^{(i)}_w \notin \axYbi \right). 
\end{align} 
We now bound the probability that the true inner-codeword $\X^{(i)}_w$ does not belong to the conditionally typical set $\axYbi$.
\begin{align}
&\mathbb{P}_{\X^{(i)}_w,\Yb^{(i)}}\left(\X^{(i)}_w \notin \axYbi \right) \\
&= \mathbb{P}_{\X^{(i)}_w,\Yb^{(i)}} \left(f_{10}^{xy}(\X^{(i)}_w,\Yb^{(i)}) \notin \left[(1-\Delta_{10}^{xy})\rho \pb, (1+\Delta_{10}^{xy})\rho \pb\right] \bigcup f_{11}^{xy}(\X^{(i)}_w,\Yb^{(i)}) \notin \left[(1-\Delta_{11}^{xy})\rho (1-\pb), (1+\Delta_{11}^{xy})\rho (1-\pb)\right]\right) \\
& \le \mathbb{P}_{\X^{(i)}_w,\Yb^{(i)}} \left(f_{10}^{xy}(\X^{(i)}_w,\Yb^{(i)}) \notin \left[(1-\Delta_{10}^{xy})\rho \pb, (1+\Delta_{10}^{xy})\rho \pb\right] \right) + \mathbb{P}_{\X^{(i)}_w,\Yb^{(i)}} \left( f_{11}^{xy}(\X^{(i)}_w,\Yb^{(i)}) \notin \left[(1-\Delta_{11}^{xy})\rho (1-\pb), (1+\Delta_{11}^{xy})\rho (1-\pb)\right]\right) \\
&\le 2 \exp \left(-\frac{1}{3} k_1k_2\pb (\log n)^{\frac{1}{3}}\right) + 2 \exp \left(-\frac{1}{3} k_1k_2(1-\pb) (\log n)^{\frac{1}{3}}\right) \label{eq:jan_12} \\
& \le 4\exp \left(-\frac{1}{3} k_1k_2\pb (\log n)^{\frac{1}{3}}\right),
\end{align}
where inequality~\eqref{eq:jan_12} is due to the Chernoff bound and the fact that $\Delta_{10}^{xy} = \Delta_{11}^{xy} = (\log n)^{-1/3}$.
By applying Markov's inequality, we obtain that with probability at least $1 - 4 (\log{n})\cdot \exp \left(-\frac{1}{3} k_1k_2\pb (\log n)^{1/3}\right)$ over the inner code design, the randomly chosen inner code $\ci$ satisfies $$\sum_{\yb^{(i)}\in \{0,1\}^B}P(\yb^{(i)}|\x_w^{(i)})\cdot  \mathbbm{1}\left\{\yb^{(i)} \in \aybi,\x_w^{(i)} \notin \axybi \right\} \le 1/(\log{n}).$$ 		\qed

\begin{claim}[\bf Term in~\eqref{eq:err3}] \label{claim:err2}
With probability at least $1-2^{-\mathcal{O}\left((\log n)^{3/4}\right)}$ over the inner code design, for the randomly chosen inner code $\ci$, the probability that Bob receives a typical $\yb^{(i)}$ and there exists another inner-codeword $\x^{(i)}_{\tilde{w}}$ ($\tilde{w}^{(i)} \ne w^{(i)}$) falling into the conditionally typical set $\axybi$ is bounded from above as 
\begin{align}
\sum_{\yb^{(i)}\in \{0,1\}^B}P(\yb^{(i)}|\x_w^{(i)})\cdot  \mathbbm{1}\left\{\exists \tilde{w}^{(i)}\ne w^{(i)} \mbox{\ s.t.\ } \yb^{(i)} \in \aybi, \x^{(i)}_{\tilde{w}} \in \axybi\right\} \le \frac{1}{\log n}. 
\end{align}
\end{claim}
 
\noindent{\emph{Proof:}} We first note that 
\begin{align}
& \mathbb{E}_{\Ci}\left[ \sum_{\yb^{(i)}\in \{0,1\}^B}P(\yb^{(i)}|\x_w^{(i)})\cdot  \mathbbm{1}\left\{\exists \tilde{w}^{(i)}\ne w^{(i)} \mbox{\ s.t.\ } \yb^{(i)} \in \aybi, \x^{(i)}_{\tilde{w}} \in \axybi\right\} \right] \\
& \le \sum_{\tilde{w}^{(i)}\ne w^{(i)}} \mathbb{E}_{\Ci}\left[ \sum_{\yb^{(i)}\in \{0,1\}^B}P(\yb^{(i)}|\x_w^{(i)})\cdot  \mathbbm{1}\left\{\yb^{(i)} \in \aybi, \x^{(i)}_{\tilde{w}} \in \axybi\right\} \right],
\end{align}
and for any $\tilde{w}^{(i)} \ne w^{(i)}$,
\begin{align}
&\mathbb{E}_{\Ci}\left[ \sum_{\yb^{(i)} \in \{0,1\}^B}P(\yb^{(i)}|\x_{w}^{(i)})\cdot  \mathbbm{1}\left\{\yb^{(i)} \in \aybi, \x^{(i)}_{\tilde{w}} \in \axybi\right\} \right]  \label{eq:1-54} \\
&= \sum_{\x^{(i)}_{\tilde{w}} \in \{0,1\}^B} \sum_{\yb^{(i)} \in \aybi}P_{\X}(\x^{(i)}_{\tilde{w}}) P(\yb^{(i)}|\x_{w}^{(i)})\cdot  \mathbbm{1}\left\{ \x^{(i)}_{\tilde{w}} \in \axybi\right\}  \label{eq:1-55} \\
&= \sum_{\yb^{(i)} \in \aybi}P(\yb^{(i)}|\x_{w}^{(i)}) \cdot \mathbb{P}_{\X^{(i)}_{\tilde{w}}}\left(\X^{(i)}_{\tilde{w}} \in \axybi \right). \label{eq:1-56}
\end{align}
For any typical $\yb^{(i)}$, the probability that a single inner-codeword falls into the conditionally typical set $\axybi$ is bounded from above as 
\begin{align}
\mathbb{P}_{\X^{(i)}_{\tilde{w}}}\left(\X^{(i)}_{\tilde{w}} \in \axybi \right) &=\sum_{(f_{10}^{xy},f_{11}^{xy})\in \mathcal{F}^{xy}_{\B}} \mathbb{P}_{\X^{(i)}_{\tilde{w}}}\left( \X^{(i)}_{\tilde{w}} \in \txybi(f^{xy}_{10},f^{xy}_{11})\right) \label{eq:errr2-0} \\
&= \sum_{(f_{10}^{xy},f_{11}^{xy})\in \mathcal{F}^{xy}_{\B}} \binom{\npri \left( f_{01}^{xy} + f_{11}^{xy}\right)}{\npri f_{11}^{xy}}\rho^{\npri f_{11}^{xy}}(1-\rho)^{\npri f_{01}^{xy}} \binom{\npri \left( f_{00}^{xy} + f_{10}^{xy}\right)}{\npri f_{10}^{xy}}\rho^{\npri f_{10}^{xy}}(1-\rho)^{\npri f_{00}^{xy}} \label{eq:errr2-1} \\
&\le \sum_{(f_{10}^{xy},f_{11}^{xy})\in \mathcal{F}^{xy}_{\B}} 2^{-k_1\sqrt{\n}(\log{\n}) \left[\I(\x^{(i)}_{\tilde{w}}; \yb^{(i)})+\mathbb{D}(\x^{(i)}_{\tilde{w}} \parallel \rho)\right]} \label{eq:errr2-2} \\
&\le 2 \rho \pb \frac{\npri}{(\log{n})^{1/3}}\cdot 2 \rho (1-\pb) \frac{\npri}{(\log{n})^{1/3}}\cdot 2^{-k_1\sqrt{\n}(\log{\n}) \left[\I(\x^{(i)}_{\tilde{w}}; \yb^{(i)})+\mathbb{D}(\x^{(i)}_{\tilde{w}} \parallel \rho)\right]} \label{eq:errr2-3}\\
&= 4k_1^2k_2^2\pb(1-\pb)(\log{\n})^{4/3} \cdot {\n}^{-2k_1\epsilon_d \sqrt{\pw(1-\pw)}\frac{1-2\pb}{1-2\pw}\log{\left(\frac{1-\pb}{\pb}\right)}+\mathcal{O}\left((\log n)^{-1/3}\right)}. \label{eq:errr2-4}
\end{align}
Equation~\eqref{eq:errr2-0} decomposes the conditionally typical set $\axybi$ into the typical conditional type classes $\txybi(f^{xy}_{10},f^{xy}_{11})$ that comprise it. To obtain Equation~\eqref{eq:errr2-1}, we use standard counting arguments to calculate the probability that $\X^{(i)}_{\tilde{w}}$ falls into the type class $\txybi(f^{xy}_{10},f^{xy}_{11})$ given a typical $\yb^{(i)}$. Equation~\eqref{eq:errr2-2} follows from the Stirling's approximation, as well as $\I(\x^{(i)}_{\tilde{w}}; \yb^{(i)}) \triangleq \sum_{(a,b) \in \left\{0,1\right\} \times \left\{0,1\right\}}f^{xy}_{ab}\log{\frac{f^{xy}_{ab}}{f^x_{a}\cdot f^y_{b}}} $ and $\mathbb{D}(\x^{(i)}_{\tilde{w}} \parallel \rho) \triangleq f^x_{0}\log{\frac{f^x_{0}}{1-\rho}}+f^x_{1}\log{\frac{f^x_{1}}{\rho}}$. Equation~\eqref{eq:errr2-3} follows since the number of typical conditional type classes is bounded from above by\footnote{Recall that we set $\Delta_{10}^{xy} = \Delta_{11}^{xy} = (\log{n})^{-1/3}$, which specify the ``width'' of the conditionally typical set $\axybi$.}
\begin{align}
2 \rho \pb \frac{\npri}{(\log{n})^{1/3}}\cdot 2 \rho (1-\pb) \frac{\npri}{(\log{n})^{1/3}} = 2 \pb \frac{k_2}{\sqrt{n}} \frac{k_1\sqrt{n}\log n}{(\log{n})^{1/3}} \cdot 2 (1-\pb) \frac{k_2}{\sqrt{n}} \frac{k_1\sqrt{n}\log n}{(\log{n})^{1/3}} = 4k_1^2k_2^2\pb(1-\pb)(\log{\n})^{4/3}.  \label{eq:num_class}
\end{align}
In Equation~\eqref{eq:errr2-4}, we use the fact that
\begin{align}
\I(\x^{(i)}_{\tilde{w}}; \yb^{(i)})+\mathbb{D}(\x^{(i)}_{\tilde{w}} \parallel \rho) &= \rho (1-2\pb) \log{\left(\frac{1-\pb}{\pb}\right)}+\mathcal{O}\left(n^{-1/2}(\log n)^{-1/3}\right) \label{eq:feng1} \\
&= 2 \epsilon_d \sqrt{\frac{\pw (1-\pw)}{n}}\frac{1-2\pb}{1-2\pw} \log{\left(\frac{1-\pb}{\pb}\right)} + \mathcal{O}\left(n^{-1/2}(\log n)^{-1/3}\right), \label{eq:feng3}
\end{align}
where Equation~\eqref{eq:feng1} is formally proved in Appendix~\ref{app:max3}, and Equation~\eqref{eq:feng3} follows since we set $\rho = k_2/\sqrt{n} = \frac{2\epsilon_d\sqrt{\pw(1-\pw)}}{(1-2\pw)\sqrt{n}}$. Returning now to Equations~\eqref{eq:1-54}-\eqref{eq:1-56}, we have
\begin{align}
\mathbb{E}_{\Ci}\left[ \sum_{\yb^{(i)} \in \{0,1\}^B}P(\yb^{(i)}|\x_{\tilde{w}}^{(i)})\cdot  \mathbbm{1}\left\{\yb^{(i)} \in \aybi, \x^{(i)}_{\tilde{w}} \in \axybi\right\} \right] \le 4k_1^2k_2^2\pb(1-\pb)(\log{\n})^{4/3} \cdot {\n}^{-2k_1\epsilon_d \sqrt{\pw(1-\pw)}\frac{1-2\pb}{1-2\pw}\log{\left(\frac{1-\pb}{\pb}\right)}+\mathcal{O}\left((\log n)^{-1/3}\right)}.
\end{align}
Since the size of inner codebook is $n^{\rin}$ (as defined in Section~\ref{sec:code}), where 
\begin{align}
\rin = r_u(1-(\log{n})^{-1/4})k_1/\lambda =2k_1\epsilon_d \sqrt{\pw(1-\pw)}\frac{1-2\pb}{1-2\pw}\log{\left(\frac{1-\pb}{\pb}\right)} \frac{1-(\log{n})^{-1/4}}{1-(28/\log n)}, \label{eq:rate}
\end{align}
the probability (averaged over the inner code design) that there exists another inner-codeword $\x^{(i)}_{\tilde{w}}$ falling into the conditionally typical set is bounded from above as
\begin{align}
&\mathbb{E}_{\Ci}\left[ \sum_{\yb^{(i)}\in \{0,1\}^B}P(\yb^{(i)}|\x_w^{(i)})\cdot  \mathbbm{1}\left\{\exists \tilde{w}^{(i)}\ne w^{(i)} \mbox{\ s.t.\ } \yb^{(i)} \in \aybi, \x^{(i)}_{\tilde{w}} \in \axybi\right\} \right] \label{eq:feng4} \\
&\le n^{\rin} \cdot \mathbb{E}_{\Ci}\left[ \sum_{\yb^{(i)} \in \{0,1\}^B}P(\yb^{(i)}|\x_{\tilde{w}}^{(i)})\cdot  \mathbbm{1}\left\{\yb^{(i)} \in \aybi, \x^{(i)}_{\tilde{w}} \in \axybi\right\} \right] \\
&\le  n^{\rin} \cdot  4k_1^2k_2^2\pb(1-\pb)(\log{\n})^{4/3}\cdot {\n}^{-2k_1\epsilon_d \sqrt{\pw(1-\pw)}\frac{1-2\pb}{1-2\pw}\log{\left(\frac{1-\pb}{\pb}\right)}+\mathcal{O}\left((\log n)^{-1/3}\right)} \\ 
&= 4k_1^2k_2^2\pb(1-\pb)(\log{\n})^{4/3}\cdot {\n}^{\rin -2k_1\epsilon_d \sqrt{\pw(1-\pw)}\frac{1-2\pb}{1-2\pw}\log{\left(\frac{1-\pb}{\pb}\right)}+\mathcal{O}\left((\log n)^{-1/3}\right)}  \\
&= 4k_1^2k_2^2\pb(1-\pb)(\log{\n})^{4/3}\cdot 2^{-\left(2k_1\epsilon_d \sqrt{\pw(1-\pw)}\frac{1-2\pb}{1-2\pw}\log{\left(\frac{1-\pb}{\pb}\right)}\right)(\log n)^{3/4}+\mathcal{O}\left((\log n)^{2/3}\right)} \label{eq:xuehui1}\\
&=2^{\log\left(4k_1^2k_2^2\pb(1-\pb)\right)+\log \left((\log n)^{4/3}\right)}\cdot 2^{-\left(2k_1\epsilon_d \sqrt{\pw(1-\pw)}\frac{1-2\pb}{1-2\pw}\log{\left(\frac{1-\pb}{\pb}\right)}\right)(\log n)^{3/4}+\mathcal{O}\left((\log n)^{2/3}\right)} \label{eq:xuehui2}\\
&= 2^{-\mathcal{O}\left((\log n)^{3/4}\right)}. \label{eq:feng5}
\end{align}
Equation~\eqref{eq:xuehui1} holds since
\begin{align*}
&{\n}^{\rin-2k_1\epsilon_d \sqrt{\pw(1-\pw)}\frac{1-2\pb}{1-2\pw}\log{\left(\frac{1-\pb}{\pb}\right)}+\mathcal{O}\left((\log n)^{-1/3}\right)} \\
&= n^{\left(2k_1\epsilon_d \sqrt{\pw(1-\pw)}\frac{1-2\pb}{1-2\pw}\log{\left(\frac{1-\pb}{\pb}\right)}\right)\cdot \left(\frac{1-(\log n)^{-1/4}}{1-(28/\log n)}-1\right)+\mathcal{O}\left((\log n)^{-1/3}\right)}\\
&= n^{-\left(2k_1\epsilon_d \sqrt{\pw(1-\pw)}\frac{1-2\pb}{1-2\pw}\log{\left(\frac{1-\pb}{\pb}\right)}\right)\cdot \left((\log n)^{-1/4}-\mathcal{O}\left(\frac{28}{\log n}\right)\right)+\mathcal{O}\left((\log n)^{-1/3}\right)}\\
 &= 2^{-\left(2k_1\epsilon_d \sqrt{\pw(1-\pw)}\frac{1-2\pb}{1-2\pw}\log{\left(\frac{1-\pb}{\pb}\right)}\right)(\log n)^{3/4}+\mathcal{O}\left((\log n)^{2/3}\right)}.
\end{align*}
By applying Markov's inequality, we obtain that with probability at least $1-2^{-\mathcal{O}\left((\log n)^{3/4}\right)}$ over the inner code design, the chosen inner code $\ci$ satisfies 
\begin{align}
\sum_{\yb^{(i)} \in \{0,1\}^B}P(\yb^{(i)}|\x_w^{(i)})\cdot  \mathbbm{1}\left\{\exists \tilde{w}^{(i)}\ne w^{(i)} \mbox{\ s.t.\ } \yb^{(i)} \in \aybi, \x^{(i)}_{\tilde{w}} \in \axybi\right\} \le \frac{1}{\log n}
\end{align}
This completes the proof of Claim~\ref{claim:err2}. \qed
%If we set $\rin$ to be $\rin_u \left(1-\frac{1}{(\log{n})^{1/2}}\right)$, then we have $1 - {\n}^{-(\rin_u - \rin)} = 1- 2^{(\log{n})^{1/2}}$.

\vskip 0.2cm
\noindent{\underline{\textbf{Upper bound on $P_{e,(2)}$:}}} When Alice's transmission status $\T = 0$, the probability of error can be decomposed as 
\begin{align}
\mathbb{P}{\left(\What^{(i)} \ne 0 | \T = 0\right)} &= \sum_{\yb^{(i)} \in \{0,1\}^B} P(\yb^{(i)}|\x^{(i)} = \underline{0}) \cdot \mathbbm{1}\left\{\left(\yb^{(i)} \notin \left(\aaybi \cup \aybi\right)\right) \text{ or } \left(\yb^{(i)} \in \aybi \text{ and } \exists w^{(i)}: \x^{(i)}_w \in \axybi \right) \right\}\\ 
&\le \sum_{\yb^{(i)} \in \{0,1\}^B} P(\yb^{(i)}|\x^{(i)} = \underline{0}) \cdot \mathbbm{1}\left\{\yb^{(i)} \notin \aaybi \right\} \label{eq:err2-1} \\
&+ \sum_{\yb^{(i)}\in \{0,1\}^B} P(\yb^{(i)}|\x^{(i)} = \underline{0}) \cdot \mathbbm{1}\left\{\yb^{(i)} \in \aybi \text{ and } \exists w^{(i)}: \x^{(i)}_w \in \axybi \right\}. \label{eq:err2-2}
\end{align}
The term in~\eqref{eq:err2-1} corresponds to the probability that Bob receives an atypical $\yb^{(i)}$ (with respect to $\T = 0$). The term in~\eqref{eq:err2-2} corresponds to the probability that Bob receives a $\yb^{(i)} \in \aybi$ but there exists an inner-codeword $\x^{(i)}$ falling into the conditionally typical set $\axybi$. In Claims~\ref{claim:err4} and~\ref{claim:err5}, we show that the terms in~\eqref{eq:err2-1} and~\eqref{eq:err2-2} decrease to zero as $n$ grows without bound.

\begin{claim}[\bf Term in~\eqref{eq:err2-1}] \label{claim:err4}
When Alice's transmission status $\T=0$, the probability that Bob receives an atypical $\yb^{(i)}$ (with respect to $\T=0$) is bounded from above as 
$$\sum_{\yb^{(i)} \in \{0,1\}^B} P(\yb^{(i)}|\x^{(i)} = \underline{0}) \cdot \mathbbm{1}\left\{\yb^{(i)} \notin \aaybi \right\} \le 2\exp \left(-\frac{k_1\pb}{3} {n}^{\delta} \log{\n}\right).$$
\end{claim}

\noindent{\emph{Proof:}} The typical set $\aaybi$ when $\T=0$ is centered at $\pb$ with width $\Delta^{y}_{1}={\n}^{-1/4+\delta/2}$ (recall that $\Delta^{y}_{1}$ is the parameter, defined in Section~\ref{sec:def}, specifying the ``width'' of the typical set $\aaybi$). We then use the Chernoff bound to calculate the probability of receiving an atypical $\yb^{(i)}$ as follows:
\begin{align}
\sum_{\yb^{(i)} \in \{0,1\}^B} P(\yb^{(i)}|\x^{(i)} = \underline{0}) \cdot \mathbbm{1}\left\{\yb^{(i)} \notin \aaybi \right\} &= \mathbb{P}_{\zb}\left(f^{y}_{1}(\Yb^{(i)}) \notin \left[\pb \left(1 \pm \Delta^{y}_{1}\right)\right] \right)  \\
&\le 2\exp \left(-\frac{k_1\pb}{3} {n}^{\delta} \log{\n}\right). \label{eq:sub}
\end{align}
   \qed

\begin{claim}[\bf Term in~\eqref{eq:err2-2}] \label{claim:err5}
With probability at least $1-2^{-\mathcal{O}\left((\log n)^{3/4}\right)}$ over the inner code design, for the chosen inner code $\ci$, the probability that Bob receives a typical $\yb^{(i)}$ as well as there exists an inner-codeword falling into the conditionally typical set $\axybi$ is bounded from above as
$$\sum_{\yb^{(i)} \in \{0,1\}^B} P(\yb^{(i)}|\x^{(i)} = \underline{0}) \cdot \mathbbm{1}\left\{\yb^{(i)} \in \aybi \text{ and } \exists w^{(i)}: \x^{(i)}_w \in \axybi \right\} \le \frac{1}{\log n}.$$
\end{claim}

\noindent{\emph{Proof:}} We first note that 
\begin{align}
&\mathbb{E}_{\ci}\left(\sum_{\yb^{(i)} \in \{0,1\}^B} P(\yb^{(i)}|\x^{(i)} = \underline{0}) \cdot \mathbbm{1}\left\{\yb^{(i)} \in \aybi \text{ and } \exists w^{(i)}: \x^{(i)}_w \in \axybi \right\}\right) \\
&\le \mathbb{E}_{\ci}\left(\sum_{w^{(i)}} \sum_{\yb^{(i)} \in \aybi } P(\yb^{(i)}|\x^{(i)} = \underline{0}) \cdot \mathbbm{1}\left\{\x^{(i)}_w \in \axybi \right\} \right) \\
&= \sum_{w^{(i)}} \sum_{\x^{(i)}_w \in \{0,1\}^B}\sum_{\yb^{(i)} \in \aybi}P_{\X}(\x^{(i)}_w)P(\yb^{(i)}|\x^{(i)} = \underline{0}) \cdot \mathbbm{1}\left\{\x^{(i)}_w \in \axybi \right\} \\
&= n^{\rin}\cdot \sum_{\yb^{(i)} \in \aybi}P(\yb^{(i)}|\x^{(i)} = \underline{0}) \cdot  \mathbb{P}_{\X^{(i)}_{w}}\left(\X^{(i)} \in \axybi \right).  \label{eq:appear}
\end{align}
As noted in~\eqref{eq:errr2-4}, for any $\yb^{(i)} \in \aybi$, the probability that a single inner-codeword falls into the conditionally typical set is bounded from above as
\begin{align}
\mathbb{P}_{\X^{(i)}_{w}}\left(\X^{(i)} \in \axybi \right) \le 4k_1^2k_2^2\pb(1-\pb)(\log{\n})^{4/3} \cdot {\n}^{-2k_1\epsilon_d \sqrt{\pw(1-\pw)}\frac{1-2\pb}{1-2\pw}\log{\left(\frac{1-\pb}{\pb}\right)}+\mathcal{O}\left((\log n)^{-1/3}\right)}. \label{eq:home3}
\end{align}
The rest of the proof is the same as that of Claim~\ref{claim:err2}. Combining~\eqref{eq:home3} and the size of the inner-codebook $n^{\rin}$ (as shown in~\eqref{eq:rate}), we have  
\begin{align}
\mathbb{E}_{\ci}\left(\sum_{\yb^{(i)} \in \{0,1\}^B} P(\yb^{(i)}|\x^{(i)} = \underline{0}) \cdot \mathbbm{1}\left\{\yb^{(i)} \in \aybi \text{ and } \exists w^{(i)}: \x^{(i)}_w \in \axybi \right\}\right) \le 2^{-\mathcal{O}\left((\log n)^{3/4}\right)}.
\end{align}
Finally, by the Markov inequality, we obtain that with probability at least $1-2^{-\mathcal{O}\left((\log n)^{3/4}\right)}$ over the inner code design, the randomly chosen inner code $\ci$ satisfies
$$\sum_{\yb^{(i)} \in \{0,1\}^B} P(\yb^{(i)}|\x^{(i)} = \underline{0}) \cdot \mathbbm{1}\left\{\yb^{(i)} \in \aybi \text{ and } \exists w^{(i)}: \x^{(i)}_w \in \axybi \right\} \le \frac{1}{\log n}.$$
This completes the proof of Claim~\ref{claim:err5}. \qed

\vskip 0.2cm
Having proved Claims~\ref{claim:err1}-\ref{claim:err5}, it turns out that the probability of decoding error of one single chunk follows directly. A summary of Claims~\ref{claim:err1}-\ref{claim:err5} is presented in Table~\ref{table:reliability}. For notational convenience we define $\zeta_{prob} = 5 (\log{n})\cdot \exp \left(-\frac{1}{3} k_1k_2\pb (\log n)^{1/3}\right)$, and then we have the following lemma.

\begin{table}\centering\caption{Summary of probability of error of one single chunk}\label{table:reliability}
	\begin{tabular}{|c|l|l|}
		\hline
		Claim & $(\dagger)$ Probability of error contributed by the corresponding Claim & Probability of the inner code satisfying $(\dagger)$ 	\\
		\hline Claim~\ref{claim:err1} & at most $1/n$ & $1-2\exp{\left(-\left(\frac{k_1(\rho *\pb)}{3} n^{\delta}+ (\ln 2)\right)(\log n) \right)}$  \\
		\hline Claim~\ref{claim:err3} & at most $1/(\log n)$ & $1 - 4 (\log{n})\cdot \exp \left(-\frac{1}{3} k_1k_2\pb (\log n)^{1/3}\right)$\\ 
		\hline Claim~\ref{claim:err2} & at most $1/(\log n)$ &$1-2^{-\mathcal{O}\left((\log n)^{3/4}\right)}$ \\
		\hline Claim~\ref{claim:err4} & at most $2\exp \left(-\frac{k_1\pb}{3} {n}^{\delta} \log{\n}\right)$ & $1$ \\
		\hline Claim~\ref{claim:err5} & at most $1/(\log n)$ &$1-2^{-\mathcal{O}\left((\log n)^{3/4}\right)}$  \\
		\hline
	\end{tabular}
\end{table}

\begin{lemma}\label{lem:err}
With probability at least $1 - \zeta_{prob}$ over the inner code design, for the randomly chosen inner code $\ci$, the probability of error is bounded from above as 
$$\mathbb{P} \left(\What^{(i)} \ne \W^{(i)} \right) < \frac{6}{\log n}.$$
\end{lemma}
\noindent{\emph{Proof:}} By Table~\ref{table:reliability} and the union bound, for sufficiently large $n$, we prove that with probability at least $1 - \zeta_{prob}$ over the inner code design and for sufficiently large $n$, the probability of error of the chosen inner code is bounded from above as
\begin{align}
\mathbb{P} \left(\What^{(i)} \ne \W^{(i)} \right) &=  \mathbb{P} \left(\What^{(i)} \ne W^{(i)} \big| \T = 1\right) +  \mathbb{P}{\left(\What^{(i)} \ne 0 \big| \T = 0\right)} \\
&\le  \frac{3}{\log n}+ n^{-1} +  2\exp \left(-\frac{k_1\pb}{3} {n}^{\delta} \log{\n}\right) \label{eq:5claim}\\
&\le \frac{6}{\log n}. 
\end{align} 
Inequality~\eqref{eq:5claim} basically follows from Claims~\ref{claim:err1}-\ref{claim:err5}. This completes the proof of Lemma~\ref{lem:err}.  \qed

\subsection{Probability of error of the concatenated code}
\begin{lemma} \label{lemma:last}
With probability at least $1 - \exp{\left(-\frac{L}{3}\zeta_{prob}\right)}$ over the concatenated code design, for the randomly chosen code $\cc$, the overall probability of error $P_{err}$ is bounded from above as 
\begin{align}
P_{err} \le \exp{\left(-2\sqrt{n}/(k_1(\log{n})^2)\right)}.
\end{align}
\end{lemma}
\noindent{{\it Proof:}} Lemma~\ref{lem:err} shows that with probability at least $1 - \zeta_{prob}$ over the inner code design, the probability of error of a randomly chosen inner code $\ci$ satisfies $\mathbb{P} \left(\What^{(i)} \ne \W^{(i)} \right) < 6/(\log{\n})$. An inner code (for chunk $i$) is said to be a {\it good inner code (for chunk $i$)} if the probability of error over the channel noise is bounded from above by $6/(\log{\n})$, and is said to be a {\it bad inner code (for chunk $i$)} otherwise. Let $\Lambda_1$ and $\Lambda_2$ be the number of chunk errors induced by good and bad inner codes respectively, that the RS outer code will need to correct. In the following we focus on the impact of good and bad inner codes on number of chunk in error.

\noindent{{\bf (i) Impact of good inner codes on number of chunk in error:}} Since the number of good inner codes is at most $L$, and the probability of error of good inner codes is bounded from above by $6/(\log{n})$, it then follows that the expected number of chunk in error induced by good inner codes, $\mathbb{E}(\Lambda_1)$, is bounded from above by $6L/(\log{\n})$. By the Chernoff bound, with probability at least $1 - \exp{\left(-2L/(\log{n})\right)}$ over the code design, the number of chunk in error induced by good inner codes is bounded from above by $12L/(\log{n})$.

\noindent{{\bf (ii) Impact of bad inner codes on number of chunk in error:}} Note that the probability of generating a bad inner code is at most $\zeta_{prob}$, hence the expected number of bad inner codes is bounded from above by $L \zeta_{prob}$. Since the inner codes are generated independently, by the Chernoff bound, with probability at least $1 - \exp{\left(-\frac{L}{3}\zeta_{prob}\right)}$ over the code design, the number of bad inner codes is bounded from above by $2L \zeta_{prob}$, which implies the number of chunk in error induced by bad inner codes, $\Lambda_2$, is bounded from above by $2L \zeta_{prob}$.

\noindent{{\bf (iii) Concentration of overall inner codes in error:}} A concatenated code $\cc$ is said to be a {\it decent} code if the number of bad inner codes of $\cc$ is no more than $2L \zeta_{prob}$. From (ii) we know that with probability at least $1 - \exp{\left(-\frac{L}{3}\zeta_{prob}\right)}$ over the code design, a randomly chosen code $\cc$ from the concatenated code ensemble is decent. Conditioned on the event that a decent code $\cc$ is chosen, it then follows from (i) that with probability at least $1-\exp{\left(-2L/(\log{n})\right)}$, the number of chunk in error induced by good inner codes is bounded from above by $12L/(\log{n})$, and hence the number of overall inner codes in error is bounded from above as
\begin{align}
\Lambda_1 + \Lambda_2 \le \frac{12L}{\log{n}}+ 2L \zeta_{prob} \le \frac{14L}{\log{n}},
\end{align} 
for sufficiently large $n$.
Our outer Reed-Solomon code is able to correct $14L/(\log{n})$ errors, since the number of parity chunks is $28L/(\log{n})$. Therefore, with probability at least\footnote{We note that $\exp{\left(-2L/(\log{n})\right)}$ is decaying faster than $\exp{\left(-\frac{L}{3}\zeta_{prob}\right)}$. } $1 - \exp{\left(-\frac{L}{3}\zeta_{prob}\right)}$ over the concatenated code design, for the randomly chosen code $\cc$, the overall probability of error $P_{err}$ is bounded from above as
\begin{align}
P_{err} &\le \mathbb{P} \left((\Lambda_1+\Lambda_2) > \frac{14L}{\log{n}}\right) \\
&\le \exp{\left(-\frac{2L}{\log{n}}\right)} \\
&= \exp{\left(-\frac{2\sqrt{n}}{k_1(\log{n})^2}\right)}.
\end{align} 
This completes the proof of Lemma~\ref{lemma:last}, as well as the proof of covertness of our codes, as in Property 2) in Theorem 1. \qed

\section{Conclusion and Future Directions}
\label{sec:conclusion}
In this paper we put forth the first computationally efficient codes for simultaneously covert and reliable communication over BSCs. Our coding scheme, which is proved to be both covert and reliable, achieves the best known throughput and ensures that the computational complexity for both encoding and decoding is polynomial in the number of transmitted message bits. Though both the exponent of the complexity and the blocklength for this performance to kick in are relatively high, it is still a proof-of-concept first attempt to show the existence of such computationally efficient codes for covert communication. In fact, getting the truly practical codes where the gap to covert throughput scales as the inverse of polynomial is still worthy exploring. 

Having designed the coding scheme for BSCs, one would expect to generalize the concatenated-style codes to other DMCs and AWGN channels. Though a detailed analysis is needed, it is plausible that a non-trivial combination of the code proposed in~\cite{7407378} and our code results in corresponding concatenated-style codes for DMCs and AWGN channels. Besides such generalizations of our code constructions to arbitrary DMCs and AWGN channels, another direction is to study different metrics for covertness, \emph{a la}~\cite{tahmasbi2017second}.

%%%%%%%%%%%%%%%%%%%%%%%%%%%%%%%%%%%%%%%%%%%%%%%%%%%%%%%%%%%%%%%%%%%%%%%%%%%%%%%%%%%%%%%%%%%%%%%%%%%%%%%%%%%%%%%%%%%%%%%%%%%%%%%%%%%%%%%%%%%%%%%%%%%%%%%%%%%%

\vskip 0.5cm

\section*{Acknowledgement}
The authors would like to thank Andrej Bogdanov, Xuan Guang, Tongxin Li and Pak Hou Che for their valuable suggestions.

\vskip 0.5cm

\appendices
\section{} \label{app:chernoff}
The Chernoff bound~\cite{chernoff1952measure} is widely used in this work. Since there are many different versions of the Chernoff bound in the literature, and each version has a slightly different formulation, in this Appendix we explicitly state the version of the Chernoff bound~\cite{chernoff1952measure} used throughout this work.

Suppose $Q_1, \ldots, Q_n$ are independent (but not necessarily identically distributed) random variables taking values in $\left\{0,1\right\}$. We define $Q$ as $Q_1 + \cdots + Q_n$, and denote the expectation of $Q$ by $\mathbb{E}(Q)$. Then for any $0 < \epsilon < 1$, 
\begin{align}
&\mathbb{P}(Q \ge (1+\epsilon)\mathbb{E}(Q)) \le \exp\left(-\frac{\epsilon^2 \mathbb{E}(Q)}{3}\right), \\
&\mathbb{P}(Q \le (1-\epsilon)\mathbb{E}(Q)) \le \exp\left(-\frac{\epsilon^2 \mathbb{E}(Q)}{2}\right) \le \exp\left(-\frac{\epsilon^2 \mathbb{E}(Q)}{3}\right).
\end{align}

%%%%%%%%%%%%%%%%%%%%%%%%%%%%%%%%%%%%%%%%%%%%%%%%%%%%%%%%%%%%%%%%%%%%%%%%%%%%%%%%%%%%%%%%%%%%%%%%%%%%%%%%%%%%

\section{} \label{app:max}
We aim to calculate the value of $\I(\x^{(i)}; \yw^{(i)}) + \mathbb{D}(\x^{(i)} \parallel\rho)$ when $f^z_{1} \in (\rho*\pw)(1\pm \Delta^z_{1})$, $f^{xz}_{10} \in \rho \pw(1\pm \Delta^{xz}_{10})$, and $f^{xz}_{11} \in \rho(1-\pw)(1\pm \Delta^{xz}_{11})$. By definition, The first term $\I(\x^{(i)}; \yw^{(i)})$ can be expressed as 
\begin{align}
\I(\x^{(i)}; \yw^{(i)}) &= \sum_{(j,j^\prime) \in \left\{0,1\right\} \times \left\{0,1\right\}}f^{xz}_{jj^\prime}\log{\frac{f^{xz}_{jj^\prime}}{f^x_{j}\cdot f^z_{j^\prime}}} \\
&=f^{xz}_{00}\log{\left(\frac{f^{xz}_{00}}{(1-f^x_{1})(1-f^z_{1})}\right)} + f^{xz}_{01}\log{\left(\frac{f^{xz}_{01}}{(1-f^x_{1})f^z_{1}}\right)} + f^{xz}_{10}\log{\left(\frac{f^{xz}_{10}}{f^x_{1}(1-f^z_{1})}\right)}+
f^{xz}_{11}\log{\left(\frac{f^{xz}_{11}}{f^x_{1}f^z_{1}}\right)} \\
&=(1-f^z_{1}-f^{xz}_{10})\log{\left(\frac{1-f^z_{1}-f^{xz}_{10}}{(1-f^{xz}_{10}-f^{xz}_{11})(1-f^z_{1})}\right)} + (f^z_{1}-f^{xz}_{11})\log{\left(\frac{f^z_{1}-f^{xz}_{11}}{(1-f^{xz}_{10}-f^{xz}_{11})f^z_{1}}\right)} \\
&+ f^{xz}_{10}\log{\left(\frac{f^{xz}_{10}}{(f^{xz}_{10}+f^{xz}_{11})(1-f^z_{1})}\right)}+
f^{xz}_{11}\log{\left(\frac{f^{xz}_{11}}{(f^{xz}_{10}+f^{xz}_{11})f^z_{1}}\right)}.
\end{align} 
It then follows that the partial derivative of $\I(\x^{(i)}; \yw^{(i)})$ with respect to $f^z_{1}$ equals
\begin{align}
\frac{\partial \I(\x^{(i)}; \yw^{(i)})}{\partial f^z_{1}} &= -\log{\left(\frac{1-f^z_{1}-f^{xz}_{10}}{(1-f^{xz}_{10}-f^{xz}_{11})(1-f^z_{1})}\right)} -\frac{f^{xz}_{10}}{1-f^z_{1}} + \log{\left(\frac{f^z_{1}-f^{xz}_{11}}{(1-f^{xz}_{10}-f^{xz}_{11})f^z_{1}}\right)}+\frac{f^{xz}_{11}}{f^z_{1}}+\frac{f^{xz}_{10}}{1-f^z_{1}}-\frac{f^{xz}_{11}}{f^z_{1}} \\
&= \log{\left(\frac{f^z_{1}-f^{xz}_{11}}{(1-f^{xz}_{10}-f^{xz}_{11})f^z_{1}} \cdot \frac{(1-f^{xz}_{10}-f^{xz}_{11})(1-f^z_{1})}{1-f^z_{1}-f^{xz}_{10}} \right)}  \\
&= \log{\left(\frac{(f^z_{1}-f^{xz}_{11})(1-f^z_{1})}{f^z_{1}(1-f^z_{1}-f^{xz}_{10})}\right)}. \label{eq:partial_f*1}
\end{align}
Similarly, the partial derivative of $\I(\x^{(i)}; \yw^{(i)})$ with respect to $f^{xz}_{10}$ equals
\begin{align}
\frac{\partial \I(\x^{(i)}; \yw^{(i)})}{\partial f^{xz}_{10}} = \log{\left(\frac{f^{xz}_{10}(1-f^{xz}_{10}-f^{xz}_{11})}{(f^{xz}_{10}+f^{xz}_{11})(1- f^z_{1}-f^{xz}_{10})}\right)}, \label{eq:partial_f10}
\end{align} 
and the partial derivative of $\I(\x^{(i)}; \yw^{(i)})$ with respect to $f^{xz}_{11}$ equals
\begin{align}
\frac{\partial \I(\x^{(i)}; \yw^{(i)})}{\partial f^{xz}_{11}} = \log{\left(\frac{f^{xz}_{11}(1-f^{xz}_{10}-f^{xz}_{11})}{(f^{xz}_{10}+f^{xz}_{11})(f^z_{1}-f^{xz}_{11})}\right)}. \label{eq:partial_f11}
\end{align} 
The second term $\mathbb{D}(\x^{(i)} \parallel \rho)$ can be expressed as
\begin{align}
\mathbb{D}(\x^{(i)} \parallel \rho) &= f^x_{0}\log{\frac{f^x_{0}}{1-\rho}}+f^x_{1}\log{\frac{f^x_{1}}{\rho}} \\
&= (1-f^{xz}_{10}-f^{xz}_{11})\log{\frac{(1-f^{xz}_{10}-f^{xz}_{11})}{1-\rho}}+(f^{xz}_{10}+f^{xz}_{11})\log{\frac{(f^{xz}_{10}+f^{xz}_{11})}{\rho}}.
\end{align}
The partial derivative of $\mathbb{D}(\x^{(i)} \parallel \rho)$ with respect to $f^{xz}_{10}$ equals
\begin{align}
\frac{\partial \mathbb{D}(\x^{(i)} \parallel \rho)}{\partial f^{xz}_{10}} &= \log{\left(\frac{(1-\rho)(f^{xz}_{10}+f^{xz}_{11})}{\rho(1-f^{xz}_{10}-f^{xz}_{11})}\right)},
\end{align}
and the partial derivative of $\mathbb{D}(\x^{(i)} \parallel \rho)$ with respect to $f^{xz}_{11}$ also equals
\begin{align}
\frac{\partial \mathbb{D}(\x^{(i)} \parallel \rho)}{\partial f^{xz}_{11}} &= \log{\left(\frac{(1-\rho)(f^{xz}_{10}+f^{xz}_{11})}{\rho(1-f^{xz}_{10}-f^{xz}_{11})}\right)}.
\end{align}
Therefore, the partial derivative of $\I(\x^{(i)}; \yw^{(i)})+\mathbb{D}(\x^{(i)} \parallel \rho)$ with respect to $f^{xz}_{10}$ is given as
\begin{align}
\frac{\partial [\I(\x^{(i)}; \yw^{(i)})+\mathbb{D}(\x^{(i)} \parallel \rho)]}{\partial f^{xz}_{10}} &= \log{\left(\frac{f^{xz}_{10}(1-f^{xz}_{10}-f^{xz}_{11})}{(f^{xz}_{10}+f^{xz}_{11})(1- f^z_{1}-f^{xz}_{10})}\right)} + \log{\left(\frac{(1-\rho)(f^{xz}_{10}+f^{xz}_{11})}{\rho(1-f^{xz}_{10}-f^{xz}_{11})}\right)}  \\
&= \log{\left(\frac{f^{xz}_{10}(1-\rho)}{\rho (1-f^z_{1}-f^{xz}_{10})}\right)}.  \label{eq:partial1}
\end{align}
Note that the value of term~\eqref{eq:partial1} is negative when $f^{xz}_{10} < \rho(1-\pw)-(1-2\pw)\rho^2$ and $f^{xz}_{10} \in \rho \pw(1 \pm \Delta^{xz}_{10})$. The analysis of $\I(\x^{(i)}; \yw^{(i)})+\mathbb{D}(\x^{(i)}\parallel \rho)$ with respect to $f^{xz}_{10}$ is as follows:
\begin{itemize}
\item If $\rho\pw(1 + \Delta^{xz}_{10}) \le \rho(1-\pw)-(1-2\pw)\rho^2$, the value of $\I(\x^{(i)}; \yw^{(i)})+\mathbb{D}(\x^{(i)}\parallel \rho)$ decreases monotonically as $f^{xz}_{10}$ increases, and hence $\I(\x^{(i)}; \yw^{(i)})+\mathbb{D}(\x^{(i)}\parallel \rho)$ achieves maximum when $f^{xz}_{10} = \rho \pw(1 - \Delta^{xz}_{10})$.
\item If $\rho \pw(1 + \Delta^{xz}_{10}) > \rho(1-\pw)-(1-2\pw)\rho^2$, the value of $\I(\x^{(i)}; \yw^{(i)})+\mathbb{D}(\x^{(i)}\parallel \rho)$ first decreases and then increases as $f^{xz}_{10}$ increases, and hence $\I(\x^{(i)}; \yw^{(i)})+\mathbb{D}(\x^{(i)}\parallel \rho)$ achieves maximum when $f^{xz}_{10} = \rho \pw(1 - \Delta^{xz}_{10})$ or $f^{xz}_{10} = \rho \pw(1 + \Delta^{xz}_{10})$.
\end{itemize}
Similarly, the partial derivative of $\I(\x^{(i)}; \yw^{(i)})+\mathbb{D}(\x^{(i)} \parallel \rho)$ with respect to $f^{xz}_{11}$ is given as
\begin{align}
\frac{\partial [\I(\x^{(i)}; \yw^{(i)})+\mathbb{D}(\x^{(i)} \parallel \rho)]}{\partial f^{xz}_{11}} &= \log{\left(\frac{f^{xz}_{11}(1-f^{xz}_{10}-f^{xz}_{11})}{(f^{xz}_{10}+f^{xz}_{11})(f^z_{1}-f^{xz}_{11})}\right)} + \log{\left(\frac{(1-\rho)(f^{xz}_{10}+f^{xz}_{11})}{\rho(1-f^{xz}_{10}-f^{xz}_{11})}\right)}  \\
&= \log{\left(\frac{f^{xz}_{11}(1-\rho)}{\rho (f^z_{1}-f^{xz}_{11})}\right)}.  \label{eq:partial2}
\end{align}
Note that the value of term~\eqref{eq:partial2} is positive when $f^{xz}_{11} > \rho \pw + (1-2\pw)\rho^2$ and $f^{xz}_{11} \in \rho (1-\pw)(1 \pm \Delta^{xz}_{11})$. The analysis of $\I(\x^{(i)}; \yw^{(i)})+\mathbb{D}(\x^{(i)}\parallel \rho)$ with respect to $f^{xz}_{11}$ is as follows:
\begin{itemize}
\item If $\rho (1-\pw)(1 - \Delta^{xz}_{11}) \ge \rho \pw + (1-2\pw)\rho^2$, the value of $\I(\x^{(i)}; \yw^{(i)})+\mathbb{D}(\x^{(i)}\parallel \rho)$ increases monotonically as $f^{xz}_{11}$ increases, and hence $\I(\x^{(i)}; \yw^{(i)})+\mathbb{D}(\x^{(i)}\parallel \rho)$ achieves maximum when $f^{xz}_{11} = \rho (1-\pw)(1 + \Delta^{xz}_{11})$.
\item If $\rho (1-\pw)(1 - \Delta^{xz}_{11}) < \rho \pw + (1-2\pw)\rho^2$, the value of $\I(\x^{(i)}; \yw^{(i)})+\mathbb{D}(\x^{(i)}\parallel \rho)$ first decreases and then increases as $f^{xz}_{11}$ increases, and hence $\I(\x^{(i)}; \yw^{(i)})+\mathbb{D}(\x^{(i)}\parallel \rho)$ achieves maximum when $f^{xz}_{11} = \rho (1-\pw)(1 + \Delta^{xz}_{11})$ or $f^{xz}_{11} = \rho (1-\pw)(1 - \Delta^{xz}_{11})$.
\end{itemize}
Moreover, it is worthwhile noting that the value of $f^z_{1}$ has negligible impact on the value of $\I(\x^{(i)}; \yw^{(i)})+\mathbb{D}(\x^{(i)} \parallel \rho)$, since
\begin{align}
(f^z_{1}-\rho *\pw) \frac{\partial \I(\x^{(i)}; \yw^{(i)})}{\partial f^z_{1}}\bigg|_{\rho*\pw,\rho \pw, \rho (1-\pw)} = \mathcal{O}(n^{-1}),
\end{align}
while the value of $\I(\x^{(i)}; \yw^{(i)})+\mathbb{D}(\x^{(i)} \parallel \rho)$ scales as $\mathcal{O}(n^{-1/2})$. For simplicity, we set $f^z_{1}$ to be $\rho*\pw$, since $f^z_{1} \in (\rho *\pw)(1\pm \Delta^z_{1})$. 

Therefore, the maximal value of $\I(\x^{(i)}; \yw^{(i)})+\mathbb{D}(\x^{(i)}\parallel \rho)$ is attained at one of the four ``corner'' points, \emph{i.e.}, $(f^z_{1}, f^{xz}_{10}, f^{xz}_{11}) = (\rho*\pw, \rho\pw(1 \pm \Delta^{xz}_{10}),\rho(1-\pw)(1 \pm \Delta^{xz}_{11}))$.

%%%%%%%%%%%%%%%%%%%%%%%%%%%%%%%%%%%%%%%%%%%%%%%%%%%%%%%%%%%%%%%%%%%%%%%%%%%%%%%%%%%%%%%%%%%%%%%%%%%%%%%%%%%%

\vspace{8pt}
\section{} \label{app:max2} 
In Appendix~\ref{app:max} we have shown that the maximal value of $\I(\x^{(i)}; \yw^{(i)})+\mathbb{D}(\x^{(i)} \parallel \rho)$ is attained at one of the four ``corner'' points, \emph{i.e.}, $f^z_{1}=\rho *\pw, f^{xz}_{10}=\rho \pw(1 \pm \Delta^{xz}_{10}), f^{xz}_{11}=\rho (1-\pw)(1 \pm \Delta^{xz}_{11})$. We now prove that there exists an explicitly computable constant $c_1$ such that for sufficiently large $n$,
\begin{align}
 -k_1\sqrt{n}\left[\I(\x^{(i)}; \yw^{(i)}) + \mathbb{D}(\x^{(i)} \parallel\rho)\right] \ge k_1\cdot \left( \max_{j\in \{1,2,3,4\}} \{g_i(\pw,\epsilon_d,\Delta^{xz}_{10},\Delta^{xz}_{11})\}+c_1n^{-1/2}\right).
 \end{align}
We first note that
\begin{align}
&-k_1\sqrt{\n} \left[\ixyw+\mathbb{D}\left(\x^{(i)} \parallel \rho \right) \right]  \\
&= k_1\sqrt{n}\left[f_{00}^{xz}\log\frac{f_{0}^x f_{0}^z}{f_{00}^{xz}}+f_{01}^{xz}\log\frac{f_{0}^x f_{1}^z}{f_{01}^{xz}}+f_{10}^{xz}\log\frac{f_{1}^x f_{0}^z}{f_{10}^{xz}}+f_{11}^{xz}\log\frac{f_{1}^x f_{1}^z}{f_{11}^{xz}}-f_{0}^x\log\frac{f_{0}^x}{1-\rho}-f_{1}^x\log\frac{f_{1}^x}{\rho}\right] \label{eq:ver1}\\
&= k_1\sqrt{n}\left[f_{00}^{xz}\log\frac{f_{0}^z}{f_{00}^{xz}}+f_{01}^{xz}\log\frac{f_{1}^z}{f_{01}^{xz}}+f_{10}^{xz}\log\frac{f_{0}^z}{f_{10}^{xz}}+f_{11}^{xz}\log\frac{f_{1}^z}{f_{11}^{xz}}+f_{0}^x\log f_{0}^x+f_{1}^x\log f_{1}^x-f_{0}^x\log\frac{f_{0}^x}{1-\rho}-f_{1}^x\log\frac{f_{1}^x}{\rho}\right]  \\
&=k_1\sqrt{n}\left[(f_{01}^{xz}+f_{11}^{xz})H\left(\frac{f_{11}^{xz}}{f_{01}^{xz}+f_{11}^{xz}} \right)+(f_{00}^{xz}+f_{10}^{xz})H\left(\frac{f_{10}^{xz}}{f_{00}^{xz}+f_{10}^{xz}} \right)+(f^{xz}_{10}+f^{xz}_{11})\log{\rho}+(f^{xz}_{00}+f^{xz}_{01})\log{(1-\rho)}\right]. \label{eq:night1}
\end{align}
We now calculate the value of term~\eqref{eq:night1} at one ``corner'' point, (\emph{i.e.}, $f^z_{1} = \rho *\pw, \ f^{xz}_{10} = \rho \pw (1 - \Delta^{xz}_{10})$, \ $f^{xz}_{11} = \rho (1-\pw) (1 + \Delta^{xz}_{11}), \ f^{xz}_{01}=f^z_{1}-\rho (1-\pw)(1+\Delta^{xz}_{11})$, and $f^{xz}_{00}=1-f^z_{1}-\rho \pw(1-\Delta^{xz}_{10})$), which equals
\begin{footnotesize}
\begin{align}
&k_1\sqrt{n}\left[(f_{01}^{xz}+f_{11}^{xz})H\left(\frac{f_{11}^{xz}}{f_{01}^{xz}+f_{11}^{xz}} \right)+(f_{00}^{xz}+f_{10}^{xz})H\left(\frac{f_{10}^{xz}}{f_{00}^{xz}+f_{10}^{xz}} \right)+(f^{xz}_{10}+f^{xz}_{11})\log{\rho}+(f^{xz}_{00}+f^{xz}_{01})\log{(1-\rho)}\right]  \\
&=k_1\sqrt{n}\Bigg[f^z_{1}H\left(\frac{k_2(1-\pw)(1+\Delta^{xz}_{11})}{\sqrt{n}f^z_{1}}\right) + (1-f^z_{1})H\left(\frac{k_2\pw(1-\Delta^{xz}_{10})}{\sqrt{n}(1-f^z_{1})}\right) \notag \\
&\ \ \ \ \ \ \ + \frac{k_2\pw(1-\Delta^{xz}_{10})+k_2(1-\pw)(1+\Delta^{xz}_{11})}{\sqrt{n}}\log{\left(\frac{k_2}{\sqrt{n}}\right)} + \frac{\sqrt{n}-\left(k_2\pw(1-\Delta^{xz}_{10})+k_2(1-\pw)(1+\Delta^{xz}_{11})\right)}{\sqrt{n}}\log{\left(\frac{\sqrt{n}-k_2}{\sqrt{n}}\right)} \Bigg]  \\
&=k_1\sqrt{n}\Bigg[-f^z_{1}\frac{k_2(1-\pw)(1+\Delta^{xz}_{11})}{\sqrt{n}f^z_{1}}\log{\left(\frac{k_2(1-\pw)(1+\Delta^{xz}_{11})}{\sqrt{n}f^z_{1}}\right)} - f^z_{1}\frac{\sqrt{n}f^z_{1}-k_2(1-\pw)(1+\Delta^{xz}_{11})}{\sqrt{n}f^z_{1}}\log{\left(\frac{\sqrt{n}f^z_{1}-k_2(1-\pw)(1+\Delta^{xz}_{11})}{\sqrt{n}f^z_{1}}\right)} \notag \\
&\ \ \ \ \ \ \ \ \ \ \ \ \ \ \  -(1-f^z_{1})\frac{k_2\pw(1-\Delta^{xz}_{10})}{\sqrt{n}(1-f^z_{1})}\log{\left(\frac{k_2\pw(1-\Delta^{xz}_{10})}{\sqrt{n}(1-f^z_{1})}\right)} - (1-f^z_{1})\frac{\sqrt{n}(1-f^z_{1})-k_2\pw(1-\Delta^{xz}_{10})}{\sqrt{n}(1-f^z_{1})}\log{\left(\frac{\sqrt{n}(1-f^z_{1})-k_2\pw(1-\Delta^{xz}_{10})}{\sqrt{n}(1-f^z_{1})}\right)}  \notag \\
&\ \ \ \ \ \ \ \ \ \ \ \ \ \ \  +\frac{k_2\pw(1-\Delta^{xz}_{10})+k_2(1-\pw)(1+\Delta^{xz}_{11})}{\sqrt{n}}\log{\left(\frac{k_2}{\sqrt{n}}\right)} + \frac{\sqrt{n}-\left(k_2\pw(1-\Delta^{xz}_{10})+k_2(1-\pw)(1+\Delta^{xz}_{11})\right)}{\sqrt{n}}\log{\left(\frac{\sqrt{n}-k_2}{\sqrt{n}}\right)} \Bigg]  \\
&= k_1 \Bigg[ k_2(1-\pw)(1+\Delta^{xz}_{11})\log{\left(\frac{\sqrt{n}f^z_{1}-k_2(1-\pw)(1+\Delta^{xz}_{11})}{(\sqrt{n}-k_2)(1-\pw)(1+\Delta^{xz}_{11})}\right)} + k_2\pw(1-\Delta^{xz}_{10})\log{\left(\frac{\sqrt{n}(1-f^z_{1})-k_2\pw(1-\Delta^{xz}_{10})}{(\sqrt{n}-k_2)\pw (1-\Delta^{xz}_{10})}\right)} \notag \\
&\ \ \ \ \ \ \ \ \ \ \ \ \ \ \  +\sqrt{n}f^z_{1}\log{\left(\frac{\sqrt{n}f^z_{1}(1-f^z_{1})-f^z_{1}k_2\pw(1-\Delta^{xz}_{10})}{\sqrt{n}f^z_{1}(1-f^z_{1})-(1-f^z_{1})k_2(1-\pw)(1+\Delta^{xz}_{11})}\right)} + \sqrt{n}\log{\left(\frac{(\sqrt{n}-k_2)(1-f^z_{*1})}{\sqrt{n}(1-f^z_{1})-k_2\pw(1-\Delta^{xz}_{10})}\right)} \Bigg]  \\
&= k_1 \Bigg[ k_2(1-\pw)(1+\Delta^{xz}_{11})\log{\left(\frac{\pw}{(1-\pw)(1+\Delta^{xz}_{11})}\right)} + k_2\pw(1-\Delta^{xz}_{10})\log{\left(\frac{1-\pw}{\pw(1-\Delta^{xz}_{10})}\right)} \notag \\
& \ \ \ \ \ \ \ \ \ \ \ \ \ \ \ \ \ \ \ \ \ \ \ \ \ \ \ \ \ \ \ \ \ \ \ \ \ \ \ \ \ \ \ \ \ \ \ \ \ \ \ +k_2\left((1-\pw)(1+\Delta^{xz}_{11})-1\right)\log{e} + k_2\pw(1-\Delta^{xz}_{10})\log{e} + \mathcal{O}(n^{-1/2})\Bigg]. \label{eq:complex2}
\end{align}
\end{footnotesize} 
Note that the term in~\eqref{eq:complex2} equals $k_1 \cdot \left( g_1(\pw, \epsilon_d, \Delta^{xz}_{10}, \Delta^{xz}_{11}) + \mathcal{O}(n^{-1/2})\right)$,
where the auxiliary multivariable function $g_1(u, v, w, t)$, defined in Section~\ref{sec:result}, has the form 
\begin{align}
g_1(u,v,w,t) = k_2(u,v)\Bigg[u(1-w)\left(\log{\left(\frac{1-u}{u(1-w)}\right)}+\log{e}\right) + (1-u)(1+t)\left(\log{\left(\frac{u}{(1-u)(1+t)}\right)}+\log{e}\right)  - \log{e} \Bigg].
\end{align}
Similarly, we also calculate the values of $-k_1\sqrt{n}\left(\I(\x^{(i)}; \yw^{(i)}) + \mathbb{D}(\x^{(i)} \parallel\rho)\right)$ at the other three ``corner'' points, and it turns out that these values can respectively be characterized by $k_1\left(g_2(\pw, \epsilon_d, \Delta^{xz}_{10}, \Delta^{xz}_{11})+\mathcal{O}(n^{-1/2})\right),k_1\left(g_3(\pw, \epsilon_d, \Delta^{xz}_{10}, \Delta^{xz}_{11})+\mathcal{O}(n^{-1/2})\right)$ and $k_1\left(g_4(\pw, \epsilon_d, \Delta^{xz}_{10}, \Delta^{xz}_{11})+\mathcal{O}(n^{-1/2})\right)$. Since the maximal value of $\I(\x^{(i)}; \yw^{(i)})+\mathbb{D}(\x^{(i)} \parallel \rho)$ is attained at the ``corner'' points, we conclude that there exists an explicitly computable constant $c_1$ such that for sufficiently large $n$, 
\begin{align}
 -k_1\sqrt{n}\left[\I(\x^{(i)}; \yw^{(i)}) + \mathbb{D}(\x^{(i)} \parallel\rho)\right] \ge k_1\cdot \left( \max_{j\in \{1,2,3,4\}} \{g_i(\pw,\epsilon_d,\Delta^{xz}_{10},\Delta^{xz}_{11})\}+c_1n^{-1/2}\right).
 \end{align}

%%%%%%%%%%%%%%%%%%%%%%%%%%%%%%%%%%%%%%%%%%%%%%%%%%%%%%%%%%%%%%%%%%%%%%%%%%%%%%%%%%%%%%%%%%%%%%%%%%%%%%%%%%%%

\section{ } \label{app:series}
Recall that the auxiliary function $h(i)$, first defined in~\eqref{eq:hi}, has the form
\begin{align}
h(i) = \binom{k_1\sqrt{\n}\log{\n}}{i}\left(\frac{k_2\pw}{\sqrt{\n}}\right)^{i}\left(1-\frac{k_2\pw}{\sqrt{\n}}\right)^{k_1\sqrt{\n}\log{\n}-i}.
\end{align} 
Let $i_0 = k_1k_2\pw (\log{\n})\left(1+\Delta_{10}^{xz}\right)$. Then, via Stirling's approximation~\cite[pp. 50-53]{feller1968introduction}, we can bound $h(i_0)$ from above as 
\begin{align}
h(i_0) &= \binom{k_1\sqrt{\n}\log{\n}}{i_0}\left(\frac{k_2\pw}{\sqrt{\n}}\right)^{i_0}\left(1-\frac{k_2\pw}{\sqrt{\n}}\right)^{k_1\sqrt{\n}\log{\n}-i_0}  \\
& \le \frac{1}{\sqrt{2\pi i_0}} 2^{i_0\log{\left(\frac{ek_1\sqrt{\n}\log{\n}}{i_0}\right)}}2^{i_0\log{\left(\frac{k_2\pw}{\sqrt{\n}}\right)}}2^{\left(k_1\sqrt{\n}\log{\n}-i_0\right)\log{\left(\frac{\sqrt{\n}-k_2\pw}{\sqrt{\n}}\right)}}  \\
& = \frac{1}{\sqrt{2\pi i_0}} 2^{i_0\log{\left(\frac{e\sqrt{\n}}{k_2\pw(1+\Delta_{10}^{xz})}\right)}}2^{i_0\log{\left(\frac{k_2\pw}{\sqrt{\n}}\right)}}2^{\left(k_1\sqrt{\n}\log{\n}-i_0\right)\log{\left(\frac{\sqrt{\n}-k_2\pw}{\sqrt{\n}}\right)}}  \\
& = \frac{1}{\sqrt{2\pi i_0}} \n^{k_1k_2\pw \left(1+\Delta_{10}^{xz}\right)\log{\left(\frac{e}{1+\Delta_{10}^{xz}}\right)}+\left[\sqrt{\n}-k_2\pw\left(1+\Delta_{10}^{xz}\right)\right]k_1\log{\left(\frac{\sqrt{\n}-k_2\pw}{\sqrt{\n}}\right)}} \\
& = \frac{1}{\sqrt{2\pi k_1k_2\pw (\log{\n})\left(1+\Delta_{10}^{xz}\right)}} \n^{k_1k_2\pw \left(1+\Delta_{10}^{xz}\right)\log{\left(\frac{e}{1+\Delta_{10}^{xz}}\right)}-\frac{k_1k_2\pw \sqrt{\n}\log{e}}{\sqrt{\n}-k_2\pw}+\mathcal{O}\left(\n^{-1/2}\right)}, \label{eq:thai} \\
& =\frac{1}{\sqrt{2\pi k_1k_2\pw (\log{\n})\left(1+\Delta_{10}^{xz}\right)}} \n^{k_1k_2\pw \left(\left(1+\Delta_{10}^{xz}\right)\log{\left(\frac{e}{1+\Delta_{10}^{xz}}\right)}-\log{e}\right) +\mathcal{O}(n^{-1/2})}, \label{eq:szlib}
\end{align}
where equality~\eqref{eq:thai} follows from $\log{\left(\frac{\sqrt{\n}-k_2\pw}{\sqrt{\n}}\right)} = -\frac{k_2\pw \log{e}}{\sqrt{\n}-k_2\pw}+\mathcal{O}(n^{-1})$, by applying Taylor's series expansion. For sufficiently large $n$, the term~\eqref{eq:szlib} is bounded from above by
\begin{align}
\frac{1}{\sqrt{2\pi k_1k_2\pw (\log{\n})\left(1+\Delta_{10}^{xz}\right)}} \n^{k_1k_2\pw \left(\left(1+\Delta_{10}^{xz}\right)\log{\left(\frac{e}{1+\Delta_{10}^{xz}}\right)}-\log{e}\right) + \delta/2}, \label{eq:thai2}
\end{align}
where $\delta = 0.01$ is the slackness parameter first defined in Section~\ref{sec:result}.
Note that the ratio between two successive terms is 
\begin{align}
\frac{h(i+1)}{h(i)} = \frac{k_1\sqrt{\n}\log{\n}-i}{i+1}\cdot \frac{k_2\pw}{\sqrt{\n}-k_2\pw}.
\end{align}
Hence for $i \ge i_0 = k_1k_2\pw (\log{\n})\left(1+\Delta_{10}^{xz}\right)$, we have 
\begin{align}
\frac{h(i+1)}{h(i)} \le \frac{h(i_0+1)}{h(i_0)} = \frac{k_1\sqrt{\n}\log{\n}-k_1k_2\pw (\log{\n})\left(1+\Delta_{10}^{xz}\right)}{k_1k_2\pw (\log{\n})\left(1+\Delta_{10}^{xz}\right)+1}\cdot \frac{k_2\pw}{\sqrt{\n}-k_2\pw} = \frac{1}{1+\Delta_{10}^{xz}}\left(1+\mathcal{O}(n^{-1/2})\right),
\end{align} 
and there exists an explicitly computable constant $c_3$ such that for sufficiently large $n$,
\begin{align}
\frac{h(i+1)}{h(i)} \le \frac{1}{1+\Delta_{10}^{xz}}\left(1+c_3n^{-1/2}\right).
\end{align}
This implies that the tail of the series $\left\{h(i)\right\}$ can be bounded from above by a geometric series as follows:
\begin{align}
\sum_{i = i_0}^{k_1\sqrt{\n}\log{\n}}h(i) &\le h(i_0) \left[1+\left(\frac{1+c_3n^{-1/2}}{1+\Delta_{10}^{xz}}\right)+\cdots+\left(\frac{1+c_3n^{-1/2}}{1+\Delta_{10}^{xz}}\right)^{k_1\sqrt{\n}\log{\n}-i_0} \right]  \\
&\le h(i_0) \left[\sum_{j=0}^{\infty} \left(\frac{1+c_3n^{-1/2}}{1+\Delta_{10}^{xz}}\right)^j \right]  \\
&= h(i_0)\left(\frac{1+\Delta_{10}^{xz}}{\Delta_{10}^{xz}-c_3n^{-1/2}}\right).
\end{align} 
Substituting in the bound on $h(i_0)$ from Equation~\eqref{eq:thai2} gives us
\begin{align}
\sum_{i = i_0}^{k_1\sqrt{\n}\log{\n}}h(i) &\le  \sqrt{\frac{1+\Delta_{10}^{xz}}{2\pi k_1k_2\pw \left(\Delta_{10}^{xz}-c_3n^{-1/2}\right)^2(\log{\n})}}\cdot \n^{k_1k_2\pw \left(\left(1+\Delta_{10}^{xz}\right)\log{\left(\frac{e}{1+\Delta_{10}^{xz}}\right)}-\log{e}\right)+\delta/2} \\
&\le n^{-k_1k_2\pw f(\Delta^{xz}_{10})+\delta/2},
\end{align}
hence proving the term~\eqref{eq:exp1} in Section~\ref{sec:deniability-A}. Once the term~\eqref{eq:exp1} is proved, one can also show that the term~\eqref{eq:exp2} is bounded from above by $n^{-k_1k_2\pw f(\Delta^{xz}_{10})+\delta/2}$. Let $i_0^{\prime} = k_1k_2\pw(\log n)(1-\Delta_{10}^{xz})$, by a similar argument that we omit here, for sufficiently large $n$, 
\begin{align}
h(i_0^{\prime}) \le \frac{1}{\sqrt{2\pi k_1k_2\pw (\log{\n})\left(1-\Delta_{10}^{xz}\right)}} \n^{k_1k_2\pw \left(\left(1-\Delta_{10}^{xz}\right)\log{\left(\frac{e}{1-\Delta_{10}^{xz}}\right)}-\log{e}\right)+\delta/2}.
\end{align}
The summation of the first $i_0^{\prime}$ terms can further be bounded from above as
\begin{align}
\sum_{i=0}^{i_0^{\prime}}h(i) &\le i_0^{\prime}\cdot \frac{1}{\sqrt{2\pi k_1k_2\pw (\log{\n})\left(1-\Delta_{10}^{xz}\right)}} \n^{k_1k_2\pw \left(\left(1-\Delta_{10}^{xz}\right)\log{\left(\frac{e}{1-\Delta_{10}^{xz}}\right)}-\log{e}\right)+\delta/2}  \\
&= \sqrt{\frac{k_1k_2\pw (\log n)(1-\Delta_{10}^{xz})}{2 \pi}}\n^{k_1k_2\pw \left(\left(1-\Delta_{10}^{xz}\right)\log{\left(\frac{e}{1-\Delta_{10}^{xz}}\right)}-\log{e}\right)+\delta/2}  \\
& < n^{k_1k_2\pw \left(\left(1+\Delta_{10}^{xz}\right)\log{\left(\frac{e}{1+\Delta_{10}^{xz}}\right)}-\log{e}\right)+\delta/2}  \label{eq:thai3} \\
&= n^{-k_1k_2\pw f(\Delta_{10}^{xz})+\delta/2}.  
\end{align}
Inequality~\eqref{eq:thai3} follows since for $0 < \Delta_{10}^{xz} < 1$,
\begin{align}
\left(1-\Delta_{10}^{xz}\right)\log{\left(\frac{e}{1-\Delta_{10}^{xz}}\right)} < \left(1+\Delta_{10}^{xz}\right)\log{\left(\frac{e}{1+\Delta_{10}^{xz}}\right)},
\end{align}
and hence for sufficiently large $n$,
\begin{align}
\sqrt{\frac{k_1k_2\pw (\log n)(1-\Delta_{10}^{xz})}{2 \pi}}\n^{k_1k_2\pw \left(\left(1-\Delta_{10}^{xz}\right)\log{\left(\frac{e}{1-\Delta_{10}^{xz}}\right)}-\log{e}\right)+\delta/2} < n^{k_1k_2\pw \left(\left(1+\Delta_{10}^{xz}\right)\log{\left(\frac{e}{1+\Delta_{10}^{xz}}\right)}-\log{e}\right)+\delta/2}.
\end{align}
Similarly, one can also prove that for sufficiently large $n$, the terms~\eqref{eq:exp3} and~\eqref{eq:exp4} in Section~\ref{sec:deniability-A} respectively satisfy
\begin{align}
\sum_{i = k_1k_2(1-\pw)(\log{n})(1+\Delta^{xz}_{11})}^{k_1\sqrt{\n}\log{\n}}h(i) &\le  n^{-k_1k_2(1-\pw) f(\Delta^{xz}_{11})+\delta/2}, \\
\sum_{i=0}^{k_1k_2(1-\pw)(\log{n})(1-\Delta^{xz}_{11})} h(i) &\le n^{-k_1k_2(1-\pw) f(\Delta^{xz}_{11})+\delta/2}.
\end{align}

%%%%%%%%%%%%%%%%%%%%%%%%%%%%%%%%%%%%%%%%%%%%%%%%%%%%%%%%%%%%%%%%%%%%%%%%%%%%%%%%%%%%%%%%%%%%%%%%%%%%%%%%%%%%

\section{} \label{app:matrix}
Suppose the generator matrix $G_{\LL \times L}$ of a general Reed-Solomon code has the form (recall that $\LL = \lambda L$)
\begin{equation}
\begin{bmatrix}
    1 & 1 & 1 & \dots  & 1 \\
    \mu_1 & \mu_2 & \mu_3 & \dots  & \mu_L \\
    \vdots & \vdots & \vdots & \ddots & \vdots \\
    \mu_1^{\LL-1} & \mu_2^{\LL-1} & \mu_3^{\LL-1} & \dots  & \mu_L^{\LL-1}
\end{bmatrix},
\end{equation}
where $\mu_1, \mu_2, \ldots, \mu_L \in \mathbb{F}$ are all distinct. The systematic inner-message vector $\smv = [w^{(1)}, w^{(2)}, \ldots, w^{(\LL)}]$ of the Reed-Solomon code is uniformly distributed over $\mathbb{F}^{\LL}$, and whole inner-message vector $\mv = \smv\cdot G_{\LL \times L} = [w^{(1)}, w^{(1)}, \ldots, w^{(L)}] \in \mathbb{F}^{L}$. The code is denoted by $C_{RS} = \left\{\mv: \mv = \smv\cdot G_{\LL \times L}, \ \forall \smv \in \mathbb{F}^{\LL}\right\}$. 

The generator matrix $G^\prime_{\LL \times L}$ of a systematic Reed-Solomon code can be obtained by performing Gaussian eliminations on $G_{\LL \times L}$, \emph{i.e.}, $G^\prime_{\LL \times L} = A^{-1}\cdot G_{\LL \times L} = 
\begin{bmatrix}
  \  I_{\LL \times \LL} \  \big| \ P \ \
\end{bmatrix}$, where $A^{-1}$ is an invertible matrix and $I_{\LL \times \LL}$ is an identity matrix. The systematic Reed-Solomon code with $G^\prime_{\LL \times L}$ is denoted by 
\begin{align}
C_{SRS}&=\left\{\mv': \mv' = \smv\cdot G^\prime_{\LL \times L}, \ \forall \smv \in \mathbb{F}^{\LL}\right\}  \notag \\
&=\left\{\mv': \mv' = \smv\cdot A^{-1}G_{\LL \times L}, \ \forall \smv \in \mathbb{F}^{\LL}\right\} \notag \\
&=\left\{\mv': \mv' = \smv\cdot G_{\LL \times L}, \ \forall \smv \in \mathbb{F}^{\LL}\right\} \label{eq:do}
\end{align}
where Equation~\eqref{eq:do} holds since the linear mapping $A^{-1}$ is bijective. Note that the systematic code $C_{SRS}$ with generator matrix $G^\prime_{\LL \times L}$ is same as $C_{RS}$, hence in the following it suffices to study $C_{RS}$ and its corresponding generator matrix $G_{\LL \times L}$. 

Let $\hat{G}_{\LL \times l_2}$ be a $\LL \times l_2$ matrix consisting of the last $l_2$ columns of $G_{\LL \times L}$, with the form
\begin{equation}\hat{G}_{\LL \times l_2} \triangleq
\begin{bmatrix}
1 & 1 & \dots  & 1 \\
\mu_{\LL+1} & \mu_{\LL+2}  & \dots  & \mu_{L} \\
\vdots & \vdots & \ddots & \vdots \\
\mu_{\LL+1}^{\LL-1} & \mu_{\LL+2}^{\LL-1} &  \dots  & \mu_{L}^{\LL-1}
\end{bmatrix}. \label{eq:hat}
\end{equation}
For any specific parity inner-message vector $\pmv = [w^{(\LL+1)}, \ldots, w^{(L)}]$, the systematic inner-message vectors $\smv$ that could cause it satisfies 
\begin{align}
\smv \cdot \hat{G}_{\LL \times l_2} = \pmv,
\end{align}
hence the set $\mathcal{S}(\pmv)$, defined in Section~\ref{sec:reliability}, can be expressed as $\{\smv: \smv \cdot \hat{G}_{\LL \times l_2} = \pmv \}$. By noting the null space of the Vandermonde matrix $\hat{G}_{\LL \times l_2}$ is $(\LL-l_2)$-dimensional, we have $\big|\mathcal{S}(\pmv)\big| = |\mathbb{F}|^{\LL-l_2}$.

\section{} \label{app:mds}
As noted in Appendix~\ref{app:matrix}, the systematic RS code $C_{SRS}$ with generator matrix $G^{\prime}_{\LL \times L}$ is the same as the RS code $C_{RS}$ with generator matrix $C_{RS}$, hence we stick to $C_{RS}$ in the following analysis. For a fixed parity inner-message vector $\pmv$, we calculate how many pairs of $(\smv,\smv')$ satisfying $\smv \in \mathcal{S}(\pmv)$, $\smv' \in \mathcal{S}(\pmv)$ have distance $t$ ($t \ge l_2+1$ since the minimum distance of $C_{RS}$ is $l_2+1$). 

Let $\underline{\delta}= [\delta^{(1)},\delta^{(2)}, \ldots, \delta^{(t)}, 0, \ldots, 0]$ be a length-$\LL$ vector of weight $t$, where $\delta^{(i)} \ne 0$ $(\forall i \in \{1,2,\ldots, t\})$.
We first fix a $\smv = [w^{(1)}, w^{(2)}, \ldots, w^{(l_1)}] \in \mathcal{S}(\pmv)$, and consider the number of $\smv' \in \mathcal{S}(\pmv)$ such that $\smv$ and $\smv'$ differ in the first $t$ locations. Such $\smv'$ can be expressed as 
\begin{align}
\smv' = \smv + \underline{\delta} = [w^{(1)}+\delta^{(1)}, w^{(2)}+\delta^{(2)}, \ldots, w^{(t)}+\delta^{(t)}, w^{(t+1)}, \ldots, w^{(\LL)}]. 
\end{align}  
Since $\smv \cdot \hat{G}_{\LL \times l_2} = \pmv$ (recall that $\hat{G}_{\LL \times l_2}$ is defined in~\eqref{eq:hat}) and
\begin{align}
\smv' \cdot \hat{G}_{\LL \times l_2} = (\smv + \underline{\delta})\cdot \hat{G}_{\LL \times l_2} = \smv \cdot \hat{G}_{\LL \times l_2} + \underline{\delta} \cdot \hat{G}_{\LL \times l_2} =\pmv,
\end{align} 
we obtain $\underline{\delta}\cdot \hat{G}_{\LL \times l_2} = \underline{0}$. Let $\tilde{\underline{\delta}}$ be a length-$t$ vector consisting of the first $t$ element in $\underline{\delta}$, and $\tilde{G}_{t \times l_2}$ be the first $t$ rows of $\hat{G}_{\LL \times l_2}$, \emph{i.e.},
\begin{equation}\tilde{G}_{t \times l_2} \triangleq
\begin{bmatrix}
1 & 1 & \dots  & 1 \\
\mu_{\LL+1} & \mu_{\LL+2}  & \dots  & \mu_{L} \\
\vdots & \vdots & \ddots & \vdots \\
\mu_{\LL+1}^{t-1} & \mu_{\LL+2}^{t-1} &  \dots  & \mu_{L}^{t-1}
\end{bmatrix}.
\end{equation}
Our goal is to calculate the size of $\{\tilde{\underline{\delta}}: \tilde{\underline{\delta}} \cdot \tilde{G}_{t \times l_2} = \underline{0} \text{ and } \delta^{(i)} \ne 0, \forall i \in \{1,2,\ldots, t\}  \}$. To do so, we treat the matrix $\tilde{G}_{t \times l_2}$ as a {\it parity-check matrix} of a linear code $\tilde{C}$. One can verify the linear code $\tilde{C}$ is a Maximum Distance Separable (MDS) code since the length $l(\tilde{C}) = t$, the dimension $k(\tilde{C})= t - l_2$, and the minimum distance $d_{min}(\tilde{C})= l_2 + 1$ (by noting that any $l_2$ rows are linearly independent). For any MDS code, the weight distribution of $\tilde{C}$ is 
\begin{align}
A_i = \binom{l(\tilde{C})}{i} (|\mathbb{F}|-1) \sum_{j=0}^{i-d_{min}(\tilde{C})}(-1)^j \binom{i-1}{j} |\mathbb{F}|^{i-d_{min}(\tilde{C}) - j}, 
\end{align} 
where $A_i$ is the number of codewords in $\tilde{C}$ of Hamming weight $i$ ($d_{min}(\tilde{C}) \le i \le l(\tilde{C})$). By substituting $l(\tilde{C}) = t$, $d_{min}(\tilde{C}) = l_2+1$, $|\mathbb{F}| = n^{\rin}$, and $i = t$, we obtain
\begin{align}
\Big\{\tilde{\underline{\delta}}: \tilde{\underline{\delta}} \cdot \tilde{G}_{t \times l_2} = \underline{0} \text{ and } \delta^{(i)} \ne 0, \forall i \in \{1,2,\ldots, t\}  \Big\} = (n^{\rin}-1) \sum_{j=0}^{t-l_2-1}(-1)^j \binom{t-1}{j} n^{\rin(t-j-l_2-1)}. \label{eq:fina}
\end{align}
It is worth noting that the above analysis only considers for a fixed $\smv \in \mathcal{S}(\pmv)$, the number of $\smv'$ that differs from $\smv$ in the first $t$ coordinates. We still need to multiply~\eqref{eq:fina} by $\binom{l_1}{t}$ (the number of different subsets of $\{1,2,\ldots, l_1\}$ of size $t$) and $\nu$ (the number of systematic inner-message vectors that belong to $\mathcal{S}(\pmv)$). Therefore, for a fixed $\pmv$, the number of  $(\smv,\smv')$ satisfying $\smv \in \mathcal{S}(\pmv)$, $\smv' \in \mathcal{S}(\pmv)$ have distance $t$ ($t \ge l_2+1)$ equals
\begin{align}
\nu \cdot \binom{l_1}{t}(n^{\rin}-1)\sum_{i=0}^{t-l_2-1}(-1)^i \binom{t-1}{i}\left(n^{\rin}\right)^{t-i-l_2-1} \ \text{ if } t \ge l_2 + 1.
\end{align}

%%%%%%%%%%%%%%%%%%%%%%%%%%%%%%%%%%%%%%%%%%%%%%%%%%%%%%%%%%%%%%%%%%%%%%%%%%%%%%%%%%%%%%%%%%%%%%%%%%%%%%%%%%%%

\section{} \label{app:max3}
In this Appendix we show that
\begin{align}
\I(\x^{(i)}_{\tilde{w}}; \yb^{(i)})+\mathbb{D}(\x^{(i)}_{\tilde{w}} \parallel \rho) = \rho (1-2\pb) \log{\left(\frac{1-\pb}{\pb}\right)}+\mathcal{O}\left(n^{-1/2}(\log n)^{-1/3}\right), \label{eq:chun}
\end{align}
when $(\rho *\pb)(1-\Delta^y_{1})\le f^y_{1} \le (\rho *\pb)(1+\Delta^y_{1}), f^{xy}_{10} \in \rho \pb (1 \pm \Delta^{xy}_{10})$ and $f^{xy}_{11} \in \rho (1-\pb)(1 \pm \Delta^{xy}_{11})$, where $\Delta^y_{1} = n^{-1/4+\delta/2}$ and $\Delta^{xy}_{10} = \Delta^{xy}_{11} = (\log n)^{-1/3}$. By applying Taylor's series expansion with center at $(\rho * \pb, \rho \pb, \rho (1-\pb))$, the empirical mutual information $\I(\x^{(i)};\yb^{(i)})$ equals\footnote{Note that the second and higher order derivative terms are bounded by $\mathcal{O}\left(n^{-1/2}(\log n)^{-1/3}\right)$.} 
\begin{align}
\I(\x^{(i)}_{\tilde{w}}; \yb^{(i)}) &= \I(\x^{(i)}_{\tilde{w}}; \yb^{(i)})\big|_{(\rho * \pb, \rho \pb, \rho (1-\pb))} + (f^y_{1}-\rho *\pb) \frac{\partial \I(\x^{(i)}_{\tilde{w}}; \yb^{(i)})}{\partial f^y_{1}}\bigg|_{(\rho * \pb, \rho \pb, \rho (1-\pb))} \notag \\
& + (f^{xy}_{10}-\rho \pb) \frac{\partial \I(\x^{(i)}_{\tilde{w}}; \yb^{(i)})}{\partial f^{xy}_{10}}\bigg|_{(\rho * \pb, \rho \pb, \rho (1-\pb))} + (f^{xy}_{11}-\rho (1-\pb)) \frac{\partial \I(\x^{(i)}_{\tilde{w}}; \yb^{(i)})}{\partial f^{xy}_{11}}\bigg|_{(\rho * \pb, \rho \pb, \rho (1-\pb))} + \mathcal{O}\left(n^{-1/2}(\log n)^{-1/3}\right).
\end{align}
The value of $\I(\x^{(i)}_{\tilde{w}}; \yb^{(i)})$ at the center point $(\rho * \pb, \rho \pb, \rho (1-\pb))$ equals
\begin{align}
\I(\x^{(i)}_{\tilde{w}}; \yb^{(i)})|_{(\rho * \pb, \rho \pb, \rho (1-\pb))} &= \left(\mathbb{H}(\yb^{(i)}) - \mathbb{H}(\yb^{(i)}|\x^{(i)}_{\tilde{w}}) \right)\Big|_{(\rho * \pb, \rho \pb, \rho (1-\pb))}  \\
&= \mathbb{H}(\rho * \pb)-\mathbb{H}(\pb) \notag \\
&= \mathbb{D}(\pb \parallel \rho * \pb) + \rho (1-2\pb) \log \left(\frac{1-\rho * \pb}{\rho * \pb}\right)  \label{eq:jan_12_1} \\
&= \rho (1-2\pb) \log \left(\frac{1-\pb+\mathcal{O}(n^{-1/2})}{\pb+\mathcal{O}(n^{-1/2})}\right) + \mathcal{O}(n^{-1}) \label{eq:jan_12_2} \\
&= \rho (1-2\pb) \log \left(\frac{1-\pb}{\pb} \right) + \mathcal{O}(n^{-1}).  
\end{align} 
Equation~\eqref{eq:jan_12_2} follows from the Reverse Pinsker's inequality~\cite{berend2012reverse}, \emph{i.e.},
\begin{align}
\mathbb{D}(\pb \parallel \rho * \pb) \le \frac{\rho^2 (1-2\pb)^2}{2\pb (1-\pb) \ln 2} = \mathcal{O}(n^{-1}).
\end{align} 
Similar to Equations~\eqref{eq:partial_f*1}-\eqref{eq:partial_f11} in Appendix~\ref{app:max}, we obtain the partial derivatives of $\I(\x^{(i)}_{\tilde{w}};\yb^{(i)})$ in $f^y_{1}$, $f^{xy}_{10}$ and $f^{xy}_{11}$ as follows.
\begin{align}
\frac{\partial \I(\x^{(i)}_{\tilde{w}}; \yb^{(i)})}{\partial f^y_{1}} &= \log{\left(\frac{(f^y_{1}-f^{xy}_{11})(1-f^y_{1})}{f^y_{1}(1-f^y_{1}-f^{xy}_{10})}\right)},  \\
\frac{\partial \I(\x^{(i)}_{\tilde{w}}; \yb^{(i)})}{\partial f^{xy}_{10}} &= \log{\left(\frac{f^{xy}_{10}(1-f^{xy}_{10}-f^{xy}_{11})}{(f^{xy}_{10}+f^{xy}_{11})(1- f^y_{1}-f^{xy}_{10})}\right)}, \\
\frac{\partial \I(\x^{(i)}_{\tilde{w}}; \yb^{(i)})}{\partial f^{xy}_{11}} &= \log{\left(\frac{f^{xy}_{11}(1-f^{xy}_{10}-f^{xy}_{11})}{(f^{xy}_{10}+f^{xy}_{11})(f^y_{1}-f^{xy}_{11})}\right)}. 
\end{align}
The values of these partial derivatives centered at $(\rho * \pb, \rho \pb, \rho (1-\pb))$ are given by
\begin{align}
&\frac{\partial \I(\x^{(i)}_{\tilde{w}};\yb^{(i)})}{\partial f^y_{1}}\bigg|_{(\rho * \pb, \rho \pb, \rho (1-\pb))} = \log \left( \frac{\pb (1-\rho * \pb)}{(1-\pb)(\rho * \pb)}\right) = \mathcal{O}(n^{-1/2}), \\
&\frac{\partial \I(\x^{(i)}_{\tilde{w}};\yb^{(i)})}{\partial f^{xy}_{10}}\bigg|_{(\rho * \pb, \rho \pb, \rho (1-\pb))} = \log \left( \frac{\pb}{1-\pb}\right) = \mathcal{O}(1), \\
&\frac{\partial \I(\x^{(i)}_{\tilde{w}};\yb^{(i)})}{\partial f^{xy}_{11}}\bigg|_{(\rho * \pb, \rho \pb, \rho (1-\pb))} = \log \left( \frac{1-\pb}{\pb}\right) = \mathcal{O}(1).
\end{align}  
Since $\Delta^y_{1} = n^{-1/4+\delta/2}$ and $\Delta^{xy}_{10} = \Delta^{xy}_{11} = (\log n)^{-1/3}$, we have
\begin{align}
(f^y_{1}-\rho *\pb) \frac{\partial \I(\x^{(i)}_{\tilde{w}};\yb^{(i)})}{\partial f^y_{1}}\bigg|_{(\rho * \pb, \rho \pb, \rho (1-\pb))} &= \mathcal{O}(n^{-3/4+\delta/2}), \\
(f^{xy}_{10}-\rho \pb) \frac{\partial \I(\x^{(i)}_{\tilde{w}};\yb^{(i)})}{\partial f^{xy}_{10}}\bigg|_{(\rho * \pb, \rho \pb, \rho (1-\pb))} &=  \mathcal{O}\left(n^{-1/2}(\log n)^{-1/3}\right), \\
(f^{xy}_{11}-\rho (1-\pb)) \frac{\partial \I(\x^{(i)}_{\tilde{w}};\yb^{(i)})}{\partial f^{xy}_{11}}\bigg|_{(\rho * \pb, \rho \pb, \rho (1-\pb))} &= \mathcal{O}\left(n^{-1/2}(\log n)^{-1/3}\right).
\end{align}
Up to now, we have already proved the empirical mutual information 
\begin{align}
\I(\x^{(i)}_{\tilde{w}};\yb^{(i)}) = \rho (1-2\pb) \log{\left(\frac{1-\pb}{\pb}\right)}+\mathcal{O}\left(n^{-1/2}(\log n)^{-1/3}\right).
\end{align}
The last step is to show the empirical KL divergence $\mathbb{D}(\x^{(i)}_{\tilde{w}} \parallel \rho) = \mathcal{O}\left(n^{-1/2}(\log n)^{-1/3}\right)$. Since $f^{xy}_{10} = \rho \pb (1 \pm \Delta^{xy}_{10})$, $f^{xy}_{11} = \rho (1-\pb) (1 \pm \Delta^{xy}_{11})$ and $\Delta^{xy}_{10} = \Delta^{xy}_{11} = (\log n)^{-1/3}$, the Hamming weight of $\x^{(i)}_{\tilde{w}}$ falls into the range $\left[\rho \left(1 - \Delta^{xy}_{10}\right), \rho \left(1 + \Delta^{xy}_{10}\right)\right]$. One can show that 
\begin{align}
\mathbb{D}\left(\rho (1 + \Delta^{xy}_{10}) \parallel \rho\right) &= \left(1-\rho\left(1+\Delta^{xy}_{10}\right) \right)\log \left( \frac{1-\rho(1+\Delta^{xy}_{10})}{1-\rho}\right) + \rho\left(1+\Delta^{xy}_{10}\right) \log \left( \frac{\rho(1+\Delta^{xy}_{10})}{\rho}\right) \\
&= \left(1-\rho\left(1+\Delta^{xy}_{10}\right)\right) \log \left( 1 - \frac{\rho \Delta^{xy}_{10}}{1-\rho}\right) + \rho\left(1+\Delta^{xy}_{10}\right) \log \left(1+\Delta^{xy}_{10}\right) \\
&= \left(1-\rho\left(1+\Delta^{xy}_{10}\right)\right) \left( -\frac{\rho \Delta^{xy}_{10}}{1-\rho} - \frac{\left(\frac{\rho \Delta^{xy}_{10}}{1-\rho}\right)^2}{2} - \cdots \right) + \rho\left(1+\Delta^{xy}_{10}\right) \left( \Delta^{xy}_{10} - \frac{\left(\Delta^{xy}_{10}\right)^2}{2} + \cdots \right) \\
& = \mathcal{O}\left(n^{-1/2}(\log n)^{-1/3}\right).
\end{align} 
Similarly, $\mathbb{D}(\rho (1 - \Delta^{xy}_{10}) \parallel \rho)$ also scales as $\mathcal{O}\left(n^{-1/2}(\log n)^{-1/3}\right)$. Finally, when $f^{xy}_{10} = \rho \pb (1 \pm \Delta^{xy}_{10})$, $f^{xy}_{11} = \rho (1-\pb) (1 \pm \Delta^{xy}_{11})$, we have
\begin{align}
\mathbb{D}(\x^{(i)}_{\tilde{w}} \parallel \ \rho) \le \max \left\{\mathbb{D}(\rho (1 - \Delta^{xy}_{10}) \parallel \rho), \mathbb{D}(\rho (1 + \Delta^{xy}_{10}) \parallel \rho)\right\} = \mathcal{O}\left(n^{-1/2}(\log n)^{-1/3}\right).
\end{align}

\vskip 0.5cm
\bibliographystyle{IEEEtran}
\bibliography{ISIT2016_long}
\end{document}